 \documentclass[preprint2]{aastex}

\newcommand{\mqd}{M\,82}
\newcommand{\etal}{{\em et al.\,}}
\newcommand{\micr}{$\,\mu$m}
\newcommand{\brg}{Br$\gamma$}
\newcommand{\hh}{$\rm H_{2}$}
\newcommand{\hebrg}{\ion{He}{1}~2.058/\brg}
\newcommand{\hebrdix}{\ion{He}{1}~1.701/Br10}
\newcommand{\water}{$\rm H_{2}O$}
\newcommand{\waterph}{$H_{2}O_{\rm ph}$}
\newcommand{\coph}{$CO_{\rm ph}$}
\newcommand{\ewcoa}{$W_{2.29}$}
\newcommand{\ewcoh}{$W_{1.62}$}
\newcommand{\ewsi}{$W_{1.59}$}
\newcommand{\cohsi}{$\log (W_{1.62}/W_{1.59})$}
\newcommand{\cohcok}{$\log (W_{1.62}/W_{2.29})$}
\newcommand{\loglkllyc}{$\log (L_{K}/L_{\rm Lyc})$}
\newcommand{\dilh}{$D_{1.6}$}
\newcommand{\dilk}{$D_{2.3}$}
\newcommand{\masssol}{$\rm ~M_{\odot}$}
\newcommand{\teff}{$T_{\rm eff}$}
\newcommand{\teffob}{$T_{\rm eff}^{\rm OB}$}
\newcommand{\lbol}{$L_{\rm bol}$}
\newcommand{\lir}{$L_{\rm IR}$}
\newcommand{\llyc}{$L_{\rm Lyc}$}
\newcommand{\lk}{$L_{K}$}
\newcommand{\mlk}{$M^{\star}/L_{K}$}
\newcommand{\mdyn}{$M_{\rm dyn}$}
\newcommand{\snrate}{$\nu_{\rm SN}$}

\slugcomment{Accepted for publication in the Astrophysical Journal}
\shorttitle{Near- and mid-infrared study of M\,82}
\shortauthors{F\"orster Schreiber \etal}

\begin{document}

\title{Near-infrared Integral Field Spectroscopy and Mid-infrared Spectroscopy
       of the Starburst Galaxy M\,82
  \footnote{Based on observations with {\em ISO\/}, an ESA project with
  instruments funded by ESA Member States (especially the PI
  countries: France, Germany, the Netherlands, and the United Kingdom)
  and with the participation of ISAS and NASA. The SWS is a joint
  project of SRON and MPE.}
}

\author{N. M. F\"orster Schreiber\altaffilmark{1}}
\affil{Max-Planck-Institut f\"ur extraterrestrische Physik,
       Postfach 1312, D-85741 Garching, Germany \\
   and CEA/DSM/DAPNIA/Service d'Astrophysique,
       C.E. Saclay, F-91191 Gif sur Yvette CEDEX, France}
\email{forster@strw.leidenuniv.nl}
\author{R. Genzel, D. Lutz, D. Kunze}
\affil{Max-Planck-Institut f\"ur extraterrestrische Physik,
       Postfach 1312, D-85741 Garching, Germany}
\email{genzel@mpe.mpg.de, lutz@mpe.mpg.de, dietmar.kunze@otn.lfk.dasa.de}
\and
\author{A. Sternberg}
\affil{School of Physics and Astronomy, Tel Aviv University,
       Ramat Aviv, Tel Aviv 69978, Israel}
\email{amiel@wise.tau.ac.il}

\altaffiltext{1}{Current address: Leiden Observatory,
PO Box 9513, 2300 RA Leiden, The Netherlands}

\begin{abstract}

We present new infrared observations of the central regions of the
starburst galaxy \mqd.  The observations consist of near-infrared
integral field spectroscopy in the $H$- and $K$-band obtained with
the MPE 3D instrument, and of $\rm \lambda = 2.4 - 45~\mu m$
spectroscopy from the Short Wavelength Spectrometer (SWS)
on board the {\em Infrared Space Observatory\/}. 
These measurements are used, together with data from the literature, to
(1) re-examine the controversial issue of extinction,
(2) determine the physical conditions of the 
interstellar medium (ISM) within the star-forming regions, and
(3) characterize the composition of the stellar populations.
Our results provide a set of constraints for detailed starburst
modeling which we present in a companion paper.

We find that purely foreground extinction cannot reproduce the global
relative intensities of H recombination lines from optical to radio
wavelengths.  A good fit is provided by a homogeneous mixture of
dust and sources, and with a visual extinction of $A_{V} = 52~{\rm mag}$.
The SWS data provide evidence for deviations from commonly assumed 
extinction laws between 3\micr\ and 10\micr.
The fine-structure lines of Ne, Ar, and S detected with SWS imply an
electron density of $\approx~300~{\rm cm^{-3}}$, and abundance
ratios Ne/H and Ar/H nearly solar and S/H about one-fourth solar.
The excitation of the ionized gas indicates an average effective temperature
for the OB stars of 37400~K, with little spatial variation across the starburst
regions.  We find that a random distribution of closely packed gas clouds
and ionizing clusters, and an ionization parameter of $\approx 10^{-2.3}$
represent well the star-forming regions on spatial scales ranging from
a few tens to a few hundreds of parsecs.
From detailed population synthesis and the mass-to-$K$-light ratio,
we conclude that the near-infrared continuum emission across the starburst
regions is dominated by red supergiants with average effective temperatures
ranging from 3600~K to 4500~K, and roughly solar metallicity.  
Our data rule out significant contributions from older,
metal-rich giants in the central few tens of parsecs of \mqd.

\end{abstract}


\nopagebreak

\section{INTRODUCTION} \label{Sect-intro}

\mqd\ is one of the nearest (3.3~Mpc; \citealt{Fre88};
$\rm 1^{\prime\prime} = 16~pc$), brightest, and best-studied starburst galaxies.
It has long been considered as the archetype of this class of objects,
and has often been used as test laboratory for starburst theories
(see {\em e.g.} \citealt{Tel88} and \citealt{Rie93} for reviews).
The central 500~pc of \mqd\ harbours the most active starburst 
regions and has thus received particular attention in the past.
Most of \mqd 's infrared luminosity of $3 \times 10^{10}~{\rm L_{\odot}}$
originates from this ``starburst core,'' which is severely obscured
at optical and ultraviolet wavelengths.

The qualitative picture of \mqd\ features the following.  A prominent
nucleus, a stellar disk, and a kiloparsec-long stellar bar are
revealed by near-infrared observations 
\citep[{\em e.g. \rm}][]{Tel91, McL93, Lar94}.  The molecular gas resides
mainly in a rotating ring or tightly wound spiral arms and in an inner spiral 
arm at radii of $\rm \approx 400~pc$ and $\rm \approx 125~pc$, respectively
\citep*[{\em e.g. \rm}][]{She95, Sea98, Nei98}.
The \ion{H}{2} regions are concentrated in a smaller rotating ring-like
structure of radius $\rm \approx 85~pc$, and along the stellar bar at
larger radii \citep[{\em e.g. \rm}][]{Lar94, Ach95}.
{\em HST\/} observations have resolved over a hundred compact 
and luminous ``super star clusters'' across the central kiloparsec
\citep*{OCo95, Gri00}.
An important series of young compact radio supernova remnants extends along
the galactic plane over 600~pc \citep*[{\em e.g. \rm}][]{Kro85, Mux94, Ped99}
and a bipolar outflow traces a starburst wind out to several kiloparsecs
\citep*[{\em e.g. \rm}][]{Bre95, Sho98, Leh99, Cap99}.

The triggering of starburst activity in \mqd\ is generally
attributed to tidal interaction with its massive neighbour M\,81
$\sim 10^{8}$ years ago or, alternatively, to the stellar bar
which may itself have been induced by the interaction 
\citep*[{\em e.g. \rm}][]{Got77, Lo87, Yun93, Yun94, Tel91, Ach95}.
Beginning with the seminal paper by \citet{Rie80}, several authors
have applied evolutionary synthesis modeling to understand the nature and
evolution of starburst activity in \mqd\ 
\citep[{\em e.g. \rm}][]{Ber92, Rie93, Doa93, Sat97}.
However, despite extensive studies, crucial issues remain open 
concerning in particular the composition of the stellar population
and spatial variations thereof, the initial mass function of the
stars formed in the starburst, and the spatial and temporal evolution
of starburst activity.

In order to address the above issues quantitatively, we have obtained 
new observations of the central regions of \mqd\ consisting of
near-infrared $H$- and $K$-band integral field spectroscopy with the
Max-Planck-Institut f\"ur extraterrestrische Physik (MPE) 3D instrument 
\citep{Wei96}, and mid-infrared spectroscopy from the Short
Wavelength Spectrometer \citep[SWS;][]{deG96} on board the
{\em Infrared Space Observatory\/} \citep[{\em ISO\/\rm; }][]{Kes96}.
The 3D data provide {\em detailed information on small spatial scales\/}
from key features tracing the stellar populations and the interstellar
medium (ISM) while the SWS data cover the {\em entire\/} 
$\rm \lambda = 2.4 - 45~\mu m$ range containing a wealth of additional
starburst signatures.  The 3D and SWS data allow us to apply a variety 
of new and essential diagnostics which we use, together with results 
from the literature, to address several controversial issues and reveal
additional aspects of \mqd.

In this paper, we focus on characterizing the physical conditions
of the ISM (gas density, abundances, extinction), the composition of
the stellar populations (hot massive stars and cool evolved stars),
and the relative distribution of the gas clouds and stellar clusters.
In a companion paper (F\"orster Schreiber {\em et al.} 2000,
in preparation; hereafter paper 2),
we will apply recent starburst models to our results to constrain
quantitatively the star formation parameters and the
detailed starburst history in the central regions of \mqd.

The paper is organized as follows. 
Section~\ref{Sect-obs} describes the observations and data reduction
procedure and section~\ref{Sect-res} presents the results.
The physical conditions of the ISM are derived in 
section~\ref{Sect-neb_analysis} together with the composition of the
massive star population, while the composition of the cool evolved
stellar population is investigated in section~\ref{Sect-pop_synthesis}.
Section~\ref{Sect-conclu} summarizes and discusses the results of our analysis.

\section{OBSERVATIONS AND DATA REDUCTION} \label{Sect-obs}

\subsection{Near-infrared Data} \label{Sub-NIRobs}

We observed \mqd\ at near-infrared wavelengths using the MPE integral field
spectrometer 3D \citep{Wei96}.  A first data set was obtained at the
3.5~m telescope in Calar Alto, Spain, on 1995 January 13, 14, 16, and 21.
The observations were completed at the 4.2~m William-Herschel-Telescope
in La Palma, Canary Islands, on 1996 January 6.
3D slices the focal plane into 16 parallel ``slits''
which are imaged and dispersed in wavelength by a grism onto
a 256$\times$256~HgCdTe~NICMOS~3 array, providing simultaneously
the entire $H$- or $K$-band spectrum of each spatial pixel.
Nyquist-sampled spectra are achieved by dithering the spectral
sampling by half a pixel on alternate data sets, which are
interleaved in wavelength after the data reduction.
For the \mqd\ observations, the instrument setup provided a pixel
scale of $\rm 0.5^{\prime\prime}\,pixel^{-1}$ with a field of view of
$8^{\prime\prime} \times 8^{\prime\prime}$, and a spectral resolution
of $R \equiv \lambda/\Delta\lambda \approx 1015$ and $R \approx 830$
in the $H$- and $K$-band, respectively.

In order to sample representative starburst regions in \mqd,
we selected for the observations an area approximately 
$\rm 260~pc \times 160~pc$ (corresponding to 
$16^{\prime\prime}\times 10^{\prime\prime}$), nearly parallel to the
plane of the galaxy.  The regions mapped include the nucleus and
extend to the west up to the inner edge of the molecular ring
(see figure~\ref{fig-3DSWSFOV}).
The 3D raster consists of four different fields, with adjacent 
fields overlapping by $\approx 3^{\prime\prime}$.
Each field was observed in an {\em object-sky-sky-object\/} sequence,
with the off-source frames taken on blank portions
of the sky $2^{\prime} - 4^{\prime}$ away in the north and south directions.
The total on-source integration times per field and wavelength channel
were $8 - 15$ and 10 minutes for the $H$- and $K$-band data, respectively.
Typical single-frame exposures were $\rm 60 - 100~s$.
For atmospheric calibration, we observed B-, late-F, or early-G
dwarf stars each night, before and/or after \mqd.
The seeing varied from 1\arcsec\ to 1.5\arcsec\ during the observations.
Table~\ref{tab-3Dlog} summarizes the log of the observations.

We carried out the data reduction using the 3D data analysis package,
within the GIPSY environment \citep{vdH92}.
We first performed correction for the non-linear signal response of the
detector, dark current and background subtraction, spatial
and spectral flat-fielding, wavelength calibration, rearrangement
of the data in a three-dimensional cube, and bad pixel correction
following standard procedures as described by \citet{Wei96}.
We reduced the reference stars data in the same manner.
We then corrected the \mqd\ data cubes for atmospheric transmission 
by division with calibration spectra obtained as follows.

In the $K$-band, we first ratioed the stars spectra with a black-body
curve of the appropriate temperature given their spectral type.
We removed the intrinsic stellar features by division with the
normalized spectrum of the \ion{G3}{5} star from the \citet{KH86}
atlas, convolved to the resolution of 3D.  This template has similar
absorption line strengths as our \ion{F5}{5} and \ion{G0}{5} calibrators
except for the \brg\ line, which we removed by linear interpolation.
In the $H$-band, no appropriate template spectra were digitally available,
so we constructed composite calibration spectra as follows.  We generated
synthetic transmission spectra between 1.55\micr\ and 1.75\micr\ 
at the proper zenithal distance using the program ATRAN \citep{Lor92}.
At the edges of the $H$-band, the transmission is very sensitive
to the actual atmospheric conditions.  Since our reference stars
have no important features bluewards of 1.55\micr\ and redwards of
1.75\micr\ (within the range observed with 3D), we ratioed these portions
of their spectrum with the appropriate blackbody curve and connected
them to the synthetic spectra.

The residuals from intrinsic features of the reference stars
are less than 1\%.  Those from telluric features do not exceed
$3\% - 4\%$ for most of the $H$- and $K$-band but amount up to
20\% between 1.9\micr\ and 2.0\micr\ and at both edges of the 
$H$-band due to spatial and temporal variability of the
atmospheric transmission.
By fine-tuning the airmass for the calibration spectra
with the help of ATRAN, we reduced the spurious features
to less than 1\% in the $K$-band and middle of the $H$-band,
and less than 10\% at the edges of the $H$-band.

For proper mosaicking, we smoothed the resulting data cubes with
a two-dimensional gaussian profile to a common spatial resolution of
1.5\arcsec.  The small scale structure is thus smeared out in the higher
resolution fields but no important spatial features are lost.
We combined the data cubes to produce the final mosaic, adjusting
the mean broad-band flux in the overlapping areas to a common level.
We performed absolute flux calibration based on the broad-band
photometric measurements of \citet{Rie80}.
We estimate the uncertainties on the absolute fluxes
to be 15\% in the $H$-band and 10\% in the $K$-band.  They
include possible systematic errors in the relative spatial and spectral
flux distribution which may occur in the background subtraction,
in the correction for telluric absorption, and in the mosaicking
(see also section~\ref{Sub-NIRimages}).

\subsection{Mid-infrared Data} \label{Sub-MIRobs}

We obtained the mid-infrared spectrum of \mqd\ with the SWS 
\citep{deG96} on board {\em ISO\/} \citep{Kes96} during revolution 116
on 1996 March 12.  The grating scan mode AOT SWS01 was used to cover
the entire SWS range from 2.4\micr\ to 45\micr.  The slowest scan speed
was selected to obtain the highest spectral resolution possible in this
mode ($R \sim 500 - 1000$).  In addition, individual lines were scanned
with the grating line profile mode AOT SWS02 to get full resolving power 
($R \sim 1000-2000$) and improved sensitivity for the key lines used in
the data interpretation.

The SWS rectangular aperture was centered on the western mid-infrared 
emission peak (actual pointing at
$\alpha_{1950}$: $\rm 09^{h} 51^{m} 42\fs 20$, 
$\delta_{1950}$: $+69^{\circ} 55^{\prime} 00\farcs 74$). 
The major axis of the aperture was oriented at a position angle
of 64\fdg 5, nearly parallel to the galactic plane of \mqd.
The aperture size varies from $14^{\prime\prime} \times 20^{\prime\prime}$
at short wavelengths to $20^{\prime\prime} \times 33^{\prime\prime}$
at long wavelengths.  The SWS field of view thus includes the
regions mapped with 3D, extending out to a maximum distance of
about 350~pc from the nucleus (see figure~\ref{fig-3DSWSFOV}).
The SWS full scan took in total $\rm 1^{h} 50^{m}$.
The on-source integration time for the individual line scans varied 
between 100~s for most lines and 600~s for the weakest lines.

We reduced the data with the SWS Interactive Analysis package (SIA), using
special interactive extensions for glitch tail removal, dark subtraction, 
up-down correction, flat-fielding, and defringing.  As the combined spectra 
of the 12 detectors of each SWS band are oversampled, we rebinned 
the final data to the proper instrumental resolution.  We performed the
calibration using the calibration tables as of 1998 February 15, 
equivalent to OLP version 7.0.
The overall accuracy of the line and continuum fluxes is estimated
to be $\approx 10\% - 20\%$ \citep{Sch96, Sal97}.

\section{RESULTS} \label{Sect-res}

\subsection{Near-infrared Spectra} \label{Sub-NIRspec}

We selected three individual regions within the 3D field of view for a
detailed analysis: the nucleus, as well as two regions covering the \brg\
sources located approximately 10\arcsec\ and 5\arcsec\ southwest from the 
nucleus and which we designate as M82:Br1 and M82:Br2, respectively.
For convenience, we will refer to the latter regions as ``B1'' and ``B2''
in the rest of this paper.  B1 is close to the molecular ring
and B2 coincides with the western mid-infrared emission peak.  These three
positions were chosen because they sample, along the galactic plane of
\mqd, representative regions with different spectral properties indicative
of different stellar populations as described below.

We extracted the $H$- and $K$-band spectra of the individual 
regions from the 3D data cubes using synthetic apertures of
$\rm 2.25^{\prime\prime} \times 2.25^{\prime\prime}$,
corresponding to $\rm 35~pc \times 35~pc$.   
They are shown in figure~\ref{fig-3Dspec} together with 
those for the entire 3D field of view.
All spectra exhibit the typical signatures of starburst activity,
including H and He recombination lines tracing photoionized nebulae,
ro-vibrational lines from warm \hh, [\ion{Fe}{2}] transition lines 
originating primarily in iron-enriched shocked material, and numerous
atomic and molecular absorption features produced in the atmosphere
of cool stars.  However, the intensity of the
features relative to the continuum varies with position.
The emission lines become stronger along the sequence
$\rm Nucleus \rightarrow B2 \rightarrow B1$ while
the stellar absorption features show the opposite trend.

Table~\ref{tab-3Ddata} summarizes various continuum and line measurements
obtained for the individual regions and for the 3D field of view.
We computed the $H$- and $K$-band flux densities by averaging over
the spectra of each region in the ranges $\rm \lambda = 1.52 - 1.78~\mu m$
and $\rm \lambda = 1.96 - 2.42~\mu m$, respectively.  Throughout this
paper, we will adopt the photometric system of \citet{Wam81}.

We measured the emission line fluxes by integrating over the line profile
after subtracting a linear continuum fitted to adjacent line-free spectral
regions.  Due to the substantial continuum structure, these intervals
were carefully selected by inspection of template spectra of K and M stars
from existing stellar atlases \citep*{KH86, OMO93, DBJ96, FS00a} so that
the linear interpolation adequately represented the underlying continuum.
We tested the validity of this procedure by integrating
the line fluxes after subtracting several template spectra
of stars within $\rm \pm~3$ spectral classes of the representative
type for each region (determined in section~\ref{Sect-pop_synthesis}).
We adopted the average fluxes, with uncertainties from the continuum
subtraction corresponding to one standard deviation of the
multiple measurements.  At the spectral resolution of 3D,
the Br11 and Br12 lines at 1.6807\micr\ and 1.6407\micr\ are blended with
the [\ion{Fe}{2}] lines at 1.6769\micr\ and 1.6435\micr, respectively.
In these cases, we fitted double gaussian profiles to the
features.  The emission lines contribute at most 1.5\% to the broad-band
flux densities for all four regions.

We measured the equivalent widths (EWs) of the \ion{Si}{1} absorption feature
at 1.59\micr, and of the $\rm ^{12}CO$\,(6,3) and $\rm ^{12}CO$\,(2,0)
bandheads at 1.62\micr\ and 2.29\micr\ (denoted hereafter
\ewsi, \ewcoh, and \ewcoa)
according to the definitions and corrections for resolution effects
given by \citet{FS00a}.  Such corrections were necessary
in order to compare the EWs for population synthesis purposes 
(section~\ref{Sect-pop_synthesis}) with stellar data from existing
libraries given for $R \sim 1600$ and $R \sim 2500$ in the 
$H$- and $K$-band, respectively.  
The Br14 emission line partially fills the \ion{Si}{1}
feature in all spectra.  In order to remove this contamination, 
we scaled the profile of the Br13 line 
according to the intrinsic ratio $\rm Br14/Br13 = 0.80$ 
and subtracted it at the position of Br14.  This ratio is
for case B recombination \citep{Hum87},
assuming the electron density and temperature
determined in section~\ref{Sect-neb_analysis} 
($n_{\rm e} = 300~{\rm cm^{-3}}$ and $T_{\rm e} = 5000~{\rm K}$).
Differential extinction between these lines can be neglected
because they are so close in wavelength.

\subsection{Near-infrared Images} \label{Sub-NIRimages}

We obtained maps of the broad-band and line emission as well as
of the EWs from the 3D data cubes by applying to each pixel the
procedure described above for the spectra. 
Figure~\ref{fig-3Dimages} presents selected images and ratio maps.

The spatial distributions of the $H$- and $K$-band emission
obtained with 3D agree well with maps in the literature
\citep[{\em e.g. \rm}][]{Die86, Tel91, Lar94, Sat95}.
The emission is generally centrally concentrated
about the nucleus and small-scale structure is apparent, in particular 
the so-called ``secondary peak'' 8\arcsec\ west from the nucleus.
We assessed the accuracy of the flux calibration and mosaicking
by comparing the $H$- and $K$-band flux densities measured at different
positions between the 3D data and the large-scale images at 1\arcsec\
resolution from \citet{Sat97}, which were obtained with a single
detector array.  The relative flux densities between various regions
agree typically to 15\% in the $H$-band and 10\% in the $K$-band.
The differences may be due in part to the different spatial resolution
of the data sets.

The \brg\ and \ion{He}{1}~2.058\micr\ line emission exhibit clumpy
morphologies and follow each other very well.  The ratio map,
however, reveals spatial variations in the \hebrg\ line ratio.
The contrast between the distribution of the ionized gas and broad-band
emission is well delineated by the logarithmic map of the ratio of
$K$-band to Lyman continuum luminosities, \loglkllyc
\footnote{
For the $K$-bandpass $\rm \lambda = 1.9 - 2.5~\mu m$ \citep{Wam81},
$L_{K} = 1.87 \times 10^{19} D^{2} f_{K}$ where \lk\ is in
$\rm L_{\odot}$ (taken as $\rm 3.85 \times 10^{26}~W$),
$D$ is the distance in Mpc, and $f_{K}$ is the flux 
density in $\rm W\,m^{-2}\,\mu m^{-1}$.  For case B recombination 
\citep{Hum87} with $n_{\rm e} = 300~{\rm cm^{-3}}$,
$T_{\rm e} = 5000~{\rm K}$, and an average Lyman continuum photon energy 
of 15~eV, $L_{\rm Lyc} = 4.68 \times 10^{22} D^{2} F_{\rm Br\gamma}$ where 
\llyc\ is in $\rm L_{\odot}$ and the \brg\ line flux $F_{\rm Br\gamma}$ is 
in $\rm W\,m^{-2}$.}.
The \brg\ image obtained with 3D agrees well with previously published
maps \citep*{Wal92, Lar94, Sat95}.  Table~\ref{tab-Brgcomp} compares
the \brg\ line flux integrated in circular apertures centered on the 
nucleus with measurements from the literature.  The discrepancies
between the various results are very large: up to a factor of 3.5.
They are probably attributable to uncertainties in absolute calibration
and continuum subtraction.  In addition, positioning errors and differences
in spatial resolution for linemaps could lead to appreciably different
fluxes given the clumpy morphology of the emission.  We have however
no obvious explanation in the case of measurements at higher spectral
resolution from well-registered linemaps, as for \citet{Lar94} and
\citet{Sat95}.  Both the spectral resolution and coverage as well as the
quality of the 3D spectra allow a better continuum subtraction than in
previous studies, supporting the accuracy of our \brg\ fluxes.

The \ewcoa\ and \ewcoh\ maps reveal very deep absorption features around the
nucleus and the ``secondary peak,'' and progressively shallower features at
B2 and B1.  The \cohcok\ map is rather uniform but indicates an enhancement
of \ewcoh\ relative to \ewcoa\ around B1.  The \ewcoa\ image is consistent
with the CO index map obtained at low spectral resolution by \citet{Sat97}
within the regions observed with 3D.

The images demonstrate that the nucleus and B1
have the most extreme properties within the 3D field of view.
The 3D maps and spectra indicate important spatial variations on
relatively small scales in the composition of the stellar population,
and support that the most recent star formation activity 
--- as traced by the \brg\ emission for example ---
has taken place outside the nucleus, as noted previously 
\citep[{\em e.g. \rm}][]{Tel91, McL93, Sat97}.

\subsection{Mid-infrared Spectra} \label{Sub-MIRspec}

Figure~\ref{fig-SWS01} shows the SWS full scan spectrum of \mqd.
Several H recombination lines from the Brackett, Pfund, and Humphreys
series are detected as well as pure rotational lines from \hh\ and numerous
fine-structure lines from various atoms mostly in low ionization stages.
Broad emission features commonly attributed to polycyclic aromatic
hydrocarbons (PAHs), a broad dip centered near 10\micr, and a rising 
continuum at $\rm \lambda \gtrsim 10~\mu m$ likely due to very small dust 
grains are conspicuous \citep*[{\em e.g. \rm}][]{Gil75, Wil77, 
Leg89, All89, Des90, Roc91}.  
According to \citet{Stu00}, the minimum around 10\micr\ may not be
due entirely to absorption by interstellar silicate grains as is
usually assumed; the presence of strong flanking PAH emission complexes
superposed over a weak continuum may contribute to the apparent dip.
The higher signal-to-noise (S/N) ratio line scans are plotted in 
figure~\ref{fig-SWS02}.

We measured the emission line fluxes from both the
full scan SWS01 spectrum and the individual line scan SWS02 spectra.
We fitted gaussian profiles to the features after subtraction of the
continuum baseline obtained by fitting a line to adjacent line-free portions
of the spectrum.  Table~\ref{tab-SWSdata} gives the line fluxes.  In order
to compare the data obtained with the different SWS apertures, we also scaled 
the fluxes of the H recombination and ionic fine-structure lines to the
smallest $14^{\prime\prime} \times 20^{\prime\prime}$ aperture.  We estimated
the scaling factors from the [\ion{Ne}{2}] 12.8\micr\ map of \citet{Ach95};
they are 0.8 and 0.7 for the $14^{\prime\prime} \times 27^{\prime\prime}$ 
and $20^{\prime\prime} \times 33^{\prime\prime}$ apertures, respectively, 
with 10\% uncertainty.  The Br$\alpha$ 4.051\micr, [\ion{Ar}{3}] 8.99\micr,
and [\ion{S}{4}] 10.51\micr\ maps obtained by these authors
exhibit morphologies similar to that of the [\ion{Ne}{2}] emission,
justifying the use of the same scaling factors for all the hydrogen and
fine-structure lines considered.  No beam-size scaling was applied to
the \hh\ lines because of the lack of information on the spatial
distribution of the \hh\ emission.  At the resolution of the SWS,
the Br$\beta$ line (2.625\micr) is blended with the \hh~1-0~$O$(2)
line at 2.626\micr.  However, from the strength of the
\hh~1-0~$Q$(3) at 2.423\micr,
we estimate that the \hh~1-0~$O$(2) line contributes at most 30\% to the
measured flux \citep{Bla87, Ste89}.

\section{NEBULAR ANALYSIS OF M\,82} \label{Sect-neb_analysis}

In this section, we present our nebular analysis of \mqd.
We first derive the physical parameters critical for the
interpretation of the nebular line emission: 
extinction, electron density, gas-phase abundances,
and ionization parameter appropriate for the star-forming regions.
We then use photoionization models to constrain the average 
effective temperature of the OB stars.

\subsection{Interstellar Extinction} \label{Sub-Av_gas}

The issue of extinction towards the starburst regions of \mqd\ has long
been controversial.  In particular, extinctions in visual magnitudes 
$A_{V}$ ranging from a few to about 15~mag have been obtained under
the assumption of a uniform foreground screen model or from optical and
near-infrared diagnostics, while $A_{V} \approx 20 - 60~{\rm mag}$ have
been inferred for a mixed model or from diagnostics at longer wavelengths
(see {\em e.g.} \citealt{McL93}, \citealt{Sat97}, and references therein).
Such differences may substantially affect the derived intrinsic properties
(fluxes, luminosities).  For instance, the correction factors at 2.2\micr\
range from 1.6 to 4.1 assuming a uniform foreground dust screen with 
$A_{V} = 5 - 15~{\rm mag}$, and from 2.2 to 5.7 assuming a mixed model
with $A_{V} = 20 - 60~{\rm mag}$ (see below for the computation of the
correction factors).
The discrepancies in extinction estimates can be understood in view of the
large uncertainties in beam-size corrections involved in several studies, 
of those in the interpretation of some diagnostics (such as the broad dip
around 10\micr\ discussed above), of the spatially non-uniform extinction
across the disk of \mqd, and of the large optical depths preventing
radiation at shorter wavelengths to escape from the most obscured regions
\citep[{\em e.g. \rm}][]{Pux91, Tel91, McL93, Lar94, Sat95, Sat97, Stu00}.

\subsubsection{Global Extinction towards the Ionized Gas} 
               \label{Sub-global_Av_gas}

One major hindrance in previous studies has been the lack of data in the
mid-infrared regime.  {\em ISO\/}-SWS observations have now filled this gap,
with a consistent set including several H recombination lines
between $\rm 3\,\mu m$ and $\rm 10\,\mu m$.
H recombination lines are excellent ``standard candles'' for extinction
determinations because their intrinsic line emissivities are well determined
theoretically.  Moreover, the derived extinction parameters are fairly
insensitive to the choice of electron density and temperature since the
relative line emissivities vary slowly with these properties.  For instance,
the results from \citet{Hum87} imply that the emissivities of the
lines considered below change by 5\% on average (18\% at most) between
$n_{\rm e} = 10^{2}~{\rm cm^{-3}}$ and $\rm 10^{4}~cm^{-3}$, for
$T_{\rm e} = 5000~{\rm K}$.  They vary by 20\% on average (33\% at most)
between $T_{\rm e} = 5000~{\rm K}$ and $\rm 10^{4}~K$, for
$n_{\rm e} = 10^{2}~{\rm cm^{-3}}$.  The intrinsic Lyman continuum photon
emission rates $Q^{0}_{\rm Lyc}$ derived from the dereddened line fluxes depend
very weakly on $n_{\rm e}$, and vary only slightly with $T_{\rm e}$ since the
total H recombination coefficient $\alpha_{\rm B} \propto T_{\rm e}^{-0.81}$.
For the examples above, $Q^{0}_{\rm Lyc}$ increases on average 
by $\approx 5\%$ and 25\%, respectively.

We combined our SWS data with H line measurements from the radio to the
optical regimes, obtained mainly in large apertures (diameter 
$\gtrsim 20^{\prime\prime}$), and with appropriate beam-size corrections.
Since the SWS and larger apertures include the most prominent emission
regions (see the linemaps of \citealt*{McC87}, \citealt{Sat95},
\citealt{Ach95}, and \citealt{Sea96}), the derived extinction
should be representative for the bulk of ionized gas in \mqd.
A large wavelength coverage is essential for discriminating between 
various dust and sources geometries, since deviations from a simple uniform 
foreground screen model are perceptible only for diagnostics probing 
appreciably different optical depths. 

In the range $\rm \lambda \sim 1 - 10~mm$, we used
measurements for H26$\alpha$, H27$\alpha$, H41$\alpha$ 
\citep{Sea96}, H30$\alpha$ \citep*{Sea94},
H40$\alpha$ and H53$\alpha$ \citep{Pux89}.
The integrated fluxes for large apertures are consistent with predominantly
optically thin, spontaneous emission in local thermodynamical equilibrium
(LTE; see references above).  We excluded data at centimeter wavelengths,
which are potentially affected by stimulated emission and free-free 
absorption \citep*[{\em e.g. \rm}][]{Sea85, Sea96}.
At near-infrared wavelengths, we used the Brackett line fluxes 
measured with 3D.  These were complemented with the Pa$\beta$
measurements from \citet{McL93} and \citet{Sat95},
averaging the scaled fluxes together because of the large
discrepancy between the two results (as for Br$\gamma$ discussed
in section~\ref{Sub-NIRimages}).
Finally, at optical wavelengths, we used the H$\alpha$ measurement
of \citet{McC87}.  We obtained an additional estimate from the
H$\alpha$ + [\ion{N}{2}]\,$\lambda\lambda 6548, 6584$~\AA\
flux of \citet*{You88}, assuming a uniform 
[\ion{N}{2}]/H$\alpha$ ratio of 0.5 \citep{McC87}.

We scaled the line fluxes to match the SWS
$14^{\prime\prime} \times 20^{\prime\prime}$ beam in two steps.
The fluxes were first scaled to a 30\arcsec --diameter aperture centered
on the nucleus.  The beam-size corrections for the millimeter lines were
inferred from those derived by \citet{Sea94} between 19\arcsec,
21\arcsec, and 41\arcsec\ apertures and the entire emission region, with
an uncertainty of 15\%.  Those for the H$\alpha$ measurements were
estimated from the map of \citet{McC87}, with 40\% uncertainty.
For \brg\ and Pa$\beta$, we applied a scaling $\propto \theta^{1.5}$
(where $\theta$ is the aperture diameter)
derived from the \brg\ data of \citet{Sat95} in 3.8\arcsec, 8\arcsec,
and 30\arcsec\ apertures, with 10\% uncertainty.  Because of the weakness
of the $H$-band Brackett lines within $\approx 5^{\prime\prime}$ of the
nucleus, we used the fluxes integrated over the entire 3D map multiplied
by the ratio of \brg\ fluxes scaled to a 30\arcsec\ beam and measured
in the 3D field of view.  In the second step, a scaling factor
$30^{\prime\prime} \rightarrow {\rm SWS}$ of 0.5 was applied,
with 10\% uncertainty, as derived by comparing various continuum and line
fluxes from the SWS data with results reported in the literature, all of 
which presumably trace the same sources since they have similar spatial 
distributions; the references include \citet{Kle70}, \citet{Rie72}, 
\citet{Gil75}, \citet{Wil77}, \citet{Hou84}, \citet*{Tel93}, and \citet{Ach95}.
Table~\ref{tab-HIfluxes} lists the observed and scaled line fluxes.

We derived the extinction by least-squares fitting to
\begin{equation}
\frac{F_{\lambda}}{F_{\rm ref}} = 
\left(\frac{j_{\lambda}}{j_{\rm ref}}\right) 
\left(\frac{X_{\lambda}}{X_{\rm ref}}\right),
\label{Eq-fit}
\end{equation}
where $F_{\lambda}/F_{\rm ref}$ are the observed line 
fluxes relative to that of a reference line, 
$j_{\lambda}/j_{\rm ref}$ are the intrinsic ratios of line
emissivities, and $X_{\lambda}$ and $X_{\rm ref}$ are the attenuation factors
due to extinction at the wavelengths of the lines considered.  The SWS lines
constituting the largest, self-consistent data set, we chose Br$\alpha$
as the reference line.  We took the intrinsic line emissivities from
\citet{Hum87} for case B recombination with
$n_{\rm e} = 100~{\rm cm^{-3}}$ and $T_{\rm e} = 5000~{\rm K}$, 
appropriate for \mqd\ as shown in section~\ref{Sub-neTe}
\footnote{The electron density is actually in the range 
$\rm \sim 10 - 500~cm^{-3}$, but we used the tabulated values for 
$\rm 100~cm^{-3}$ since the relative line emissivities depend weakly on 
$n_{\rm e}$.}.
We adopted the extinction laws of \citet{Dra89} and 
\citet{Rie85} at infrared and optical wavelengths, respectively.
At millimeter wavelengths, the extinction can be neglected.

We considered two representative model geometries.  For a uniform
foreground screen of dust (``UFS''), the attenuation is given by
\begin{equation}
X_{\lambda}^{\rm UFS} = \frac{I_{\lambda}}{I^{0}_{\lambda}} = 
e^{-\tau_{\lambda}},
\label{Eq-UFS}
\end{equation}
where $I^{0}_{\lambda}$ and $I_{\lambda}$ are the intensities of the
incident and emergent radiation, and $\tau_{\lambda}$ is the
optical depth of the obscuring material.  For a mixed model (``MIX'')
consisting of a homogeneous mixture of dust and sources,
\begin{equation}
X_{\lambda}^{\rm MIX} = \frac{I_{\lambda}}{I^{0}_{\lambda}} = 
\frac{1 - e^{-\tau_{\lambda}}}{\tau_{\lambda}}.
\label{Eq-MIX}
\end{equation}
Here, $I^{0}_{\lambda}$ is the total intrinsic line intensity produced within
the mixed medium, and $\tau_{\lambda}$ is the total optical depth of this
medium.  The extinction in magnitudes is related to the optical depth through
$A_{\lambda} = 1.086\,\tau_{\lambda}$.

The best fit for each model is achieved with a total visual 
extinction of $A_{V}^{\rm UFS} = 4_{-4}^{+1.5}~{\rm mag}$ 
and $A_{V}^{\rm MIX} = 43 \pm 23~{\rm mag}$.
We estimated the uncertainties on $A_{V}$ from the
reduced chi-squared diagrams, with a $1 \sigma$ error
corresponding to a factor of $e^{1/2}$ from the minimum $\chi_{n}^{2}$.
Table~\ref{tab-Av} compares the best-fit extinction-corrected 
line fluxes relative to Br$\alpha$ for each geometry 
(columns labeled ``Draine'') with the intrinsic line
ratios.  Figure~\ref{fig-Avfits} illustrates the results
(plots labeled ``Draine''), as the $Q^{0}_{\rm Lyc}$ derived from each
of the extinction-corrected line fluxes assuming an average Lyman continuum
photon energy of 15~eV.  The average values of $Q^{0}_{\rm Lyc}$ are
given in table~\ref{tab-Av}.

Good extinction models are those for which equal values of $Q^{0}_{\rm Lyc}$
are inferred from each of the extinction-corrected line fluxes.
Figure~\ref{fig-Avfits} demonstrates that purely foreground extinction
provides a much less satisfactory fit to the data over the entire wavelength
range considered, as also shown by {\em e.g.} \citet{Pux91} and 
\citet{McL93}.  The best fit for this geometry results in a minimum
$\chi_{n}^{2} \approx 8$ compared to $\chi_{n}^{2} \approx 2$ for the 
mixed model.  For the latter geometry, the extinction corrections are
of 4.8 near 2\micr\ and 1.6 near 5\micr, with uncertainties of
$\approx 50\%$ and $\approx 20\%$, respectively.

Increasing $n_{\rm e}$ to $\rm 10^{4}~cm^{-3}$ or $T_{\rm e}$ to
$\rm 10^{4}~K$ results in larger $A_{V}$'s, but within the $1 \sigma$ 
uncertainties for the nominal case.  The implied values of 
$Q^{0}_{\rm Lyc}$ increase by about 15\% and 50\%, respectively.
We also performed fits excluding the H$\alpha$ measurements, which may
include a component from light escaping along the minor axis scattered
by dust grains \citep[{\em e.g. \rm}][]{OCo78, Not85}.
The H$\alpha$ flux could thus be overestimated relative to the
data at longer wavelengths, since the scattering efficiency
of interstellar dust grains generally decreases rapidly with 
increasing $\lambda$ \citep[{\em e.g. \rm}][]{Eme88}.  The results for
the mixed model are little affected, with $A_{V}^{\rm MIX} = 44~{\rm mag}$
(a substantially larger $A_{V}^{\rm UFS} = 9~{\rm mag}$ is obtained,
as expected since the dust and sources are actually mixed).
Fits excluding the millimeter lines yield
$A_{V}^{\rm MIX} = 18~{\rm mag}$, illustrating the 
importance of including unobscured lines.

\subsubsection{The $\mathit{\lambda = 3 - 10~\mu m}$ Extinction Law} 
               \label{Sub-Av_law}

Despite the good overall fit for the mixed model,
figure~\ref{fig-Avfits} indicates some deviations in the 
$\rm \lambda = 3 - 10~\mu m$ region.  The constraints imposed by the
millimeter lines prevent reduction of these deviations by a change
in absolute level of extinction.  The extinction law assumed
for these wavelengths may not be appropriate however.  Indeed, 
until recently, the $\rm 3 - 10~\mu m$ extinction law was
poorly determined because of the difficulties inherent to
ground-based observations and because the properties of the template
sources accessible so far in this range were not well-known.  The SWS
has now provided observations of numerous nebular H recombination lines
between 3\micr\ and 10\micr\ in a variety of objects.
These lines have been used to investigate the extinction law
in the direction of the Galactic Center \citep{Lut99}. The
``Galactic Center extinction law'' (hereafter simply GC law) lacks the 
pronounced minimum in the $\rm 4 - 8~\mu m$ region expected for standard
graphite-silicate dust mixtures which are usually assumed 
({\em e.g.} \citealt{Dra89} and references therein),
suggesting additional contributors to the extinction.

The SWS line fluxes in \mqd\ are much better reproduced if the GC law
is adopted.  The best fits are obtained with
$A_{V}^{\rm UFS} = 4^{+2}_{-3}~{\rm mag}$ and
$A_{V}^{\rm MIX} = 52 \pm 17~{\rm mag}$.
The results are given in table~\ref{tab-Av} and
figure~\ref{fig-Avfits} (columns and plots labeled ``GC'').
The fit for the uniform foreground screen model is still much poorer
than for the mixed geometry ($\chi_{n}^{2} \approx 8$ and $\approx 1$,
respectively).  The extinction corrections for the mixed model are 
5.8 near 2\micr\ and 2.6 near 5\micr, with uncertainties of
$\approx 30\%$ and 25\%, respectively.  Although the data do
not allow the accurate determination of the extinction law in \mqd, they 
provide evidence for deviations from the commonly used Draine law similar
to those found towards the Galactic Center given by \citet{Lut99}.

\subsubsection{Adopted Parameters} \label{Sub-adopted_Av_gas}

In view of the above analysis, we will adopt the mixed model with
$A_{\it V} = 52~{\rm mag}$ throughout this paper as representative
of the global extinction towards the bulk of the ionized gas in \mqd.
The Lyman continuum photon emission rate for a 30\arcsec --diameter aperture
is twice that for the SWS field of view (section~\ref{Sub-global_Av_gas}),
$Q^{0}_{\rm Lyc} = 1.23 \times 10^{54}~{\rm s^{-1}}$.
Our results are consistent with the $Q^{0}_{\rm Lyc}$ inferred
from the millimeter thermal free-free emission ($\rm \sim 10^{54}~s^{-1}$;
{\em e.g.} \citealt{Car91, Sea96}), and confirm the
results of \citet{McL93} who found $A^{\rm MIX}_{V} = 55~{\rm mag}$
and $Q^{0}_{\rm Lyc} = 1.05 \times 10^{54}~{\rm s^{-1}}$.
We will also adopt the GC extinction law of \citet{Lut99} 
between 3\micr\ and 10\micr, keeping the \citet{Dra89} law
for the other relevant infrared ranges and the \citet{Rie85}
law at optical wavelengths.  \citet{Sat95} demonstrated the
validity for \mqd\ of the Draine law at near-infrared wavelengths
($A_{\lambda} \propto \lambda^{-1.75}$), or of similar extinction laws	
({\em e.g.} \citealt{Lan84}, with $A_{\lambda} \propto \lambda^{-1.85}$).

\subsubsection{Local Extinction towards the Ionized Gas} 
               \label{Sub-local_Av_gas}

From the spatially resolved Brackett line emission obtained with 3D, 
we derived the extinction towards individual regions as described 
in section~\ref{Sub-global_Av_gas} using \brg\ as the reference line.
The results for selected regions are reported in table~\ref{tab-Avlocal}, 
together with the $Q^{0}_{\rm Lyc}$ computed from the \brg\ fluxes (the 
most accurately measured line), corrected for $A_{V}^{\rm MIX}$.
Due to the weakness of the $H$-band lines over significant areas,
we could only generate a partial extinction map, from the 3D linemaps
rebinned to $1^{\prime\prime} \times 1^{\prime\prime}$ pixels to 
increase the S/N ratio.

The variations in $A_{V}$ between individual regions and in our
partial extinction map agree well with those seen in the
extinction map of \citet{Sat95} obtained from \brg\ and
Pa$\beta$ measurements.  The lower $A_{V}^{\rm MIX}$ for the
3D field of view compared to the global extinction derived above probably
reflects the optical depth limitations of the near-infrared diagnostics.
For the same reasons, the derived $A_{V}^{\rm MIX}$ and $A_{V}^{\rm UFS}$
imply similar extinction corrections; for example, the differences are
35\% or less near 2\micr\ for the selected regions, and $\lesssim 20\%$ on
the scale of the rebinned pixels over the valid regions.  The uncertainties
on the extinction corrections are typically $20\% - 40\%$ for both models.
As will be discussed in section~\ref{Sub-logU}, the ionized nebulae are
likely mixed with the molecular gas and dust clouds even on scales of a few
tens of parsecs.  We will therefore adopt the results for the mixed model.

\subsection{Physical Conditions of the ISM} \label{Sub-ISM}

\subsubsection{Electron Temperature and Density} \label{Sub-neTe}

From the ratios of H recombination lines with the (thermal) continuum flux 
density in the millimeter regime, a $T_{\rm e}$ near 5000~K has been
derived for the starburst core of \mqd\ \citep{Pux89, Car91, Sea94, Sea96}.
From the results of \citet{Sea96}, based on maps of the H41$\alpha$
line emission and of the underlying continuum, we inferred that this
temperature is appropriate on smaller spatial scales for most of the
regions observed with 3D.

The SWS spectrum provides three density-sensitive ratios:
[\ion{S}{3}] 18.7\micr/33.5\micr,
[\ion{Ne}{3}] 15.6\micr/36.0\micr, and
[\ion{Ar}{3}] 8.99\micr/21.8\micr.
All three are fairly insensitive to $T_{e}$ in the range $\rm 5000 - 20000~K$.
The [\ion{S}{3}] ratio is the most reliable one because both lines are amongst
the strongest in the SWS spectrum.  It is also the most sensitive at low
$n_{\rm e}$ because the upper levels of the transitions have the lowest 
critical densities.  For the $14^{\prime\prime} \times 20^{\prime\prime}$
SWS aperture, the dereddened suplhur, neon, and argon ratios are
$0.71 \pm 0.23$, $9.0 \pm 4.2$, and $15.5 \pm 9.3$, respectively.
The uncertainties include those of the line flux measurements (from 
the relative flux calibration and continuum subtraction), and of the 
beam-size and extinction corrections.  Comparison with results
of computations of collisional excitation \citep[{\em e.g. \rm}][]{Ale99}
shows that the measured ratios lie in the low-density limit, indicating 
$n_{\rm e} \sim 10 - 600~{\rm cm^{-3}}$.

Our result is similar to the average $n_{\rm e}$ in large apertures
obtained previously by various authors using different
infrared and radio diagnostics, and assuming a single-density
model \citep{Hou84, Duf87, Sea85, Sea96, Col99}.
We also considered the [\ion{O}{3}] 52\micr/88\micr\ ratio,
which is more sensitive to $n_{\rm e}$ at low densities.
The ratio measured by \citet{Duf87} in a 48\arcsec --diameter
aperture, similar to the value for an 80\arcsec -- aperture of
\citet{Col99}, indicates $n_{\rm e} \sim 50 - 500~{\rm cm^{-3}}$.
We will adopt a value of $\rm 300~cm^{-3}$.

\subsubsection{Gas-phase Abundances} \label{Sub-abund_gas}

We determined the gas-phase abundances of Ne, Ar, and S, three of the most 
abundant heavy elements in \ion{H}{2} regions, using the lines detected
with the SWS.  These have critical densities for collisions with electrons at
$T_{\rm e} = 5000~{\rm K}$ in the range $\sim 10^{4} - 10^{6}~{\rm cm^{-3}}$
so that collisional de-excitation can be neglected.  Assuming a ``one-layer''
model with uniform density and temperature, the ionic abundances can be
computed from
\begin{equation}
\frac{F_{\lambda ({\rm X}^{+i})}}{F_{\lambda ({\rm H}^{+})}} =
\frac{n_{({\rm X}^{+i})} n_{\rm e} j_{\lambda ({\rm X}^{+i})}}
     {n_{(\rm H^{+})} n_{\rm e} j_{\lambda ({\rm H}^{+})}},
\label{Eq-abund}
\end{equation}
where $F_{\lambda ({\rm X}^{+i})}$ and
$F_{\lambda ({\rm H}^{+})}$ are the fluxes of
the ionic line of interest and of a reference H recombination line,
$n_{({\rm X}^{+i})}$ and $n_{(\rm H^{+})}$ are the densities of ions 
$\rm X^{+i}$ and $\rm H^{+}$, and $j_{\lambda ({\rm X}^{+i})}$ and
$j_{\lambda ({\rm H}^{+})}$ are the line emissivities.
In \ion{H}{2} regions, H is nearly completely ionized so that
$n_{\rm H^{+}} \approx n_{\rm H}$.  We computed the fine-structure line
emissivities in the low-density limit using the effective collisional
strengths from \citet*{Joh86}, \citet{Sar94}, \citet{But94}, 
\citet{Pel95}, and \citet*{Gal95}.  We  took Br$\alpha$ as reference
H line, with its emissivity from \citet{Hum87}.

Table~\ref{tab-abund} gives the data and results.
For ionizing stars with effective temperatures in the range
$35000~{\rm K} - 40000~{\rm K}$, as found in section~\ref{Sub-TeffOB} below,
the Ne, Ar, and S are expected to be mostly in the ionization stages
observed with SWS.  The elemental abundances are thus well approximated
by the sum of the ionic abundances determined here.
The abundances are nearly solar or slightly above for Ne and Ar, and
about one-fourth solar for S.  Similar underabundances for S have been
found in Galactic \ion{H}{2} regions \citep[{\em e.g. \rm}][]{Sim95}
and in some extragalactic starburst systems \citep[{\em e.g. \rm}][]{Gen98},
and are attributed to depletion of S onto interstellar dust grains.

Most abundance determinations for \mqd\ in the literature
indicate no large depletions or enhancements
for most elements compared to the solar neighbourhood composition 
({\em e.g.} \citealt{Gaf92}; \citealt{McL93}
and references therein; \citealt{Ach95, Lor96}).
The exception is Si, for which \citeauthor{Lor96} found a gas-phase
abundance three times larger than in Galactic nebulae, and
which they interpret as probably due to partial destruction
of silicate grains by fast supernova-driven shocks.

\subsection{Ionization Parameter} \label{Sub-logU}

For the purpose of photoionization modeling, we represented \ion{H}{2} regions
as thin gas shells surrounding central, point-like stellar clusters
(see top of figure~\ref{fig-geom}).  In such ``central cluster'' models,
the nebular conditions are specified by the distance $R$ between the
ionizing cluster and the illuminated surface of the gas shell, the H
number density $n_{\rm H}$, and the ionization parameter $U$ defined as
\begin{equation}
U \equiv \frac{Q_{\rm Lyc}}{4\pi R^{2} n_{\rm H} c}
= \frac{\phi_{\rm Lyc}}{n_{\rm H} c},
\label{Eq-U}
\end{equation}
where $Q_{\rm Lyc}$ is the production rate of Lyman continuum photons
from the stars and $c$ is the speed of light.  $U$ thus gives the number of
Lyman continuum photons impinging at the surface of the nebula per H atom.
Since H is nearly completely ionized in \ion{H}{2} regions, 
and since He is not fully ionized in \mqd\ (section~\ref{Sub-TeffOB}),
we assumed $n_{\rm H} \approx n_{\rm e}$.

In such complex and distant systems as starburst galaxies, a 
large number of \ion{H}{2} regions may coexist in a relatively small 
volume and may not be individually resolved by the observations.
Thus, in constraining $U$ from the observed properties, the shell
geometry may not be directly applicable.
The spatial distribution of the gas relative to the sources is
crucial in determining the Lyman continuum photon flux $\phi_{\rm Lyc}$
impinging on the gas, and in deriving the {\em effective\/} $U$
(hereafter $U_{\rm eff}$) to model appropriately the nebular
emission in the framework of the idealized central cluster geometry.
Due to the lack of information on small enough spatial scales, the
ionization parameter is generally poorly determined in starburst galaxies.
\mqd\ is one exception owing to its proximity.
We have used our 3D and SWS data together with data from the
literature to constrain the degree of ionization of the nebulae within
the starburst regions.  Appendix~\ref{App-logU} gives the details of 
our derivation; in the following, we restrict ourselves to
outlining the main results.

The ISM properties within the starburst regions of \mqd\ 
\citep[{\em e.g. \rm}][]{Lor96} suggest it can be
represented by a collection of clouds with, on average, a molecular
core with radius $\rm \approx 0.5~pc$ shielded by a thin neutral atomic layer,
and an ionized layer extending out to $\rm \approx 1~pc$.  These clouds
have a mean separation of $\rm \approx 2 - 7~pc$.  High resolution optical 
and near-infrared imaging reveals large numbers of young, luminous clusters
throughout the starburst core \citep[{\em e.g. \rm}][]{OCo95, Sat97}.
Assuming the OB stars reside in such 
clusters and adopting a plausible cluster luminosity function consistent
with the properties of the optically-selected population, the intrinsic
Lyman continuum photon emission rates imply an average cluster separation
of $\rm 2 - 9~pc$.

The similar cloud-cloud and cluster-cluster mean separations suggests
that the clouds and clusters are well-mixed and uniformly distributed
throughout the regions of interest, as illustrated at the bottom of
figure~\ref{fig-geom}.  For such a random distribution model, the
flux of Lyman continuum photons impinging on the nebulae is reduced
compared to the central cluster geometry due to the increase in
surface area of the gas exposed to the radiation field.
From the analysis of selected regions in appendix~\ref{App-logU}, we
determine $\log U_{\rm eff} \approx -2.3~{\rm dex}$ and conclude that
it is representative of the local nebular conditions throughout all
of the star-forming regions in the central 500~pc of \mqd.  
We also show that $U_{\rm eff}$ can be expressed in terms of
the ratio of intrinsic Lyman continuum photon emission rate and of
the molecular gas mass, with
$U_{\rm eff} \propto Q_{\rm Lyc}/M_{\rm H_{2}}$
and the proportionality factor depending only on the gas cloud
properties (see equation [\ref{Eq-Ueff_phys}]).  
Since the cloud properties are similar throughout the starburst core
of \mqd, the comparable values of $\log U_{\rm eff}$ are interpreted
as resulting from comparable star formation efficiencies as measured
by $Q_{\rm Lyc}/M_{\rm H_{2}}$.

\subsection{Young Stellar Populations} \label{Sub-TeffOB}

The SWS and 3D data provide several diagnostics sensitive to the shape
of the ionizing radiation spectrum, dominated by OB stars in \mqd.
These include ratios of mid-infrared atomic fine-structure lines 
and of near-infrared He to H recombination lines, which probe the
$\rm 13~eV - 41~eV$ energy range.  The contribution from shock-ionized
material to the line emission considered below is not likely to be
important in \mqd\ \citep{McL93, Lut98}.  In the following,
we will assume that the lines originate entirely in gas photoionized by
the OB stars.  Table~\ref{tab-TeffOB} summarizes the data and results.

\subsubsection{Mid-infrared Fine-structure Line Ratios} \label{Sub-MIRTeffOB}

Amongst the diagnostic line ratios available from the SWS data set,
we considered [\ion{Ne}{3}] 15.6\micr/[\ion{Ne}{2}] 12.8\micr,
[\ion{Ar}{3}] 8.99\micr/[\ion{Ar}{2}] 6.99\micr, and
[\ion{S}{4}] 10.5\micr/[\ion{S}{3}] 18.7\micr\ to avoid
complications due to the uncertainties of the elemental abundances.

We modeled the variations of the line ratios with effective
temperature of the stars (\teffob) using the photoionization code
CLOUDY version C90.05 \citep{Ferl96} and the stellar atmosphere models
for solar-metallicity main-sequence stars of \citet{Pau98}.
We adopted the nebular parameters derived in the previous subsections:
solar gas-phase abundances, $n_{\rm e} = 300~{\rm cm^{-3}}$, and an
effective $\log U = -2.3~{\rm dex}$.
With this value for the ionization parameter, equation~(\ref{Eq-U}) implies
a corresponding $R$ of several tens to several hundreds of parsecs, depending
on the region. Since photoionization models are little sensitive to variations
in $R$ above $\rm \sim 10~pc$, we adopted a fixed $R = 25~{\rm pc}$.
We neglected the effects of dust grains possibly present {\em within\/} the
nebulae.  Dust grains mixed with the ionized gas are not expected to affect 
significantly the ionization equilibrium of \ion{H}{2} regions
because the variations of the dust absorption cross-section
with wavelength resembles closely that of H, peaking near 17~eV
\citep[{\em e.g. \rm}][]{Mat86}.  \citeauthor{Mat86} further argues 
that this conclusion is not altered for dust properties and 
gas-to-dust ratios characteristic of the Galactic and
Magellanic Clouds diffuse ISM or of denser \ion{H}{2} regions.

Figure~\ref{fig-TeffOB} shows the theoretical predictions and the measured
ratios, corrected for extinction and beam-size differences when appropriate.
The effects of variations of $n_{\rm e}$ and $\log U$, or of adopting
a gas and dust composition typical of the Orion nebula are also indicated.
The most sensitive parameter affecting the line ratios is the 
ionization parameter.  However, varying $\log U$ in the plausible 
range from $-2$ to $-2.5~{\rm dex}$ (see appendix~\ref{App-logU})
implies relatively small differences in \teffob:
$\leq \pm 1000~{\rm K}$ for the neon and argon ratios, and $\rm -2000~K$
or $+4000~{\rm K}$ for the sulphur ratio.  Given the uncertainties of the
data (line measurements, extinction and aperture corrections) as well as of
the models ({\em e.g.} nebular parameters, stellar atmospheres, atomic data),
the agreement between the results from the three diagnostic ratios is
satisfactory.
Our high quality and consistent data set confirms results obtained in the 
past from various infrared and millimeter lines, which ranged from
30000~K to 37000~K \citep{Gil75, Wil77, Pux89, McL93, Ach95, Col99}.

\subsubsection{Near-infrared He to H Recombination Line Ratios}
               \label{Sub-NIRTeffOB}

The strongest recombination lines from singly-ionized He detected 
in the 3D spectra correspond to the $2\,^{1}S - 2\,^{1}P$ triplet
transition at 2.058\micr\ and the $3\,^{3}P - 4\,^{3}D$ singlet 
transition at 1.701\micr\  
\footnote{We thank M. and G. Rieke for drawing our attention to
the \ion{He}{1} 1.701\micr\ line.}.
Together with the nearby \brg\ and Br10 lines, respectively, they provide
\teffob\ diagnostics sensitive in the range $\rm \lesssim 55000~K$
and very little affected by extinction.  The low-excitation SWS spectrum
and the non-detection of \ion{He}{2} lines in the 3D data (for example
at 2.189\micr) rule out the presence of important populations of
Wolf-Rayet stars or other sources that are hot enough to doubly ionize
He, which would complicate the interpretation of the ratios.

The \hebrg\ line ratio has been commonly used to estimate \teffob\ in
near-infrared studies of starburst galaxies and Galactic \ion{H}{2} regions
\citep*[{\em e.g. \rm}][]{DPJ92, Doh94, Doh95}.
However, due to sensitivity to local physical conditions
and to degeneracy in \teffob\ through resonance and collisional effects
\citep[{\em e.g. \rm}][]{Rob68, Cle87, Shi93},
this ratio alone is not sufficient for constraining \teffob.
On the other hand, the 1.701\micr\ line originates from a higher quantum
state in the $n\,^{3}P - n'\,^{3}D$ series and is essentially unaffected by
self-absorption and collisional effects but the \hebrdix\ ratio saturates
above $T_{\rm eff}^{\rm OB} \approx 40000~{\rm K}$ (see references above).
Therefore, the combination of the \hebrg\ and \hebrdix\ ratios
allows one to discriminate between the high- and low-\teffob\ regimes
(see {\em e.g.} \citealt{Van96} and \citealt{Doh95} for earlier
applications).

We modeled the He to H line ratios using CLOUDY and the same model
atmospheres and nebular parameters as for the mid-infrared line ratios.
CLOUDY computes the 2.058\micr\ line but not the 1.701\micr\ line which we
derived indirectly as explained below.  While models of \hebrg\ presented
by various authors in the past generally agree very well in the range
$T_{\rm eff}^{\rm OB} \leq 40000~{\rm K}$, the results at higher \teffob\
vary importantly \citep{DPJ92, Shi93, Doh94, Doh95, LRV94}.
In this regime, the ratio is particularly sensitive to the treatment
of the \ion{He}{1} Ly$\alpha$ opacity which is one of the major
sources of uncertainties in theoretical predictions.
\citet{Ferl99} discusses the revised treatment implemented
in the CLOUDY version we have used.

For \ion{He}{1}~1.701\micr, we used the \ion{He}{1}~4471\,\AA\ flux
predicted by CLOUDY.  Both lines originate from the $4\,^{3}D$
level and their fluxes are proportional to each other since triplet 
transitions from high $n\,^{3}D$ levels are essentially
unaffected by collisional effects or scattering from the
metastable $2\,^{3}S$ level.  We scaled the resulting 
\ion{He}{1}~4471\,\AA/Br10 curve so that its saturation value
for full He ionization equals that for \hebrdix\ given by \citet{Van96}.
While $n_{\rm e}$ is fixed in running CLOUDY, $T_{\rm e}$ varies
for each \teffob; we thus interpolated the relationship of
\citeauthor{Van96} in $T_{\rm e}$ as appropriate.
Since collisions are unimportant for both He lines involved here,
we used this relationship up to $n_{\rm e} = 10^{3}~{\rm cm^{-3}}$.

The ratios measured for selected regions are plotted against 
the model predictions in figure~\ref{fig-TeffOB}. 
The \hebrdix\ ratio clearly rules out the high-\teffob\ solutions
from the \hebrg\ ratio.  For the temperatures near 36000~K inferred, 
\hebrg\ and \hebrdix\ are little affected by variations in $n_{\rm e}$ 
or $\log U$ within plausible ranges for \mqd, or by modest changes 
in the gas and dust composition.  Again, in view of the uncertainties
of the data and models (in particular the continuum subtraction for the 
$H$-band lines), the \teffob 's inferred for each region from both
ratios agree satisfactorily.

The variations in our \hebrg\ map
(figure~\ref{fig-3Dimages}) support a general though small
increase in \teffob\ from the nucleus to larger projected
radii along the galactic plane of \mqd\ to the west. 
Such radial variations have been suggested by \citet{Sat95}
from the ratio of \brg\ to 3.29\micr\ PAH feature emission.
\citet{McL93} also suggested such a trend from the lower 
temperatures derived from mid-infrared diagnostics compared to those
inferred from optical diagnostics assuming the former trace
the innermost stellar population while the latter, foreground
and thus outermost clusters.
On the other hand, \citet{Ach95} concluded that
\teffob\ is roughly constant across the starburst core 
from the lack of clear spatial variations in the excitation state
from their Br$\alpha$, [\ion{Ne}{2}] 12.8\micr, [\ion{Ar}{3}] 8.99\micr, and
[\ion{S}{4}] 10.5\micr\ maps.

We constrained quantitatively the spatial variations in \teffob\ across the
3D field of view from our \ion{He}{1} 2.058\micr\ and \brg\ linemaps, rebinned 
to $1^{\prime\prime} \times 1^{\prime\prime}$ pixels.  We neglected
extinction effects, but this introduces errors $\lesssim 5\%$ in the ratios.
The resulting \teffob 's vary from 33200~K to 37500~K, with an average of
35700~K and a relatively small dispersion of $1\sigma = 650~{\rm K}$.
The 3D data indicate therefore a roughly constant \teffob\
for the hot massive stars across the regions observed,
with only a marginal gradient with projected radius.

\subsubsection{Additional Remarks} \label{Sub-NOB}

From our data sets, the neon ratio is probably the most reliable indicator
for the absolute \teffob.  Both lines involved are strong in \mqd\ and, among
the three mid-infrared line pairs, their proximity in wavelength minimizes
most extinction effects and they were observed through the same aperture.
It is also the most sensitive of all our diagnostics, probing the largest 
range in ionizing energy (between 22~eV and 41~eV).  In addition, the inferred 
\teffob\ is little affected by the uncertainties on the physical conditions 
within the nebulae.  The \hebrg\ and \hebrdix\ are potentially less reliable
indicators for the absolute \teffob\ because directly sensitive to the He 
abundance.  On the other hand, for the ranges observed in \mqd, these ratios
are little affected by modeling uncertainties and are thus robust diagnostics
for the relative variations in \teffob\ (assuming negligible gradients in He 
abundance).  Since the smaller SWS aperture and the 3D field of view cover the
same most prominent sources, the nebular line emission from both data sets
traces essentially the same stellar populations.  In the rest of this work,
we will adopt the result from the neon ratio (37400~K) for the 3D and 
SWS fields of view and will apply a correction of $\rm +1400~K$ to the 
temperatures inferred from \hebrg, corresponding to the difference in 
\teffob\ obtained from the neon ratio and that for the 3D field of view.

From the temperature scale of \citet*{Vac96}, the inferred \teffob 's
within the 3D and SWS fields of view correspond to \ion{O8.5}{5} stars,
with $Q^{\rm O8.5\,V}_{\rm Lyc} = 10^{48.72}~{\rm s^{-1}}$.
Given the roughly constant nebular excitation and the SWS aperture
including half of the integrated emission in a 30\arcsec --diameter
region centered on the nucleus of \mqd, this spectral type is probably
representative of the dominant OB stars for the starburst core as well.
The number of equivalent \ion{O8.5}{5} stars required to produce the 
intrinsic Lyman continuum luminosity in various regions is given in 
table~\ref{tab-TeffOB}.

\section{STELLAR POPULATION SYNTHESIS OF M\,82} \label{Sect-pop_synthesis}

In this section, we apply population synthesis to our 3D data.
We constrain the spectral type and luminosity class of the evolved stars, 
and investigate their metallicity.  We also constrain the contribution
from additional continuum sources (hot dust, OB stars, nebular 
free-free and free-bound processes) as well as the extinction towards
the evolved stars.  All results are reported in table~\ref{tab-NIRcont}.

\subsection{Analysis of Selected Absorption Features} \label{Sub-EWs}

\subsubsection{The Giants-supergiants Controversy} \label{Sub-IvsIII}

The nature of the evolved stellar population in \mqd\ has long
been debated.  Various diagnostics have been used in the past,
including the near-infrared broad-band colours, the \coph\ and \waterph\
photometric indices measuring the depth of the CO bandheads longwards
of 2.3\micr\ and of the \water\ absorption feature at 1.9\micr, spectral
synthesis in the range $\rm 2.18~\mu m - 2.28~\mu m$, and measurements
of the ratio of stellar mass to intrinsic $K$-band luminosity \mlk\
\citep*{Wal88, Les90, Gaf92, Gaf93, McL93, Lan96}.
Evolutionary synthesis models have also been applied to obtain
indirect constraints \citep{Rie80, Rie93, Sat97}.

However, no consensus has been reached yet, in particular
concerning the nucleus: some of the above studies indicate the
presence of young supergiants while others provide evidence for
old, metal-rich giants as dominant sources of the near-infrared
continuum emission.  Possible causes for these discrepancies
include positioning uncertainties and dependence on aperture
size due to important spatial variations in the indicators,
difficulties inherent to measurements of \waterph\ (in a range of poor
atmospheric transmission), and the weakness of several absorption features.
More importantly, the diagnostics used so far exhibit degeneracy in
temperature and luminosity, and are affected by extinction and 
featureless continuum emission.

The 3D data allow us to apply alternative diagnostic tools.
The CO bandheads at 2.29\micr\ and 1.62\micr\ together with the
\ion{Si}{1} feature at 1.59\micr\ are particularly useful in stellar
population studies from moderate-resolution near-infrared spectra
\citep{OMO93, OOKM95, FS00a}.
They suffer much less from measurements uncertainties and provide
sensitive indicators for the effective temperature and luminosity class
of cool stars.  Moreover, their EWs (\ewcoa, \ewcoh, and \ewsi)
are independent of extinction and provide a means of constraining the
contribution, or ``dilution,'' from featureless continuum emission sources
without requiring any assumptions on their nature and physical properties.

In the following analysis, we neglect possible contributions from 
thermally-pulsing AGB stars (such as Mira variables and N-type carbon 
stars) since none of their extreme, characteristic features 
\citep[{\em e.g. \rm}][]{Joh70, Lan99}
are seen in the 3D spectra.  Moreover, evolutionary synthesis models
show that these stars never produce more than $10\% - 40\%$ of the 
integrated near-infrared light of stellar clusters of any age 
({\em e.g.} \citealt{BC93}; paper 2).

\subsubsection{Selected Regions} \label{Sub-EWs_regions}

We analyzed the EWs using the diagnostic diagrams proposed by
\citet{OMO93} and \citet{OOKM95}.  These are shown in figures~\ref{fig-3DEWs}
and \ref{fig-dilution}, where the stellar data has been obtained from
existing relevant libraries (as compiled by \citealt{FS00a}). The horizontal
bars indicate the measurements for selected regions in \mqd; those for the
3D field of view are omitted for clarity, but are very close to those for B2.
Figure~\ref{fig-3DEWs} gives the effective temperature (\teff) and
luminosity class implied by the EWs.  Figure~\ref{fig-dilution} allows
the determination of the amount of dilution near 1.6\micr\ (\dilh)
from the vertical displacement relative to the locus of stars
in the \ewcoh\ versus \cohsi\ diagram, and near 2.3\micr\ (\dilk)
from the horizontal displacement in the \ewcoh\ versus \cohcok\
diagram once \ewcoh\ is corrected for dilution.  As shown
by \citet{OOKM95}, undiluted composite stellar populations
fall on the distributions defined by the stars in these plots.

For the central 35~pc at the nucleus of \mqd, both \ewcoh\ and \cohsi\ indicate
an average effective temperature in the range $\rm 3600 - 4100~K$, implying
negligible dilution around 1.6\micr.  The \ewcoa\ is characteristic of
either giants with $T_{\rm eff} = 3200 - 3600~{\rm K}$ or supergiants with
$T_{\rm eff} = 3600 - 4200~{\rm K}$.  Therefore, the EWs can only be
reconciled for a population of supergiants and negligible dilution in both 
$H$- and $K$-band.  Using the temperature calibration of \citet{SK82},
the corresponding average spectral type is \ion{K5}{1}.
For the other selected regions, degeneracy in luminosity class and
dilution complicates the interpretation of the EWs.  In all cases,
\ewcoh\ and \cohsi\ imply negligible dilution near 1.6\micr.  For B1,
the data indicate dominant \ion{K3}{3} stars and $D_{2.3} \approx 10\%$ 
or \ion{K2}{1} stars and $D_{2.3} \approx 10\% - 40\%$.  For B2 and the 
3D field of view, the EWs are more consistent with K3-K4 supergiants and 
small amounts of dilution near 2.3\micr, but undiluted emission from K4 
giants cannot be completely ruled out.  However, we favoured the supergiants 
solutions on the basis of the ratio of stellar mass to intrinsic $K$-band
luminosity \mlk, as explained below.

\subsubsection{Spatially Detailed Analysis} \label{Sub-EWs_pixels}

We applied a similar analysis to all the regions observed with 3D, 
using maps of the \ewsi, \ewcoh, and \ewcoa\ generated from the data 
cubes rebinned to $1^{\prime\prime} \times 1^{\prime\prime}$ pixels.
We corrected the \ewsi\ map for contamination by the Br14 emission line
as described in section~\ref{Sub-NIRspec}.
The resulting spectroscopic indices for all pixels are
plotted in the diagnostic diagrams of figure~\ref{fig-dilution}.

Together with the EW maps from figure~\ref{fig-3Dimages},
figure~\ref{fig-dilution} reveals variations in the intrinsic composition
of the evolved stellar population on small spatial scales, little dilution
around 1.6\micr, and variable dilution around 2.3\micr.  
The inferred \teff 's range from 4500~K down to 3600~K,
corresponding to spectral types G9 to M0 for supergiants.
The average is 4000~K with dispersion of $1\sigma = 200~{\rm K}$,
equivalent to K4 $\pm$ two spectral sub-classes assuming supergiants.
Several regions lie on the locus of supergiants in the \ewcoh\ versus
\cohcok\ diagram; they also coincide with the brightest $K$-band sources.
Others are characterized by too small \ewcoa\ relative to \ewcoh\ compared
to normal evolved stars, implying significant dilution near 2.3\micr\ up to
$\approx 50\%$ assuming an intrinsic population of supergiants ($\approx 20\%$
for giants).  Because dilution in the $H$-band is negligible, the \ewcoh\
map constitutes essentially a \teff\ map for the evolved stars, showing
spatial variations that are more complex than simple radial gradients.
The coolest populations are found around the nucleus and along a ridge
extending up to the secondary $K$-band peak ($\approx 8^{\prime\prime}$ to
the west).  Just south from this ridge, the \teff\ increases progressively 
along the Nucleus $\rightarrow$ B2 $\rightarrow$ B1 sequence.

For most individual regions, the \cohcok\ does not allow the
discrimination between giants and supergiants due to degeneracy in 
luminosity class and dilution.  They correspond predominantly to the smoother
low-surface brightness regions in the 3D broad-band maps.  The analysis
of the \mlk\ ratio provides an additional constraint.  From the stellar
mass derived in appendix~\ref{App-constraints}, the \mlk\ ratio within the 
starburst core of \mqd\ is very low ($\approx 1.4~{\rm M_{\odot}/L_{\odot}}$ 
and $\approx 0.5~{\rm M_{\odot}/L_{\odot}}$ for the central 35~pc 
and 500~pc, respectively)
\footnote{We remind the reader that we are using the definition
$L_{K} [{\rm L_{\odot}}] = 
1.87 \times 10^{19}\,(D~[{\rm Mpc}])^{2}\,
(f_{K}~[{\rm W\,m^{-2}\,\mu m}])$
for the $K$-band covering $\rm \lambda = 1.9 - 2.5~\mu m$ in the
photometric system of \citet{Wam81}, and with
$\rm L_{\odot} = 3.85 \times 10^{26}~W$.}.
From the 3D maps and the data of \citet{Sat97}, we estimate that
the faint smooth emission component represents about 75\% of the total 
intrinsic \lk\ within the central 30\arcsec.  This implies an upper limit of
$M^{\star}/L_{K} \approx 0.6~{\rm M_{\odot}/L_{\odot}}$ for the corresponding
population assuming it contains all the mass.

The above ratios are substantially lower than found for old 
populations in elliptical galaxies and bulges of spiral galaxies
(``normal populations''), which lie typically in the range
$10 - 30~{\rm M_{\odot}/L_{\odot}}$
\citep*[{\em e.g. \rm}][]{Dev87, OOKM95, Hun99}.
Alternatively, red giants, which have a characteristic
$M/L_{K} \sim 0.1~{\rm M_{\odot}/L_{\odot}}$, would contribute
$\sim 15\%$ of the total mass if they dominate the low-surface
brightness emission.  This is inconsistent with the
typical fraction of $0.2\% - 1\%$ determined empirically
for normal populations \citep[{\em e.g. \rm}][]{Pic85}.
These arguments strongly suggest that young red
supergiants dominate the near-infrared continuum 
throughout the entire starburst core of \mqd.

The important and complex spatial variations in the composition
of the evolved stellar population revealed by 3D indicate that
studies based on data obtained through different apertures or at
a few positions only may be misleading.  In particular, from a
comparison of the \coph\ measured at the nucleus and at the
secondary $K$-band peak, \citet{Les90} and \citet{McL93}
concluded that there are no gradients in the composition of the 
stellar population across the starburst core of \mqd.  However, the
3D EW maps clearly show that these regions are ``privileged'' in 
the sense that they sample populations with very similar properties.

From the EWs at the nucleus, a dominant population of red 
supergiants is inferred down to the central $\rm \approx 20~pc$.
The spatial resolution of the 3D images prevents a reliable
investigation of the nuclear population on smaller scales.
\citet{Gaf93} compared their \mlk\ ratio in \mqd\ to that in
the Galactic Center, both within the same radius of 7.5~pc,
and argued that their similarity supports an old bulge 
population as dominant nuclear $K$-band source in \mqd.
We simply note here that the near-infrared light within
7.5~pc of the Galactic Center contains a significant
contribution from red supergiants \citep*[{\em e.g. \rm}][]{Hal89, Blu96}.

\subsubsection{The Metallicity of the Evolved Stars} \label{Sub-abund_stars}

The analysis of the EWs presented above is based on empirical
indicators valid for stars with near-solar metallicities.
This seems justified given the gas-phase abundances derived
in section~\ref{Sub-abund_gas}.  The metallicity of the evolved stars
can in fact be directly constrained using the diagnostics proposed
by \citet{OFFO97} and \citet{OO98}.  These are based on \ewcoh\ and
\ewcoa, and derived from theoretical modeling of their behaviours
with \teff, surface gravity, micro-turbulent velocity, and metallicity.

We first examine the central 35~pc of \mqd.  Assuming a population
of red giants and applying the diagnostics of \citet{OFFO97}, the
EWs imply $\rm [Fe/H] \approx -0.5$ to $\rm -0.2~dex$ for a carbon
depletion of $\rm [C/Fe] = 0.0$ to $\rm -0.5~dex$.  Larger carbon
depletions are ruled out since for $\rm [C/Fe] \leq -0.5~dex$, the OH
bands at 1.6265\micr\ become comparably deep or deeper than the 
$\rm {}^{12}CO$ (6,3) bandhead \citep{OFFO97}, which is not the
case in the 3D spectra.  Hence, the 3D data are definitely inconsistent 
with a dominant old metal-rich population in the central 35~pc at the
nucleus.  Metallicity estimates for supergiants using the diagnostics
from \citet{OO98} depend more sensitively on \teff\ and are less well
constrained.  For the derived $T_{\rm eff} \approx 3800~{\rm K}$, 
$\rm [Fe/H] \approx -0.5$ to $\rm 0.0~dex$ depending on [C/Fe].
Lower temperatures would reduce [Fe/H] by up to 0.3~dex while higher
temperatures up to 4300~K would increase it by up to 0.4~dex.
The analysis of the EWs in section~\ref{Sub-EWs_regions}
leading to a consistent interpretation together with the near-solar
abundances of the \ion{H}{2} regions support that the supergiants
in the central 35~pc of \mqd\ have roughly solar metallicity.

Similar metallicities are inferred over the entire regions mapped
with 3D, assuming a dominant population of supergiants and accounting
for the variations in \teff.  The spatial variations in the CO bandhead
EWs are not likely due to variations in [Fe/H] of the stars.  Indeed, 
the observed ranges for \ewcoh\ and \ewcoa\ would imply variations in 
the metallicity by factors of at least $4 - 5$, not plausible on scales
of $\rm \sim 10 - 100~{\rm pc}$ and over the $\rm 10 - 50~Myr$ lifetimes
of red supergiants.

\subsection{Additional Continuum Emission Sources} \label{Sub-dilution}

The possible sources responsible for the dilution of the
stellar absorption features in \mqd\ include young OB stars,
nebular free-free and free-bound processes, and dust heated
at $\rm 600 - 1000~K$ by the OB stars (``hot dust'').
We estimated the broad-band emission from OB stars using the
number of representative \ion{O8.5}{5} stars given in table~\ref{tab-TeffOB}
and the photometric properties tabulated by \citet{Vac96} and
\citet{Koo83}.  We computed the contribution of the nebular continuum
emission from the dereddened \brg\ fluxes using the relationships given
by \citet{Sat95}.  These are for case B recombination with
$n_{\rm e} = 100~{\rm cm^{-2}}$ and $T_{\rm e} = 10^{4}~{\rm K}$.  They
are little affected by the electron density, but depend more sensitively
on the electron temperature.  However, for $T_{\rm e} \approx 5000~{\rm K}$,
the nebular flux densities inferred from \brg\ would be lower
\citep[{\em e.g. \rm}][]{Joy88}.

For selected regions (table~\ref{tab-NIRcont})
and across the entire 3D field of view, OB stars and nebular processes
make a negligible contribution to the near-infrared continuum emission
\citep[see also][]{Sat95}, leaving the hot dust as most important
source of dilution.  A crude estimate of the dilution for the 3D field
of view can be obtained independently from the SWS data.  We assumed
grey body emission for the hot dust and adopted
$T_{\rm HD} = 800~{\rm K}$ and $n = 1.5$, consistent with previous work
\citep[{\em e.g. \rm}][]{Smi90, Lar94}, and also with the
negligible dilution near 1.6\micr.  With extinction-corrected continuum
flux densities of $\rm \approx 9 \times 10^{-13}~W\,m^{-2}\,\mu m^{-1}$
near 4\micr\ and $\rm \approx 3 \times 10^{-12}~W\,m^{-2}\,\mu m^{-1}$
near 2.3\micr, the predicted contribution from hot dust at the latter
wavelength is $10\% - 15\%$, consistent with the dilution inferred
from the EWs alone.

\subsection{Extinction towards the Evolved Stars} \label{Sub-Av_stars}

We finally constrained the extinction towards the evolved stellar population,
important for deriving its intrinsic luminosity since it has a very different 
spatial distribution than the ionized gas (see figure~\ref{fig-3Dimages}) and
may suffer from different levels of obscuration 
\citep[see also {\em e.g. \rm}][]{McL93}.

The extinction for selected regions was derived from
minimum $\chi^{2}$-fitting to the 3D spectra as follows.
For each region, we combined the $K$-band spectrum for the appropriate
stellar spectral type with a grey-body emission curve for the hot dust
with $T_{\rm HD} = 800~{\rm K}$ and $n = 1.5$, in the
proportions given by \dilk, and adjusted the extinction ($A_{V}$)
for the best fit to the observed spectrum.  We considered a uniform 
foreground screen and a mixed model, and adopted the extinction law
from \citet{Dra89}.  The template stellar spectra were taken from the
atlases of \citet{FS00a} and \citet{KH86}, convolved to the spectral
resolution of the \mqd\ data when appropriate.  Due to the rather
poor sampling for K supergiants, a \ion{K5}{1} template was used for all
regions except B1, for which the \ion{K0}{1} and \ion{K5}{1} spectra 
available were averaged to produce a template \ion{K2}{1} spectrum.  
We excluded the $H$-band data from this analysis because available 
libraries had too limited a wavelength coverage.  

The results are not significantly affected by the choice of template 
star within a few spectral sub-classes, by $T_{\rm HD}$ in the range 
$600 - 1000~{\rm K}$, by $n$ between 1 and 2, and by the power-law
index for the extinction law within $0.1 - 0.2~{\rm dex}$.  The assumption
that the same extinction applies to the evolved stars and to the hot dust
is of little consequences since the stars dominate the continuum emission.
We have also constrained the extinction from the $H - K$ colour excess
(after correction for dilution), using the stellar data compiled by
\citet{Koo83}.  The values obtained are in excellent agreement with 
those from the spectral fits.

The extinction towards the evolved stars is lower than towards the ionized
gas, except for the central 35~pc.  The extinction corrections differ by up 
to $\approx 35\%$ for purely foreground obscuration and up to $\approx 60\%$
for the mixed model, presumably reflecting the different distributions for the
corresponding sources relative to the obscuring dust.  The data do not allow
the distinction between different model geometries owing to the relatively
small wavelength coverage.  However, the extinction corrections are
similar to within 35\% or less, except for the central 35~pc
($\approx 50\%$).  We will adopt the results for the mixed model
because such a geometry seems more plausible for populations of
clusters of supergiants, which are likely more or less uniformly mixed
with interstellar gas and dust clouds, as for the \ion{H}{2} regions.

We derived the extinction across the 3D field of view 
from the dilution-corrected $H - K$ colour map as described above,
for rebinned $1^{\prime\prime} \times 1^{\prime\prime}$ pixels and
assuming an intrinsic population of supergiants of appropriate type
at each location.  We considered only foreground obscuration along any 
line of sight.  Comparing the flux density integrated over the rebinned,
dilution- and extinction-corrected $K$-band map to the observed integrated
flux density, the global correction factor at 2.2\micr\ for this non-uniform
foreground screen model is very close to that obtained from the fits to the
3D field of view spectrum; it corresponds to an effective
$A_{V}^{\rm MIX} = 18~{\rm mag}$ (or $A_{V}^{\rm UFS} = 8~{\rm mag}$).

Figure~\ref{fig-3Dfits} shows the results of the population/spectral
synthesis for selected regions and also compares the 3D $H$-band 
spectra corrected for extinction to the template \ion{K4}{1} 
spectrum from \citet{DBJ96} --- their only K-type supergiant for
solar metallicity.  This figure illustrates well the quality of the
fits, in particular for the numerous stellar absorption features.
The number of representive stars required to produce the 
dilution- and extinction-corrected \lk, computed using the stellar data 
compiled by \citet{SK82} and \citet{Koo83}, are reported in
table~\ref{tab-NIRcont}.  The table also gives the properties for 
the starburst core which are derived in appendix~\ref{App-constraints}
together with additional constraints (bolometric luminosity, mass, and
rate of supernova explosions) used for the starburst modeling of paper 2.

\section{SUMMARY AND DISCUSSION} \label{Sect-conclu}

We have obtained near-infrared imaging spectroscopy with the 3D
instrument and mid-infrared spectroscopy with the {\em ISO\/}-SWS of the
starburst regions of \mqd.  We have used these data, together with results
from the literature, to determine the physical conditions of the ISM
and the composition of the stellar population on spatial scales ranging 
from a few tens of parsecs to 500~pc.

The central regions of \mqd\ show complex and important structure
on scales at least as small as $\rm \approx 25~pc$, with
the tracers of \ion{H}{2} regions and those of cool evolved stars
having in addition very different spatial distributions.
However, the structural properties within the most recent star-forming
regions and the excitation state of the nebular gas are
remarkably similar throughout the entire starburst core,
on all spatial scales.  The picture which emerges is that
of a large number of closely-packed ionizing stellar clusters
and small gas clouds, with a highly homogeneized distribution
within larger-scale concentrations.  The near constancy in local
structural properties (as reflected notably in the ionization
parameter) and in dominant OB star population suggests a similar
star formation efficiency and evolutionary stage for the most
recent star-forming regions over the entire central 500~pc.

The variations in the near-infrared continuum properties
are partly attributable to spatially non-uniform extinction 
and contribution from hot dust emission, but also to variations
in the intrinsic composition of the evolved stellar population.
Together with the \mlk\ ratio, the stellar absorption features
are consistent with solar-metallicity red supergiants as main 
sources of near-infrared continuum emission throughout the 
starburst core, down to a few tens of parsecs at the nucleus.
Although the derived spectral types cover a fairly large range, they
correspond in fact to a rather narrow range in evolutionary stages.

The OB stars and red supergiants, which dominate the luminosity of \mqd,
thus trace the star formation history within the starburst core 
up to about 50~Myr ago.  The complex starburst history in \mqd\
is already obvious from the relative distributions of the ionized
gas and of the red supergiants, suggesting a time sequence in the
triggering of the bursts at different locations.
Radial evolution has been proposed in the past but some authors
favoured inside-out propagation while others favoured the opposite
scenario \citep[{\em e.g. \rm}][]{McL93, She95, Sat97, Gri00}.
We will address this issue in paper 2 by applying starburst models
to the data presented here in order to constrain quantitatively the
detailed spatial and chronological evolution of starburst activity
in \mqd.

\acknowledgements

We are grateful to the 3D-team for help with the observations.
We would like to thank G. Ferland for providing us 
with CLOUDY version C90.05 in advance of publication, A. Pauldrach and
R.-P. Kudritzki for providing their model atmospheres, C. Telesco and
D. Gezari for making their 12.4\micr\ data available to us in electronic
form, and E. Sturm for the computations of collisional excitation.
Special thanks to Henrik Spoon for help with using CLOUDY, and to
Michele Thornley, Jack Gallimore, Linda Tacconi, Lowell Tacconi-Garman,
Roberto Maiolino, and Marcia and George Rieke for stimulating discussions
as well as useful comments on various aspects of this work.
We also wish to thank the anonymous referee for further useful
comments and suggestions.
NMFS acknowledges the Fonds pour les Chercheurs et l'Aide \`a la Recherche
(Gouvernement du Qu\'ebec, Canada) for a Graduate Scholarship, 
and the Max-Planck-Institut f\"ur extraterrestrische Physik and
Service d'Astrophysique of the CEA Saclay for additional financial support.
We also thank the German-Israeli Foundation (grant-I-551-186.07/97)
for support of this work.
SWS and the {\em ISO\/} Spectrometer Data Center at MPE are supported
by DLR under grants 50 QI 8610 8 and 50 QI 9402 3.

\pagebreak

\appendix

\section{DERIVATION OF THE EFFECTIVE IONIZATION PARAMETER IN M\,82} 
         \label{App-logU}

In this appendix, we present our determination of the effective
ionization parameter for the photoionized nebulae in \mqd.
Since it depends on the flux of Lyman continuum photons impinging
on the gas (see equation [\ref{Eq-U}], section~\ref{Sub-logU}),
its derivation requires a detailed knowledge of the properties and
distribution of the ionizing stars and gas clouds.  
We selected three representative regions for this analysis:
the entire starburst core, and regions B1 and B2.
For clarity, most numerical results are reported directly in 
table~\ref{tab-logU}.

\subsection{Properties and Distribution of the Gas Clouds}
            \label{Sub-geom_gas}

Various observations of the molecular, neutral atomic, and ionized gas in
the starburst regions of \mqd\ reveal important structure on scales at
least as small as $\rm \approx 20 - 30~pc$ ({\em e.g.} 
figure~\ref{fig-3Dimages}; \citealt{Lar94, She95, Ach95, Sat95}).
Models of the ISM imply even more extreme properties, suggesting
that the gas clouds possess small and dense molecular cores with
thin neutral atomic surfaces, and are embedded in large ionized envelopes
in a highly pressurized ISM \citep*{Olo84, Lug86, Duf87, Wol90, Lor96, Stu97}.
We adopted the neutral cloud properties derived by \citet{Lor96}.
For the starburst core, we took their average cloud radius $r_{\rm cl}$ 
and mass $M_{\rm cl}$ inferred from the global properties of \mqd.
For B1 and B2, we adopted their results for the southwestern infrared 
emission lobe encompassing these regions.

Combining $M_{\rm cl}$, the mass of molecular gas $M_{\rm H_{2}}$, and the
volume $V$ in which the clouds are distributed yields their space number 
density $n_{\rm cl}$ and mean separation $d_{\rm cl - cl}$.  Since $V$ is
not well constrained, we considered two limiting cases for each region.
For the starburst core, we adopted a sphere of diameter 500~pc ({\em i.e.} 
30\arcsec), and an edge-on disk of radius and thickness of 200~pc more
consistent with the global distribution of various gas components
\citep[{\em e.g. \rm}][]{Lar94, She95, Ach95}.
For B1 and B2, we assumed spheres of radius 19.5~pc, and columns with radii
of 19.5~pc and lengths equal to the intersection along the line of sight of
the starburst disk at the corresponding projected locations (230~pc and
365~pc, respectively).  The cross-section radii were chosen to cover an
area equivalent to the $2.25^{\prime\prime} \times 2.25^{\prime\prime}$ 
square aperture used to extract the spectra from the 3D data cubes.

We estimated $M_{\rm H_{2}}$ at B1 and B2
from the CO $J\,=\,1\,\rightarrow\,0$ map of \citet{She95}, which 
has a spatial resolution comparable to the 3D data (2.5\arcsec).  We 
converted the CO intensities into \hh\ column densities and masses using
$\mathcal{N}_{\rm H_{2}}/I_{\rm CO\,1 \rightarrow 0} = 
7 \times 10^{19}~{\rm cm^{-2}\,K^{-1}\,\left(km\,s^{-1}\right)^{-1}}$ 
derived by \citet{Wil92} for regions encompassing B1 and B2.
For the entire starburst core, these authors estimated
$M_{\rm H_{2}} = 1.8 \times 10^{8}~{\rm M_{\odot}}$, the value adopted
by \citet{Lor96}.  We emphasize that \citeauthor{Wil92} derived the
$\mathcal{N}_{\rm H_{2}}/I_{\rm CO\,1 \rightarrow 0}$ conversion factor
from detailed radiative transfer calculations applied
to observations of $\rm ^{12}CO$ (up to $J = 6 \rightarrow 5$), 
$\rm ^{13}CO$, and $\rm ^{18}CO$ lines at various positions along
the galactic plane of \mqd, so that it should account properly
for the molecular gas mass at each location.
We also derived the mass of ionized gas $M_{\rm H^{+}}$ from the
$Q^{0}_{\rm Lyc}$ determined in section~\ref{Sub-Av_gas},
assuming case B recombination with appropriate electron density
and temperature for \mqd\ (see {\em e.g.} \citealt{Ost89} for details).
The radius $r_{\rm i}$ of the outer ionized surface of the clouds is set by
\begin{equation}
\frac{4\pi}{3}\,\left(r_{\rm i}^{3} - r_{\rm cl}^{3}\right) =
\frac{\Phi_{V}^{\rm H^{+}} V}{N_{\rm cl}},
\label{Eq-ri}
\end{equation}
where $N_{\rm cl}$ is the total number of clouds and $\Phi_{V}^{\rm H^{+}}$
is the volume filling factor of the ionized gas.  The $r_{\rm i}$ does
not depend on the volume considered because $\Phi_{V}^{\rm H^{+}} V$
does not.

\subsection{Properties and Distribution of the Ionizing Stellar Clusters}
            \label{Sub-geom_clusters}

The OB stars in \mqd\ are likely to reside mainly in clusters. 
Indeed, high-resolution {\em HST\/} imaging of the central regions
of \mqd\ reveals the presence of over a hundred compact, luminous 
clusters with ages estimated between $\rm \sim 10~Myr$ and $\rm 10~Gyr$ 
\citep{OCo95, Gal99, Gri00}.
Due to the high extinction in \mqd, these are probably mostly foreground,
but some may reside in the inner regions if they lie in directions of
lower extinction.  Near-infrared observations provide evidence for
the presence of such systems deep in the obscured nuclear regions of \mqd:
several distinct compact $K$-band continuum sources are seen, with
surface brightnesses, sizes, and CO bandhead strengths consistent with
young clusters of red supergiants (\citealt{Sat97}; see also
section~\ref{Sect-pop_synthesis}).  Such super star clusters
are observed in a growing number of starburst systems and may
constitute an important mode of star formation in starbursts 
\citep*[{\em e.g. \rm}][]{Hol92, OCo94, Whi95, Mao96, 
Ho96, Tac96}.

The space number density $n_{\star}$ and mean separation $d_{\star - \star}$
of the ionizing clusters can be computed from the intrinsic Lyman continuum
photon emission rates.  We assumed that the clusters follow the luminosity
function (LF)
\begin{equation}
\frac{{\rm d}N_{\star}}{{\rm d}(\log Q^{\star}_{\rm Lyc})}
\propto \left(Q^{\star}_{\rm Lyc}\right)^{-\beta},
\label{Eq-LF}
\end{equation}
with $\beta = 0.19$ in the range
$Q^{\star}_{\rm Lyc} = 10^{45} - 10^{49.5}~{\rm s^{-1}}$ and
$\beta = 1$ in the range 
$Q^{\star}_{\rm Lyc} = 10^{49.5} - 10^{53}~{\rm s^{-1}}$.   This LF is
derived by \citet{Tho00} based on the optical and Lyman continuum LF's
observed for super star clusters and \ion{H}{2} regions in a variety of
local star-forming galaxies, with extension to low luminosities from
Monte-Carlo simulations.  The upper limit in $Q^{\star}_{\rm Lyc}$ 
is consistent with the properties of the optically-selected clusters
in \mqd.  \citet{OCo95} measured intrinsic absolute $V$-band magnitudes
in the range $\rm -9.6~mag$ to $\rm -14.5~mag$.  For an unevolved stellar
population with a \citet{Sal55} initial mass function, these values 
correspond to $Q^{\star}_{\rm Lyc} = 10^{50.5} - 10^{52.7}~{\rm s^{-1}}$.
Accounting for possible evolutionary effects, the $V$-band magnitudes
are still consistent with high $Q^{\star}_{\rm Lyc}$ up to
$\sim 10^{52}~{\rm s^{-1}}$.  The lower limit of the LF corresponds
to the smallest associations susceptible of ionizing an \ion{H}{2}
region, {\em i.e.} containing at least one early-B star.

The results, summarized in table~\ref{tab-logU}, suggest that the starburst
regions of \mqd\ can be represented by a collection of closely-packed 
ionizing clusters and small gas clouds, separated on average by a few
parsecs.  The comparable mean separations inferred for the clouds and
for the clusters suggest that they have a well-mixed distribution
throughout the regions of interest, as illustrated in 
the bottom of figure~\ref{fig-geom}.
Uncertainties in the LF or in the cloud properties are not likely
to alter this picture.  Substantial differences
in $n_{\star}$ and $d_{\star - \star}$ require either a very flat LF
($\beta < 0.5$ at high luminosities), or a very high lower cutoff (near
the inflection point at $Q^{\star}_{\rm Lyc} = 10^{49.5}~{\rm s^{-1}}$).
The former case would not be consistent with the observed range 
$\beta = 0.5 - 1.0$ (see \citealt{Tho00} and references therein),
while the latter would imply the unlikely situation that only massive
clusters containing at least one 50\masssol\ star can form, excluding
smaller OB associations.  Similarly, the cloud properties would need
to be very different than those adopted.  For example, increasing the 
$M_{\rm cl}$ and $r_{\rm cl}$ for B1 and B2 to the values for the starburst 
core and vice-versa implies a variation by 36\% only for $d_{\rm cl - cl}$.

\subsection{Effective Ionization Parameter}  \label{Sub-logUeff}

Adopting therefore a random distribution model, we derived
$U_{\rm eff}$ by considering the Lyman continuum photon flux
incident on the surface area afforded by the outer ionized edge
of the clouds.  By analogy with equation~(\ref{Eq-U}),
\begin{equation}
U_{\rm eff} = \frac{Q_{\rm Lyc}}{4\pi r_{\rm i}^{2} N_{\rm cl} n_{\rm e} c}
\label{Eq-Ueff}
\end{equation}
The resulting $\log U_{\rm eff}$ are $-2.3$ to $\rm -2.4~dex$.
For comparison, we also computed values of $\log U$ applying directly 
the central cluster model with the radius of the spherical volumes 
considered above.  The $U_{\rm eff}$ values are 
lower by about an order of magnitude, due to the important increase
in surface area of the gas exposed to the Lyman continuum radiation field 
for the more realistic randomized distribution.

Variations in the input parameters over plausible ranges do not affect
importantly the derived $\log U_{\rm eff}$.  For instance, varying 
$n_{\rm e}$ in the range $\rm 10^{2} - 10^{3}~cm^{-3}$ changes the thickness
of the ionized layers but the effect on $\log U_{\rm eff}$ due to the
variation in exposed surface is partly compensated by that of $n_{\rm e}$
itself, resulting in differences smaller than $\rm \pm 0.2~dex$.  A higher
$T_{\rm e} = 10^{4}~{\rm K}$ decreases $\log U_{\rm eff}$ by about 0.15~dex.
Interverting the neutral cloud properties between the small-scale regions 
B1 and B2, and the starburst core implies $\log U_{\rm eff}$ higher and
lower by about 0.1~dex, respectively.  The $\log U_{\rm eff}$ is
fairly well constrained in the range $-2$ to $\rm -2.6~dex$; we have
adopted a representative $\rm -2.3~dex$.

Interestingly, very similar conditions are derived for individual regions 
on scales of a few tens of parsecs as well as for the 500--pc size
starburst core.  B1 and B2 contribute however only about 5\% of the 
total $Q^{0}_{\rm Lyc}$ determined for the entire starburst core.  This
indicates little variation in the degree of ionization of the photoionized
nebulae throughout \mqd.  In order to better understand
this result physically, equation~(\ref{Eq-Ueff}) can be expressed
in terms of more fundamental properties.  Neglecting the thin neutral 
atomic layer, $N_{\rm cl} = M_{\rm H_{2}} / M_{\rm cl}$ and
$M_{\rm cl} = \pi r_{\rm cl}^{2} \mathcal{N}_{\rm cl} m_{\rm H}$,
where $\mathcal{N}_{\rm cl}$ is the H column density of a
molecular cloud core and $m_{\rm H}$ is the mass of a H atom.
The mean free path for Lyman continuum photons is
determined by the neutral surface of the clouds,
$\lambda_{\rm Lyc} = (n_{\rm cl} \pi r_{\rm cl}^{2})^{-1}$.
We also introduce $\lambda_{i} = (n_{\rm cl} \pi r_{\rm i}^{2})^{-1}$
which can be interpreted as a geometrical mean free path characterizing
the path required by a Lyman continuum photon to reach the outer ionized
surface of a cloud.  Combining these relationships, equation~(\ref{Eq-Ueff}) 
can be re-written as
\begin{equation}
U_{\rm eff} = \left(\frac{m_{\rm H}}{4 n_{\rm e} c}\right)\,
\left(\frac{Q_{\rm Lyc}}{M_{\rm H_{2}}}\right)\,
\left(\frac{\lambda_{\rm i}}{\lambda_{\rm Lyc}}\right)\,
\mathcal{N}_{\rm cl}
\label{Eq-Ueff_phys}
\end{equation}
The effective ionization parameter is thus related to 
$Q_{\rm Lyc}/M_{\rm H_{2}}$ which provides a measure of the star
formation efficiency.  Consequently, for similar gas cloud properties,
the comparable $U_{\rm eff}$ at different locations and on various
spatial scales results from a near constancy in the average star formation
efficiency throughout the starburst regions of \mqd.

\section{ADDITIONAL OBSERVATIONAL CONSTRAINTS FOR M\,82} 
         \label{App-constraints}

In the following, we derive further constraints that will be used
in paper 2 for the application of starburst models to \mqd.
These include the near-infrared continuum properties for the
starburst core, as well as the bolometric luminosity, the stellar 
mass, and the rate of supernova explosions for various regions.

\subsection{Near-infrared Properties of the Starburst Core} \label{Sub-NIRcore}

\citet{Sat97} measured the depth of the CO bandheads longwards of
2.3\micr\ in a 24\arcsec --diameter aperture on the nucleus of \mqd.
From a calibration derived using the stellar atlases of \citet{KH86} 
and \citet{FS00a}, their spectroscopic index of 0.18~mag (which is
independent of extinction) corresponds to $W_{2.29} = 14.5$~\AA.
This is essentially the global value measured for the 3D field of view
accounting for 10\% dilution ($14.4$~\AA).
Dilution can be neglected for the starburst core, since about
half of the infrared emission by dust and of the nebular emission
from \ion{H}{2} regions originates outside of the 3D field of view
while this fraction is 80\% for the $K$-band emission
(see table~\ref{tab-NIRcont} and section~\ref{Sect-neb_analysis}).
Because of the similar intrinsic \ewcoa\ for the 3D field of view
and the central 24\arcsec\ as well as the relative constancy of
\mlk\ within the central 500~pc, we adopted the \ewcoh\ for the
3D field of view as representative for the starburst core.

While the observed $K$-band magnitudes determined by various authors for
the central $\rm \approx 500~pc$ of \mqd\ agree well with each other,
there are substantial discrepancies in the reported absolute intrinsic
magnitudes ($M^{0}_{K}$).  The results vary between $\rm -23.3~mag$
and $\rm -22.0~mag$ \citep{Rie80, Tel91, McL93, Sat97}.  These differences
are mainly attributable to the different extinction values and model 
geometries assumed.  We re-examined this issue using the results from our
3D and SWS observations.  Based on the morphology of the $K$-band and nebular 
line emission and on the global distributions of the corresponding sources,
we argue as \citet{Rie80} that half of the extinction towards the
bulk of ionized gas probably applies to the bulk of evolved stars.
Taking the observed $K$-band magnitude measured by \citet[][5.43~mag]{Tel91},
an $A^{\rm MIX}_{V} = 26~{\rm mag}$ implies $M^{0}_{K} = -23.2~{\rm mag}$.
Similar values are obtained with the mixed extinction model for the 3D
field of view ($-22.9~{\rm mag}$) and the effective extinction derived
from the $H - K$ map ($-23.0~{\rm mag}$).  Our results are consistent
with the lower range of values reported in the literature.
The 3D data trace projected regions close to the nucleus which are 
on average the most obscured, leading to a possible overestimate of 
the global extinction for the starburst core.  On the other hand,
large optical depths are inferred, which may indicate that the
near-infrared light does not probe the stars throughout the entire 
galaxy along the line of sight.  Therefore, we will adopt an average
$M^{0}_{K} = -23.0~{\rm mag}$, implying an effective global 
$A^{\rm MIX}_{V} = 21~{\rm mag}$ (or $A^{\rm UFS}_{V} = 9~{\rm mag}$)
towards the evolved stars in \mqd.

\subsection{Bolometric Luminosity} \label{Sub-Lbol}

Two main sources dominate the bolometric luminosity (\lbol) 
in \mqd: the hot massive stars and the cool evolved stars.
In dusty starbursts such as \mqd, the infrared luminosity (\lir)
provides a good approximation to the \lbol\ from hot stars.
For the entire starburst core, we adopted 
$L_{\rm IR} = 3 \times 10^{10}~{\rm L_{\odot}}$ \citep{Tel80}.
We estimated \lir\ for individual regions using measurements
of the mid-infrared emission which has a similar morphology
as the infrared emission out to at least 100\micr, where the 
global energy distribution of \mqd\ peaks \citep{Tel91}.
\citet{Tel93} obtained $L_{\rm IR} = 18\,L_{N}$ for the starburst core of
\mqd, where $L_{N}$ is the luminosity in the $N$-band centered at 10.8\micr,
also valid within a factor of two for a sample of 11 starburst galaxies.
We actually used the 12.4\micr\ map of \citet{Tel92}
\footnote{kindly made available to us in digital form by the authors}
which has a spatial resolution comparable to the 3D data,
scaling the fluxes according to $f_{N} = 0.4\,f_{12.4\mu {\rm m}}$
derived by these authors.  Combining the above relationships,
\begin{equation}
\frac{L_{\rm IR}}{\rm L_{\odot}} = 3.25 \times 10^{7}\,
\left(\frac{D}{\rm Mpc}\right)^{2}\,
\left(\frac{f_{\rm 12.4\,\mu m}}{\rm Jy}\right).
\label{Eq-Lir}
\end{equation}
Equation~(\ref{Eq-Lir}) includes average extinction effects as well as 
contributions in the 12.4\micr\ and $N$-bandpasses from PAH
emission features, emission lines, and the silicate absorption feature.
From our SWS and 3D data together with the fine-structure line
maps of \citet{Ach95} and the 3.3\micr\ PAH map of \citet{Sat95},
we estimate that spatial variations in the spectral features and 
in extinction introduce errors of about 35\% in applying 
equation~(\ref{Eq-Lir}) to individual regions.  Following \citet{McL93},
we included an additional 30\% of the detected \lir\ in \lbol\ to
account for light escaping in directions perpendicular to the galactic plane.

Individual giants and supergiants with temperatures
between 3500~K and 6000~K have $L_{\rm bol}/L_{K} \approx 10 - 30$,
similar to the ratio for mixed populations of cool evolved stars
\citep[{\em e.g. \rm}][]{McL93}.  We have thus estimated the \lbol\
from evolved stars using the intrinsic stellar \lk\ and assuming
a representative ratio of 20, with 50\% uncertainty.
Table~\ref{tab-Lbol} summarizes the results for selected regions.

\subsection{Mass} \label{Sub-mass}

We estimated the dynamical mass \mdyn\ from position-velocity maps
available in the literature.  These include observations of the
CO $J = 1 \rightarrow 0$ millimeter line \citep{She95},
of the [\ion{Ne}{2}] 12.8\micr\ line \citep{Ach95},
and of the [\ion{S}{3}] $\lambda$9069~\AA\ line \citep{McK93}.
We assumed a uniform mass distribution and dynamical equilibrium,
so that within projected radius $r$
\begin{equation}
M_{\rm dyn}(<r) = 3.49 \times 10^{3}\,
\left(\frac{r}{\rm arcsec}\right)\,
\left(\frac{v_{\rm rot}(r)}{\rm km\,s^{-1}}\right)^{2},
\label{Eq-Mdyn}
\end{equation}
where $v_{\rm rot}$ is the rotational velocity.  Since various observations
of the molecular and ionized gas reveal important concentrations which are
interpreted as circumnuclear rotating rings or spiral arms
\citep[{\em e.g. \rm}][]{Lar94, She95, Ach95, Sea98, Nei98}, 
we determined $v_{\rm rot}$ using the peak velocity at the locations
of these concentrations, corrected for an inclination of 80\arcdeg,
and for beam and velocity smearing.  In addition, we used the dynamical
mass obtained by \citet{Gaf93} in the central 1\arcsec\ of \mqd, from
stellar velocity dispersion measurements using the CO bandhead at 2.29\micr.
The resulting \mdyn\ versus $r$ curve is plotted in figure~\ref{fig-Mdyn}
along with the mass model proposed by \citet{Got90}.  Our results imply
larger masses for $r \lesssim 10^{\prime\prime}$, by up to a factor of 
$\approx 4$ at $r = 0.5^{\prime\prime}$.  The more recent data used here
have higher resolution than those used by \citet{Got90}, and thus probe
much better the central regions of \mqd.

To obtain the stellar mass $M^{\star}$, we subtracted the gaseous mass 
from $M_{\rm dyn}$.  The mass of ionized and molecular hydrogen 
($M_{\rm H^{+}}$ and $M_{\rm H_{2}}$) for the starburst core is
estimated in appendix~\ref{App-logU}.  Following the same procedure
as described there for B1 and B2, we computed $M_{\rm H^{+}}$ and
$M_{\rm H_{2}}$ for the central 35~pc from the intrinsic $Q_{\rm Lyc}$,
and from the CO data of \citet{She95} with the conversion factor between
the CO intensity and $\rm H_{2}$ column density from \citet{Wil92}
at the position of the nucleus.  Table~\ref{tab-Mass} gives the
various mass estimates.  The resulting $M^{\star}$ represent lower
limits since the $M_{\rm H_{2}}$ adopted for the starburst core
was determined by \citet{Wil92} in a region larger than
30\arcsec\ in diameter, and the gas observed towards the central
35~pc of \mqd\ is probably mainly located in a circumnuclear
ring at larger radius, as mentioned above.

\subsection{Rate of Supernova Explosions} \label{Sub-SNR}

Estimates of the global rate of supernova explosions (\snrate)
in \mqd\ from the properties of the compact, non-thermal synchrotron 
emission sources detected at centimeter wavelengths (size and
radio luminosity distributions, luminosity variations) 
vary in the range $\rm 0.02 - 0.1~yr^{-1}$ 
\citep[{\em e.g. \rm}][]{Kron85, vBu94, Hua94, Mux94, All98}.
For selected regions in \mqd, we applied the
relationship between \snrate\ and [\ion{Fe}{2}] 1.644\micr\
line flux derived by \citet{Van97}.  Our fluxes corrected
for the extinction inferred from the Brackett lines imply
rates of $\rm \sim 3 \times 10^{-3}~yr^{-1}$ for the central
35 pc of \mqd\ and B1, $\rm \sim 6 \times 10^{-3}~yr^{-1}$
for B2, and $\rm \sim 0.01~yr^{-1}$ for the 3D field of view.
We emphasize however that the uncertainties in \snrate 's
inferred from [\ion{Fe}{2}] fluxes are very large.
The calibration of \citet{Van97} is based on the integrated
[\ion{Fe}{2}] luminosity and radio \snrate, and depends on
uncertain assumptions about the supernova remnants (SNRs) lifetimes.
The [\ion{Fe}{2}] emission may trace a different population of
SNRs than radio observations, and its interpretation may be further
complicated by other excitation sources such as an outflowing
starburst wind \citep[{\em e.g. \rm}][]{Gre97}.
A calibration based on the average intrinsic line fluxes of
four compact [\ion{Fe}{2}] sources candidate SNRs measured
by \citet{Gre97} would imply \snrate 's about 50 times smaller!

\clearpage

\figcaption[f1.ps]
{
Regions observed in M\,82.  The 3D field of view
and {\em ISO\/}-SWS apertures are shown on a $K$-band
map \citep{For94} with superimposed contours
of the CO $J\,=\,1\,\rightarrow\,0$ emission \citep{She95}. 
The cross and the triangle indicate the positions of the nucleus
and of the western mid-infrared emission peak, respectively.
\label{fig-3DSWSFOV}
}

\figcaption[f2.ps]
{
3D $H$- and $K$-band spectra of selected regions in M\,82.
The top three panels show the spectra of the nucleus and of regions
B1 and B2 taken in $2.25^{\prime\prime} \times 2.25^{\prime\prime}$ 
apertures (see table~\ref{tab-3Ddata}) while the bottom panel shows the 
spectrum of the entire 3D field of view.
The effective resolution is $R \sim 1015$ in the 
$H$-band and $R \sim 830$ in the $K$-band.  The vertical axis is a 
linear flux density scale.  The spectra are normalized to unity in
the interval $\rm 2.2875 - 2.2910~\mu m$.  The absolute flux densities 
in $\rm W\,m^{-2}\,{\mu m}^{-1}$ can be recovered using the
multiplicative factors $6.52 \times 10^{-14}$ for the nucleus,
$2.96 \times 10^{-14}$ for B2, $1.78 \times 10^{-14}$ for B1, 
and $7.24 \times 10^{-13}$ for the 3D field of view.
The positions of various lines are indicated on the spectra of the
nucleus (stellar absorption features) and of B1 (emission lines).
\label{fig-3Dspec}
}

\figcaption[f3.ps]
{
Selected near-infrared maps of M\,82 obtained with 3D.  The greyscale
levels are indicated next to each panel, in the units given in the following.
{\em (a):\/} $K$-band emission map, with contours of the $H$-band emission
from 0.5 to 2.15 in steps of 0.15; greyscales and contours are in
units of $\rm 10^{-14}~W\,m^{-2}\,{\mu m}^{-1}\,arcsec^{-2}$.
{\em (b):\/} Br$\gamma$ emission line map, with contours of the
He~I $\rm 2.058~\mu m$ line emission from 0.5 to 1.5 in steps of 0.1;
greyscales and contours are in units of
$10^{-17}~{\rm W\,m^{-2}\,arcsec^{-2}}$.
{\em (c):\/} Map of the equivalent width of the $^{12}$CO (2,0) bandhead
at $\rm 2.29~\mu m$ ($W_{2.29}$) in units of \AA;
contours of the equivalent width of the $^{12}$CO (6,3) bandhead at 
$\rm 1.62~\mu m$ ($W_{1.62}$) from 4.8~\AA\ to 6.6~\AA, in steps of 0.3~\AA.
{\em (d):\/} $\log (L_{K}/L_{\rm Lyc})$ map.
{\em (e):\/} He~I~2.058/Br$\gamma$ line ratio map.
{\em (f):\/} $\log (W_{1.62}/W_{2.29})$ map.
The axis coordinates are relative offsets from the nucleus,
indicated by the cross. 
From left to right, the boxes in each panel indicate the apertures used to
extract the spectra of the central 35~pc of M\,82, B2, and B1 plotted in
figure~\ref{fig-3Dspec} (labeled ``N,'' ``B2,'' and ``B1'' in panel {\em a\/}).
\label{fig-3Dimages}
}


\figcaption[f4.ps]
{
{\em ISO\/}-SWS mid-infrared spectrum of M\,82 
(full scan AOT SWS01).  The spectral resolution varies from
$\sim 1000$ at short wavelengths to $\sim 500$ at long wavelengths. 
The ``jumps'' in the continuum level at $\rm 12.0~\mu m$, $\rm 27.8~\mu m$,
and $\rm 29.5~\mu m$ are caused by the increase in aperture size and the 
fluctuations in the continuum, especially in the $\rm 4 - 5~\mu m$ region, 
are mainly due to noise.
\label{fig-SWS01}
}

\figcaption[f5.ps]
{
High S/N ratio spectra of individual mid-infrared lines in M\,82, obtained 
with the {\em ISO\/}-SWS (line scans AOT SWS02).  The spectral resolution
ranges from $\sim 2000$ to $\sim 1000$ from short to long wavelengths.
\label{fig-SWS02}
}

\figcaption[f6.ps]
{
Results of the determination of the global extinction towards the
ionized gas in M\,82.  The plots show the
intrinsic Lyman continuum photon emission rates $Q^{0}_{\rm Lyc}$ for 
the SWS $14^{\prime\prime} \times 20^{\prime\prime}$ aperture, derived 
from the H recombination line fluxes corrected for the best-fit 
extinction for different geometries as indicated in each plot.
Panels to the left show the results using the \citet{Dra89}
extinction law throughout the entire infrared range while those to the
right show the results assuming the extinction law towards the Galactic
Center between $\rm 3~\mu m$ and $\rm 10~\mu m$ given by \citet{Lut99}.
The horizontal lines in each plot indicate the average $Q^{0}_{\rm Lyc}$.
\label{fig-Avfits}
}

\figcaption[f7.eps]
{
Geometries considered in the derivation of the 
effective ionization parameter in M\,82.
{\em Top:\/} a gas shell of radius $R$ surrounding a central stellar cluster,
or association of clusters.
{\em Bottom:\/} a well-mixed distribution of stellar clusters and gas clouds.
\label{fig-geom}
}


\figcaption[f8.ps]
{
$T_{\rm eff}^{\rm OB}$--sensitive ratios of mid-infrared fine-structure lines
(top three panels) and of near-infrared recombination lines
(bottom two panels).  
The various curves represent the theoretical predictions
for different sets of parameters.  The solid lines
show the results for the nebular parameters derived for M\,82:
$n_{\rm H} \approx n_{\rm e} = 300~{\rm cm^{-3}}$,
$\log U = -2.3~{\rm dex}$, $R = 25~{\rm pc}$, solar gas-phase abundances,
and no interstellar dust grains mixed with the ionized gas.
The effects of changing the ISM composition to a gas and dust mixture as
in the Orion nebula, the gas density to $n_{\rm H} = 10^{3}~{\rm cm^{-3}}$,
or $\log U$ between $-2~{\rm dex}$ and $-2.5~{\rm dex}$
are illustrated as well (see labels in each plot).
The horizontal bars show the extinction-corrected ratios
obtained with the SWS and with 3D, with vertical width
indicating the measurement uncertainties.
Different shading or filling patterns in the two bottom panels show the ratios 
for selected regions: the central 35~pc at the nucleus (empty bar and 
horizontal line marking the upper limit, labeled ``N''), B1 (dark-shaded
bar), B2 (cross-hatched bar), and the 3D field of view (light-shaded bar).
\label{fig-TeffOB}
}

\figcaption[f9.ps]
{
Spectroscopic indices sensitive to the effective temperature and 
luminosity class of cool stars.  The data for the central 35~pc
of M\,82 (labeled ``N''), B1, and B2 are indicated by the horizontal
bars, with vertical width corresponding to the measurement uncertainties.
The integrated equivalent widths for the 3D field of view are similar to 
those of B2.  The stellar data are taken from the compilation
by \citet{FS00a}; open circles, filled circles, and crosses represent
supergiants, giants, and dwarfs, respectively.
\label{fig-3DEWs}
}


\figcaption[f10.eps]
{
Diagnostic diagrams for the estimation of the amount of dilution
near $\rm 1.6~\mu m$ and $\rm 2.3~\mu m$.  The shaded areas indicate 
the loci of giants (dark shade) and supergiants (light shade),
based on the stellar data compiled by \citet{FS00a}.
The arrows indicate the effects of dilution by featureless continuum
sources.  Stellar effective temperatures corresponding to various 
intrinsic $W_{1.62}$ are labeled on the right-hand side diagrams.
{\em Top\/}: data for the central 35~pc of M\,82 (labeled ``N''),
B1, B2, and the 3D field of view (labeled ``3D'').
{\em Bottom\/}: data for individual
$1^{\prime\prime} \times 1^{\prime\prime}$ pixels from the rebinned 3D maps,
with typical uncertainties shown in the upper left corner of the diagrams.
\label{fig-dilution}
}

\figcaption[f11.ps]
{
Results of spectral synthesis for selected regions in M\,82.
The results of the fits are shown for the central 35~pc of M\,82 
(top panels), regions B1 and B2 (middle panels), and the 3D field of
view (bottom panels).  The black lines are the 3D spectra 
corrected for the best-fit mixed extinction (table~\ref{tab-NIRcont}).
The grey lines are the combination of the template stellar spectrum and
the hot dust (``HD'') emission appropriate for each region.  
The $H$-band template spectrum is the K4~I star BS\,5645
from \citet{DBJ96}.  The $K$-band template spectra, from \citet{KH86},
are the K5~I star BS\,8726, and a K2~I obtained by
averaging the spectra of BS\,8726 and of the K0~I star RW\,Cep.
\label{fig-3Dfits}
}

\figcaption[f12.ps]
{
Enclosed dynamical mass versus projected radius in M\,82
(derived in appendix~\ref{App-constraints}).
The diagonal line represents the mass model proposed by \citet{Got90}.
The horizontal line segments at projected radii of $1.3^{\prime\prime}$
and $15^{\prime\prime}$ indicate the contribution from the ionized and
molecular hydrogen gas for the central 35~pc and the entire starburst core.
\label{fig-Mdyn}
}

\clearpage

\begin{deluxetable}{ccccccl}
\tablecolumns{7}
\tablewidth{0pt}
\tablenum{1}
\tablecaption{Log of the 3D observations of M\,82 \label{tab-3Dlog}}
\tablehead{
\colhead{Date} & \colhead{Telescope\,\tablenotemark{a}} & 
\colhead{Field\,\tablenotemark{b}} & \colhead{Band} &
\colhead{$t_{\rm int}$\,\tablenotemark{c}} & \colhead{Seeing} &
\colhead{Atmospheric calibrator} \\
  &  &  &  & \colhead{(s)} & \colhead{(arcsec)} &  
}
\startdata
1995 Jan 13 & CA & 1 & $K$ & 600 & 1.5 & HD 82189 (F5 V) \\
 &  & 2 & $K$ & 600 & 1.5 & HD 82189 (F5 V) \\
1995 Jan 14 & CA & 3 & $K$ & 600 & 1 & PPM 17105 (G0 V) \\
 &  & 3 & $H$ & 480 & 1 & HD 87141 (F5 V) \\
1995 Jan 16 & CA & 4 & $K$ & 600 & 1.3 & PPM 17105 (G0 V) \\
1995 Jan 21 & CA & 1 & $H$ & 480 & 1 & HD 87141 (F5 V) \\
 &  & 4 & $H$ & 480 & 1 & HD 87141 (F5 V) \\
1996 Jan 06 & WHT & 2 & $H$ & 900 & 1 & HD 26356 (B5 V) \\
\enddata
\tablenotetext{a}
{
CA : 3.5~m telescope at Calar Alto, Spain. 
WHT : 4.2~m William-Herschel-Telescope on the Canary Islands, Spain.
}
\tablenotetext{b}
{
The right ascension and declination offsets of the fields with respect
to the nuclear position (defined by the $K$-band emission peak at
$\alpha_{1950}$: $\rm 09^{h} 51^{m} 43\fs 53$,
$\delta_{1950}$: $+69^{\circ} 55^{\prime} 00\farcs 7$; \citealt{Die86})
are the following:  field 1, ($+2^{\prime\prime}, +1.5^{\prime\prime}$);
field 2, ($+2^{\prime\prime}, -2^{\prime\prime}$);
field 3, ($-3^{\prime\prime}, -1^{\prime\prime}$);
field 4, ($-8^{\prime\prime}, -3^{\prime\prime}$). 
}
\tablenotetext{c}
{
Total on-source integration time per detector pixel.
}
\end{deluxetable}

\clearpage

\begin{deluxetable}{lcccc}
\tablecolumns{5}
\tablewidth{0pt}
\tablenum{2}
\tablecaption{Observed near-infrared properties of selected regions in M\,82
              \label{tab-3Ddata}}
\tablehead{
\colhead{Band or feature} &
\colhead{Nucleus} & \colhead{B1} & \colhead{B2} & \colhead{3D field} \\
  & \colhead{($0^{\prime\prime}, 0^{\prime\prime}$)\,\tablenotemark{a}} &
\colhead{($-10^{\prime\prime}, -4.25^{\prime\prime}$)\,\tablenotemark{a}} &
\colhead{($-5.25^{\prime\prime}, -2^{\prime\prime}$)\,\tablenotemark{a}} & 
} 
\startdata
\multicolumn{5}{c}{Broad-band flux densities\,\tablenotemark{b} \ \ (Jy)} \\
\ $H$-band & 0.076 & 0.021 & 0.032 & 1.01 \\
\ $K$-band & 0.104 & 0.030 & 0.048 & 1.20 \\
\hline
\multicolumn{5}{c}{Emission line fluxes\,\tablenotemark{c} \ \
                   ($\rm 10^{-17}~W\,m^{-2}$)} \\
\ [\ion{Fe}{2}] $a^{4}F_{9/2} - a^{4}D_{5/2}$ (1.5335\micr) & 
$< 0.11$ & $0.75 \pm 0.09$ & $0.56 \pm 0.04$ & $9.2 \pm 2.9$ \\
\ \ion{H}{1} Br13 ($n = 4 - 13$, 1.6109\micr) & 
$0.48 \pm 0.30$ & $0.90 \pm 0.08$ & $0.71 \pm 0.16$ & $13.2 \pm 4.5$ \\
\ \ion{H}{1} Br12 ($n = 4 - 12$, 1.6407\micr) & 
$0.33 \pm 0.08$ & $0.81 \pm 0.19$ & $0.75 \pm 0.45$ & $12.9 \pm 6.5$ \\
\ [\ion{Fe}{2}] $a^{4}F_{9/2} - a^{4}D_{7/2}$ (1.6435\micr) & 
$4.12 \pm 0.07$ & $3.53 \pm 0.44$ & $4.99 \pm 0.44$ & $137 \pm 12$ \\
\ [\ion{Fe}{2}] $a^{4}F_{7/2} - a^{4}D_{5/2}$ (1.6769\micr) & 
$0.52 \pm 0.37$ & $0.22 \pm 0.07$ & $0.34 \pm 0.13$ & $10.5 \pm 3.6$ \\
\ \ion{H}{1} Br11 ($n = 4 - 11$, 1.6807\micr) & 
$0.42 \pm 0.25$ & $1.44 \pm 0.28$ & $0.90 \pm 0.19$ & $14.3 \pm 4.1$ \\
\ \ion{He}{1} $3\,^{3}P - 4\,^{3}D$ (1.7008\micr) & 
$< 0.16$ & $0.50 \pm 0.12$ & $0.38 \pm 0.09$ & $6.3 \pm 2.8$ \\
\ \ion{H}{1} Br10 ($n = 4 - 10$, 1.7362\micr) & 
$1.22 \pm 0.24$ & $2.31 \pm 0.18$ & $1.76 \pm 0.17$ & $32.2 \pm 5.2$ \\
\ \ion{He}{1} $2\,^{1}S - 2\,^{1}P$ (2.0581\micr) & 
$2.53 \pm 0.28$ & $5.19 \pm 0.08$ & $4.34 \pm 0.14$ & $74.9 \pm 3.3$ \\
\ \hh\ $1 - 0~S(1)$ (2.1213\micr) & 
$1.01 \pm 0.21$ & $0.66 \pm 0.07$ & $0.98 \pm 0.11$ & $21.3 \pm 2.4$ \\
\ \ion{H}{1} Br$\gamma$ ($n = 4 - 7$, 2.1655\micr) & 
$5.25 \pm 0.55$ & $10.1 \ \pm 0.2$ & $9.04 \pm 0.25$ & $148 \pm 6$ \\
\ \hh\ $1 - 0~S(0)$ (2.2227\micr) & 
$0.67 \pm 0.14$ & $0.11 \pm 0.03$ & $0.25 \pm 0.07$ & $9.8 \pm 1.6$ \\
\ \hh\ $2 - 1~S(1)$ (2.2471\micr) & 
$0.29 \pm 0.10$ & $0.28 \pm 0.03$ & $0.20 \pm 0.06$ & $6.3 \pm 1.2$ \\
\ \hh\ $1 - 0~Q(1)$ (2.4059\micr) & 
$0.92 \pm 0.27$ & $1.04 \pm 0.22$ & $1.31 \pm 0.32$ & $13.2 \pm 3.2$ \\
\hline
\multicolumn{5}{c}{Equivalent width of stellar absorption 
features\,\tablenotemark{d} \ \ (\AA)} \\
\ \ion{Si}{1} (1.59\micr) & 3.6 & 3.3 & 3.7 & 3.4 \\
\ $\rm ^{12}CO$\,(6,3) (1.62\micr) & 5.6 & 3.4 & 4.6 & 4.8 \\
\ $\rm ^{12}CO$\,(2,0) (2.29\micr) & 15.2 & 8.4 & 12.2 & 13.0 \\
\enddata
\tablenotetext{a}
{
Position of the $2.25^{\prime\prime} \times 2.25^{\prime\prime}$
apertures defining the individual regions, relative to the nucleus.
}
\tablenotetext{b}
{
Estimated uncertainties are 15\% and 10\% for the $H$- and $K$-band flux
densities, respectively.
}
\tablenotetext{c}
{
Quoted uncertainties represent those of the continuum subtraction 
(see section~\ref{Sub-NIRspec}).
Ionic transitions are given as {\em lower level -- upper level}. 
$\rm H_{2}$ transitions are labeled
by the upper and lower vibrational quantum numbers followed by
$S(j)$ or $Q(j)$ which refer to transitions for which
$j - j^{\prime}$ equals $-2$ or 0 respectively, where $j$ and $j^{\prime}$
are the lower and upper rotational quantum numbers.
}
\tablenotetext{d}
{
Uncertainties on the equivalent widths are $\pm 0.4$~\AA\ for \ion{Si}{1},
$\pm 0.3$~\AA\ for $\rm ^{12}CO$\,(6,3), and $\pm 0.6$~\AA\ for 
$\rm ^{12}CO$\,(2,0).  The \ion{Si}{1} equivalent widths are corrected
for dilution by Br14 (see section~\ref{Sub-NIRspec}).
}
\end{deluxetable}

\clearpage

\begin{deluxetable}{clc}
\tablecolumns{3}
\tablewidth{0pt}
\tablenum{3}
\tablecaption{Comparison between various Br$\gamma$ flux measurements 
              \label{tab-Brgcomp}}
\tablehead{
\colhead{Aperture diameter\,\tablenotemark{a}} & \colhead{Observed flux} & 
\colhead{Reference} \\
\colhead{(arcsec)} & \colhead{($\rm 10^{-17}~W\,m^{-2}$)} & 
}
\startdata
3.8 & $11.3 \pm 1.8$ & 1 \\
    & $20.0 \pm 0.4$ & 2 \\
    & $5.7 \pm 0.7$ & 3 \\
    & $17.1 \pm 3.4$ & 4 \\
\hline
8 & 
$48 \pm 8$ & 1\,\tablenotemark{b} \\
    & $22 \pm 2$ & 5 \\
    & $26 \pm 3$ & 3 \\
    & $73.6 \pm 14.7$ & 4 \\
\enddata
\tablenotetext{a}
{
Aperture centered on the nucleus of \mqd.
}
\tablenotetext{b}
{
An 8\arcsec\ aperture slightly exceeds the
regions covered by the 3D \brg\ map, but we estimate
the missing flux to be less than a few percents.
}
\tablerefs
{
(1) This work; (2) \citealt{Les90}; (3) \citealt{Lar94}; 
(4) \citealt{Sat95}; (5) \citealt{Rie80}.
}
\end{deluxetable}

\clearpage

\begin{deluxetable}{llccccc}
\tablecolumns{7}
\tablewidth{0pt}
\tablenum{4}
\tablecaption{{\em ISO\/}--SWS line measurements in M\,82  \label{tab-SWSdata}}
\tablehead{
\colhead{Species} & \colhead{Transition\,\tablenotemark{a}} & 
\colhead{$\lambda_{\rm observed}$} & \colhead{FWHM} &
\colhead{Flux\,\tablenotemark{b}} & \colhead{Observation\,\tablenotemark{c}} &
\colhead{Scaled flux\,\tablenotemark{d}} \\
 & & \colhead{($\rm \mu m$)} & \colhead{($\rm \mu m$)} & 
\colhead{($\rm 10^{-15}~W\,m^{-2}$)} & \colhead{} &
\colhead{($\rm 10^{-15}~W\,m^{-2}$)}
}
\startdata
\ \hh\ & $1 - 0~Q(3)$ (2.4238\micr) & 2.42539 & 0.00214 & 
0.27 & 02 $14^{\prime\prime} \times 20^{\prime\prime}$ & ... \\
\ \ion{H}{1} & $\rm Br\beta$ ($n = 4 - 6$, 2.6252\micr) & 2.62726 & 0.00295 &
3.90 & 01 $14^{\prime\prime} \times 20^{\prime\prime}$ & ... \\
           &                             & 2.62720 & 0.00231 & 
4.10 & 02 $14^{\prime\prime} \times 20^{\prime\prime}$ & ... \\
\ \ion{H}{1} & $\rm Pf\delta$ ($n = 5 - 9$, 3.2961\micr) & 3.29913 & 0.00325 &
0.59 & 01 $14^{\prime\prime} \times 20^{\prime\prime}$ & ... \\
\ \ion{H}{1} & $\rm Pf\gamma$ ($n = 5 - 8$, 3.7395\micr) & 3.74247 & 0.00501 &
1.07 & 01 $14^{\prime\prime} \times 20^{\prime\prime}$ & ... \\
\ \ion{H}{1} & Hu14 ($n = 6 - 14$, 4.0198\micr) & 4.02301 & 0.00370 & 
0.13 & 02 $14^{\prime\prime} \times 20^{\prime\prime}$ & ... \\
\ \ion{H}{1} & $\rm Br\alpha$ ($n = 4 - 5$, 4.0512\micr) & 4.05437 & 0.00464 & 
8.75 & 01 $14^{\prime\prime} \times 20^{\prime\prime}$ & ... \\
           &                              & 4.05474 & 0.00414 & 
8.15 & 02 $14^{\prime\prime} \times 20^{\prime\prime}$ & ... \\
\ \ion{H}{1} & $\rm Pf\beta$ ($n = 5 - 7$, 4.6525\micr) & 4.65623 & 0.00447 &
1.39 & 01 $14^{\prime\prime} \times 20^{\prime\prime}$ & ... \\
\ \hh\ & $0 - 0~S(7)$ (5.5112\micr) & 5.51437 & 0.00657 & 
0.48 & 02 $14^{\prime\prime} \times 20^{\prime\prime}$ & ... \\
\ \hh\ & $0 - 0~S(5)$ (6.9095\micr) & 6.91299 & 0.01155 & 
1.08 & 01 $14^{\prime\prime} \times 20^{\prime\prime}$ & ... \\
        &   & 6.91339 & 0.00684 & 
1.15 & 02 $14^{\prime\prime} \times 20^{\prime\prime}$ & ... \\
\ [\ion{Ar}{2}] & $^{2}P_{3/2} - {}^{2}P_{1/2}$ (6.9853\micr) & 6.98912 & 
0.01083 & 26.7 & 01 $14^{\prime\prime} \times 20^{\prime\prime}$ & ... \\
\ \ion{H}{1} & $\rm Pf\alpha$ ($n = 5 - 6$, 7.4578\micr) & 7.46376 & 
0.00872 & 2.49 & 01 $14^{\prime\prime} \times 20^{\prime\prime}$ & ... \\
           &                              & 7.46417 & 0.00727 & 
2.59 & 02 $14^{\prime\prime} \times 20^{\prime\prime}$ & ... \\
\ \hh\ & $0 - 0~S(4)$ (8.0251\micr) & 8.02952 & 0.00858 & 
0.68 & 01 $14^{\prime\prime} \times 20^{\prime\prime}$ & ... \\
\ [\ion{Ar}{3}] & $^{3}P_{2} - {}^{3}P_{1}$ (8.9914\micr) & 8.99664 & 0.01096 &
4.89 & 01 $14^{\prime\prime} \times 20^{\prime\prime}$ & ... \\
           &                              & 8.99681 & 0.00824 & 
4.76 & 02 $14^{\prime\prime} \times 20^{\prime\prime}$ & ... \\
\ \hh\ & $0 - 0~S(3)$ (9.6649\micr) & 9.67182 & 0.01009 & 
0.97 & 01 $14^{\prime\prime} \times 20^{\prime\prime}$ & ... \\
\ [\ion{S}{4}] & $^{2}P_{1/2} - {}^{2}P_{3/2}$ (10.5105\micr) & 10.51646 & 
0.01468 & 1.89 & 01 $14^{\prime\prime} \times 20^{\prime\prime}$ & ... \\
           &                              & 10.51639 & 0.00825 & 
1.49 & 02 $14^{\prime\prime} \times 20^{\prime\prime}$ & ... \\
\ \hh\ & $0 - 0~S(2)$ (12.2786\micr) & 12.29064 & 0.02298 & 
2.00 & 01 $14^{\prime\prime} \times 20^{\prime\prime}$ & ... \\
        &    & 12.27736 & 0.01598 & 
1.14 & 02 $14^{\prime\prime} \times 20^{\prime\prime}$ & ... \\
\ [\ion{Ne}{2}] & $^{2}P_{3/2} - {}^{2}P_{1/2}$ (12.8136\micr) & 12.81923 & 
0.01554 & 99.1 & 01 $14^{\prime\prime} \times 27^{\prime\prime}$ & 79.3 \\
           &                              & 12.81936 & 
0.01216 & 89.2 & 02 $14^{\prime\prime} \times 27^{\prime\prime}$ & 71.4 \\
\ [\ion{Ne}{3}] & $^{3}P_{2} - {}^{3}P_{1}$ (15.5551\micr) & 15.56331 & 
0.01685 & 17.5 & 01 $14^{\prime\prime} \times 27^{\prime\prime}$ & 14.0 \\
           &                              & 15.56337 & 
0.01404 & 15.7 & 02 $14^{\prime\prime} \times 27^{\prime\prime}$ & 12.6 \\
\ \hh\ & $0 - 0~S(1)$ (17.0348\micr) & 17.04484 & 0.01486 & 
1.38 & 01 $14^{\prime\prime} \times 27^{\prime\prime}$ & ... \\
        &    & 17.04508 & 0.01228 & 
1.50 & 02 $14^{\prime\prime} \times 27^{\prime\prime}$ & ... \\
\ [\ion{S}{3}] & $^{3}P_{1}-{}^{3}P_{2}$ (18.7130\micr) & 18.72210 & 0.01734 & 
34.7 & 01 $14^{\prime\prime} \times 27^{\prime\prime}$ & 27.8 \\
           &                              & 18.72181 & 0.01384 & 
31.5 & 02 $14^{\prime\prime} \times 27^{\prime\prime}$ & 25.2 \\
\ [\ion{Ar}{3}] & $^{3}P_{1}-{}^{3}P_{0}$ (21.8293\micr) & 21.84123 & 0.05191 &
0.70 & 02 $14^{\prime\prime} \times 27^{\prime\prime}$ & 0.6 \\
\ \hh\ & $0 - 0~S(0)$ (28.2188\micr) & 28.23188 & 0.03800 & 
0.78 & 02 $20^{\prime\prime} \times 27^{\prime\prime}$ & ... \\
\ [\ion{S}{3}] & $^{3}P_{0}-{}^{3}P_{1}$ (33.4810\micr) & 33.49818 & 0.03873 & 
83.3 & 01 $20^{\prime\prime} \times 33^{\prime\prime}$ & 58.3 \\
           &                              & 33.49816 & 0.03432 & 
80.3 & 02 $20^{\prime\prime} \times 33^{\prime\prime}$ & 56.2 \\
\ [\ion{Ne}{3}] & $^{3}P_{1}-{}^{3}P_{0}$ (36.0135\micr) & 36.03283 & 0.04118 &
2.92 & 01 $20^{\prime\prime} \times 33^{\prime\prime}$ & 2.04 \\
           &                              & 36.03075 & 0.03535 & 
2.68 & 02 $20^{\prime\prime} \times 33^{\prime\prime}$ & 1.88 \\
\enddata
\tablenotetext{a}
{
Ionic transitions are given as {\em lower level -- upper level}.
$\rm H_{2}$ transitions are labeled
by the upper and lower vibrational quantum numbers followed by
$S(j)$, $Q(j)$, or $O(j)$ which refer, respectively, to transitions for which
$j - j^{\prime}$ equals $-2$, 0, or 2, where $j$ and $j^{\prime}$
are the lower and upper rotational quantum numbers.
}
\tablenotetext{b}
{
The uncertainties on the observed line fluxes are estimated
to be 20\% for most of the lines, and up to 50\% for the faintest lines
(including uncertainties of the absolute calibration, continuum
subtraction, and systematic errors).
}
\tablenotetext{c}
{
01: data from the full scan SWS01 spectrum,
02: data from the individual line scan SWS02 spectra.
The aperture size is given as well.  The SWS02 data are
adopted for the analysis because of their higher S/N ratio.
}
\tablenotetext{d}
{
H recombination and ionic fine-structure lines obtained in the 
$14^{\prime\prime} \times 27^{\prime\prime}$ and
$20^{\prime\prime} \times 33^{\prime\prime}$ apertures
are scaled down to the $14^{\prime\prime} \times 20^{\prime\prime}$ aperture
using the factors 0.8 and 0.7 respectively (see section~\ref{Sub-MIRspec}).
The uncertainties on the scaling factors are estimated to be 10\%.
}
\end{deluxetable}

\clearpage

\begin{deluxetable}{lcccc}
\tablecolumns{5}
\tablewidth{0pt}
\tablenum{5}
\tablecaption{Hydrogen recombination line fluxes in M\,82  \label{tab-HIfluxes}}
\tablehead{
\colhead{Line} & \colhead{Observed flux\,\tablenotemark{a}} & 
\colhead{Aperture\,\tablenotemark{b}} &
\colhead{Scaled flux\,\tablenotemark{c}} & 
\colhead{Reference} \\
 & \colhead{($\rm 10^{-16}~W\,m^{-2}$)} & & 
\colhead{($\rm 10^{-16}~W\,m^{-2}$)} & 
}
\startdata
H$\alpha$ 0.6563\micr & 450 & $90^{\prime\prime} \times 90^{\prime\prime}$ & 
$170 \pm 68$ & 1 \\
H$\alpha$ 0.6563\micr & 510 & $64^{\prime\prime}$ & 
$200 \pm 82$ & 2 \\
Pa$\beta$ 1.2818\micr & $86 \pm 17$ & $30^{\prime\prime}$ & 
$43 \pm 10$ & 3 \\
Br13 1.6109\micr & $1.32 \pm 0.50$ & 3D & 
$1.5 \pm 0.6$ & 4 \\
Br12 1.6407\micr & $1.29 \pm 0.70$ & 3D & 
$1.5 \pm 0.8$ & 4 \\
Br11 1.6807\micr & $1.43 \pm 0.50$ & 3D & 
$1.6 \pm 0.6$ & 4 \\
Br10 1.7362\micr & $3.22 \pm 0.70$ & 3D & 
$3.7 \pm 0.9$ & 4 \\
Br$\gamma$ 2.1655\micr & $4.8 \pm 0.8$ & $8^{\prime\prime}$ & 
$17 \pm 4$ & 4 \\
Br$\beta$ 2.6252\micr & $41.0 \pm 10.5$ & SWS &
$41.0 \pm 10.5$ & 4 \\
Pf$\delta$ 3.2961\micr & $5.9 \pm 2.0$ & SWS &
$5.9 \pm 2.0$ & 4 \\
Pf$\gamma$ 3.7395\micr & $10.7 \pm 3.1$ & SWS &
$10.7 \pm 3.1$ & 4 \\
Hu14 4.0198\micr & $1.3 \pm 0.4$ & SWS &
$1.3 \pm 0.4$ & 4 \\
Br$\alpha$ 4.0512\micr & $81.5 \pm 14.7$ & SWS &
$81.5 \pm 14.7$ & 4 \\
Pf$\beta$ 4.6525\micr & $13.9 \pm 4.1$ & SWS &
$13.9 \pm 4.1$ & 4 \\
Pf$\alpha$ 7.4578\micr & $25.9 \pm 4.7$ & SWS &
$25.9 \pm 4.7$ & 4 \\
H27$\alpha$ 0.95 mm & $(2.11 \pm 0.43) \times 10^{-2}$ & Total &
$(8.1 \pm 2.2) \times 10^{-3}$ & 5 \\
H30$\alpha$ 1.29 mm & $(3.20 \pm 0.38) \times 10^{-3}$ & $21^{\prime\prime}$ &
$(2.0 \pm 0.4) \times 10^{-3}$ & 6 \\
H40$\alpha$ 3.03 mm & $(9.22^{+1.38}_{-1.84}) \times 10^{-4}$ & 
$19^{\prime\prime}$ &  $(6.2 \pm 1.6) \times 10^{-4}$ & 7 \\
H41$\alpha$ 3.26 mm & $(1.32 \pm 0.22) \times 10^{-3}$ & Total &
$(5.1 \pm 1.3) \times 10^{-4}$ & 5 \\
H53$\alpha$ 6.98 mm & $(1.87 \pm 0.28) \times 10^{-4}$ & $41^{\prime\prime}$ &
$(8.6 \pm 2.0) \times 10^{-5}$ & 7 \\
\enddata
\tablenotetext{a}
{
Uncertainties include those of the absolute flux
calibration, continuum subtraction, and systematic errors whenever possible.
For the H$\alpha$ fluxes, none were given in the references,
but those of the beam size corrections should dominate for the scaled fluxes.
}
\tablenotetext{b}
{
All apertures except the 3D and SWS fields of view are centered
on the nucleus of \mqd. ``Total'' refers to measurements integrated over the
entire emission regions in \mqd.
}
\tablenotetext{c}
{
Line fluxes after beam-size correction to match the SWS 
$14^{\prime\prime} \times 20^{\prime\prime}$ aperture
(see section~\ref{Sub-global_Av_gas}).
The uncertainties account for those of the measurements
and of the beam size correction.
}
\tablerefs
{
(1) \citealt{McC87}; (2) \citealt{You88};
(3) Average of the results derived from the Pa$\beta$ fluxes
of \citealt{McL93} and \citealt{Sat95} (see section~\ref{Sub-global_Av_gas});
(4) this work; (5) \citealt{Sea96}; (6) \citealt{Sea94}; (7) \citealt{Pux89}.
}
\end{deluxetable}

\clearpage

\begin{deluxetable}{lccccc}
\tablecolumns{5}
\tablewidth{430pt}
\tablenum{6}
\tablecaption{Hydrogen recombination line ratios and best-fit global
              extinction in M\,82  \label{tab-Av}}
\tablehead{
\colhead{Ratio or property} & \colhead{Intrinsic} & 
\colhead{UFS, Draine} & \colhead{MIX, Draine} &
\colhead{UFS, GC} & \colhead{MIX, GC\,\tablenotemark{a}} \\
 & & \colhead{$A_{V}=4~{\rm mag}$} & \colhead{$A_{V}=43~{\rm mag}$} &
\colhead{$A_{V}=4~{\rm mag}$} & 
\colhead{$A_{V}=52~{\rm mag}$}
}
\startdata
H$\alpha$/Br$\alpha$ & $2.92 \times 10^{1}$ & $3.59 \times 10^{1}$ &
$3.60 \times 10^{1}$ & $3.39 \times 10^{1}$ & $3.02 \times 10^{1}$ \\
H$\alpha$/Br$\alpha$ & $2.92 \times 10^{1}$ & $4.23 \times 10^{1}$ &
$4.24 \times 10^{1}$ & $3.99 \times 10^{1}$ & $3.56 \times 10^{1}$ \\
Pa$\beta$/Br$\alpha$ & $1.77 \times 10^{0}$ & $1.23 \times 10^{0}$ &
$2.98 \times 10^{0}$  & $1.16 \times 10^{0}$ & $2.50 \times 10^{0}$ \\
Br13/Br$\alpha$ & $4.48 \times 10^{-2}$ & $3.10 \times 10^{-2}$ &
$6.97 \times 10^{-2}$  & $2.93 \times 10^{-2}$ & $5.84 \times 10^{-2}$ \\
Br12/Br$\alpha$ & $5.72 \times 10^{-2}$ & $3.04 \times 10^{-2}$ &
$6.73 \times 10^{-2}$  & $2.86 \times 10^{-2}$ & $5.64 \times 10^{-2}$ \\
Br11/Br$\alpha$ & $7.48 \times 10^{-2}$ & $3.16 \times 10^{-2}$ &
$6.89 \times 10^{-2}$  & $2.98 \times 10^{-2}$ & $5.78 \times 10^{-2}$ \\
Br10/Br$\alpha$ & $1.01 \times 10^{-1}$ & $7.06 \times 10^{-2}$ &
$1.51 \times 10^{-1}$  & $6.66 \times 10^{-2}$ & $1.26 \times 10^{-1}$ \\
Br$\gamma$/Br$\alpha$ & $3.17 \times 10^{-1}$ & $2.70 \times 10^{-1}$ &
$4.75 \times 10^{-1}$  & $2.55 \times 10^{-1}$ & $3.95 \times 10^{-1}$ \\
Br$\beta$/Br$\alpha$ & $5.35 \times 10^{-1}$ & $5.83 \times 10^{-1}$ &
$8.51 \times 10^{-1}$  & $5.51 \times 10^{-1}$ & $6.99 \times 10^{-1}$ \\
Pf$\delta$/Br$\alpha$ & $8.75 \times 10^{-2}$ & $7.66 \times 10^{-2}$ &
$9.03 \times 10^{-2}$  & $7.77 \times 10^{-2}$ & $9.43 \times 10^{-2}$ \\
Pf$\gamma$/Br$\alpha$ & $1.29 \times 10^{-1}$ & $1.34 \times 10^{-1}$ &
$1.42 \times 10^{-1}$  & $1.34 \times 10^{-1}$ & $1.41 \times 10^{-1}$ \\
Hu14/Br$\alpha$ & $1.45 \times 10^{-2}$ & $1.60 \times 10^{-2}$ &
$1.61 \times 10^{-2}$  & $1.60 \times 10^{-2}$ & $1.61 \times 10^{-2}$ \\
Pf$\beta$/Br$\alpha$ & $2.01 \times 10^{-1}$ & $1.66 \times 10^{-1}$ &
$1.51 \times 10^{-1}$  & $1.71 \times 10^{-1}$ & $1.72 \times 10^{-1}$ \\
Pf$\alpha$/Br$\alpha$ & $3.33 \times 10^{-1}$ & $2.92 \times 10^{-1}$ &
$2.16 \times 10^{-1}$  & $3.11 \times 10^{-1}$ & $2.89 \times 10^{-1}$ \\
H27$\alpha$/Br$\alpha$ & $2.56 \times 10^{-5}$ & $8.73 \times 10^{-5}$ &
$5.36 \times 10^{-5}$  & $8.24 \times 10^{-5}$ & $3.72 \times 10^{-5}$ \\
H30$\alpha$/Br$\alpha$ & $1.33 \times 10^{-5}$ & $2.16 \times 10^{-5}$ &
$1.32 \times 10^{-5}$  & $2.03 \times 10^{-5}$ & $9.19 \times 10^{-6}$ \\
H40$\alpha$/Br$\alpha$ & $2.48 \times 10^{-6}$ & $6.68 \times 10^{-6}$ &
$4.10 \times 10^{-6}$  & $6.31 \times 10^{-6}$ & $2.85 \times 10^{-6}$ \\
H41$\alpha$/Br$\alpha$ & $2.04 \times 10^{-6}$ & $5.50 \times 10^{-6}$ &
$3.38 \times 10^{-6}$  & $5.19 \times 10^{-6}$ & $2.34 \times 10^{-6}$ \\
H53$\alpha$/Br$\alpha$ & $4.37 \times 10^{-7}$ & $9.27 \times 10^{-7}$ &
$5.69 \times 10^{-7}$  & $8.75 \times 10^{-7}$ & $3.95 \times 10^{-7}$ \\
\hline
$F^{0}_{\rm Br\alpha}~({\rm W\,m^{-2}})$ & 
... & $9.28 \times 10^{-15}$ & $1.51 \times 10^{-14}$ &
$9.83 \times 10^{-15}$ & $2.18 \times 10^{-14}$ \\
$\log <Q^{0}_{\rm Lyc}~({\rm s^{-1}})>$\tablenotemark{b} & 
... & $53.49_{-0.41}^{+0.21}$ &
$53.70_{-0.14}^{+0.10}$ & $53.50_{-0.40}^{+0.20}$ & 
$53.79_{-0.09}^{+0.08}$ \\
\enddata
\tablecomments
{
The results reported are the
line fluxes relative to Br$\alpha$ corrected for the best-fit extinction.
The intrinsic line fluxes can be recovered using the dereddened Br$\alpha$
flux ($F_{\rm Br\alpha}^{0}$).  ``UFS'' and ``MIX'' stand for
uniform foreground screen and mixed model, respectively.
``Draine'' indicates that the extinction law of \citealt{Dra89} was assumed
throughout the infrared regime, while ``GC'' indicates that is was
amended by the Galactic Center law \citep{Lut99} between 3\micr\ and 10\micr.
}
\tablenotetext{a}
{
Adopted extinction model and $\rm 3 - 10~\mu m$ extinction law
(section~\ref{Sub-adopted_Av_gas}).
}
\tablenotetext{b}
{
$<Q^{0}_{\rm Lyc}>$ is the average Lyman continuum photon emission rate
derived from the extinction-corrected line fluxes. 
The uncertainties represent the dispersion of the individual values.
}
\end{deluxetable}

\clearpage

\begin{deluxetable}{lcccc}
\tablecolumns{5}
\tablewidth{0pt}
\tablenum{7}
\tablecaption{Extinction towards the ionized gas for selected regions in M\,82
              \label{tab-Avlocal}}
\tablehead{
\colhead{Property} & \colhead{Central 35~pc} & 
\colhead{B1} & \colhead{B2} & \colhead{3D field}}
\startdata
$A_{V}^{\rm UFS}$ (mag) & $10 \pm 5$ & $8 \pm 2$ & $11 \pm 2$ & $9 \pm 3$ \\
$A_{V}^{\rm MIX}$ (mag) & $23 \pm 10$ & $27 \pm 7$ & $45 \pm 20$ & 
$36 \pm 16$ \\
\hline
$\log \left(Q^{0}_{\rm Lyc}~[{\rm s^{-1}}]\right)$\ \tablenotemark{a} & 
$52.03_{-0.17}^{+0.12}$ & $52.37_{-0.11}^{+0.09}$ & 
$52.51_{-0.23}^{+0.15}$ & $53.64_{-0.23}^{+0.15}$ \\
\enddata
\tablenotetext{a}
{
$Q^{0}_{\rm Lyc}$ derived from the observed Br$\gamma$ 
fluxes corrected for the adopted $A_{V}^{\rm MIX}$.
}
\end{deluxetable}

\clearpage

\begin{deluxetable}{lccl}
\tablecolumns{4}
\tablewidth{0pt}
\tablenum{8}
\tablecaption{Gas-phase abundances of Ne, Ar, and S in M\,82  \label{tab-abund}}
\tablehead{
\colhead{Line} & \colhead{$F_{\lambda}$\ \tablenotemark{a}} &
\colhead{$j_{\lambda}$} & \colhead{Abundance\,\tablenotemark{b}} \\
 & \colhead{($\rm W\,m^{-2}$)} & \colhead{($\rm erg\,cm^{3}\,s^{-1}$)} & 
}
\startdata
\ Br$\alpha$ & 
$2.38 \times 10^{-14}$ & $2.29 \times 10^{-26}$ & \multicolumn{1}{c}{...} \\
\hline
\ [\ion{Ne}{2}] 12.8\micr & 
$1.37 \times 10^{-13}$ & $1.05 \times 10^{-21}$ & 
${\rm Ne^{+}/H} = 1.26 \times 10^{-4}$ \\
\ [\ion{Ne}{3}] 15.6\micr & 
$2.17 \times 10^{-14}$ & $1.89 \times 10^{-21}$ & 
${\rm Ne^{++}/H} = 1.11 \times 10^{-5}$ \\
\ [\ion{Ne}{3}] 36.0\micr & 
$2.41 \times 10^{-15}$ & $4.71 \times 10^{-22}$ & 
${\rm Ne^{++}/H} = 4.92 \times 10^{-6}$ \\
  &  &  & $\rm [Ne/H] \approx 0.08~dex$ \\
\hline
\ [\ion{Ar}{2}] 6.99\micr & 
$6.51 \times 10^{-14}$ & $1.55 \times 10^{-20}$ & 
${\rm Ar^{+}/H} = 4.03 \times 10^{-6}$ \\
\ [\ion{Ar}{3}] 8.99\micr & 
$1.70 \times 10^{-14}$ & $1.25 \times 10^{-20}$ & 
${\rm Ar^{++}/H} = 1.31 \times 10^{-6}$ \\
\ [\ion{Ar}{3}] 21.8\micr & 
$1.09 \times 10^{-15}$ & $4.39 \times 10^{-21}$ & 
${\rm Ar^{++}/H} = 2.38 \times 10^{-7}$ \\
  &  &  & $\rm [Ar/H] \approx 0.1~dex$ \\
\hline
\ [\ion{S}{3}] 18.7\micr & 
$5.21 \times 10^{-14}$ & $1.87 \times 10^{-20}$ & 
${\rm S^{++}/H} = 2.69 \times 10^{-6}$ \\
\ [\ion{S}{3}] 33.5\micr & 
$7.34 \times 10^{-14}$ & $1.48 \times 10^{-20}$ & 
${\rm S^{++}/H} = 4.79 \times 10^{-6}$ \\
\ [\ion{S}{4}] 10.5\micr & 
$5.70 \times 10^{-15}$ & $5.14 \times 10^{-20}$ & 
${\rm S^{+++}/H} = 1.07 \times 10^{-7}$ \\
  &  &  & $\rm [S/H] \approx -0.6~dex$ \\
\enddata
\tablenotetext{a}
{
Fluxes for the $14^{\prime\prime} \times 20^{\prime\prime}$ SWS aperture
corrected for $A_{V}^{\rm MIX} = 52~{\rm mag}$.
The Br$\alpha$ flux is computed from the Lyman continuum photon emission rate
(table~\ref{tab-Av}).
}
\tablenotetext{b}
{
Ionic or elemental number abundance relative to hydrogen, with
estimated uncertainties of approximately $\pm 50\%$, and
up to a factor of two for the ionic abundances derived from the
weakest lines.  The total elemental abundance is given as
$\rm [X/H] = \log (X/H)_{M\,82} - \log (X/H)_{\odot}$, with 
$\rm Ne/H = 1.17 \times 10^{-4}$, $\rm Ar/H = 3.98 \times 10^{-6}$,
and $\rm S/H = 1.62 \times 10^{-5}$ for the solar composition
(\citealt{Gre89} and \citealt{Gre93}, as quoted by \citealt{Ferl96}).
}
\end{deluxetable}

\clearpage

\begin{deluxetable}{lllll}
\tablecolumns{5}
\tablewidth{0pt}
\tablenum{9}
\tablecaption{Effective temperature and number of OB stars for selected
              regions in M\,82  \label{tab-TeffOB}}
\tablehead{
\colhead{Region} & \colhead{Diagnostic} & 
\colhead{Ratio\,\tablenotemark{a}} & \colhead{\teffob} &
\colhead{$N_{\rm O8.5\,V}$\,\tablenotemark{b}} \\
  &  &  & \colhead{(K)} & 
} 
\startdata
Starburst core\,\tablenotemark{c} & ... & ... & 37400 & $2.4 \times 10^{5}$ \\
SWS $14^{\prime\prime} \times 20^{\prime\prime}$ &
[\ion{Ne}{3}] 15.6\micr/[\ion{Ne}{2}] 12.8\micr & 
$0.16 \pm 0.04$ & $37400 \pm 400$ & $1.2 \times 10^{5}$ \\
 &  [\ion{Ar}{3}] 8.99\micr/[\ion{Ar}{2}] 6.99\micr &
$0.26 \pm 0.08$ & $33500 \pm 500$ & \\
 &  [\ion{S}{4}] 10.5\micr/[\ion{S}{3}] 18.7\micr &
$0.11 \pm 0.04$ & $39900 \pm 1000$ & \\
Central 35~pc & \ion{He}{1}~$\rm 2.058$/Br$\gamma$ &
$0.52 \pm 0.08$ & $35700 \pm 800$ & 2040 \\
 & \ion{He}{1}~$\rm 1.701$/Br10 & $< 0.13$ & $<34200$ & \\
B1 & \ion{He}{1}~$\rm 2.058$/Br$\gamma$ &
$0.55 \pm 0.02$ & $36000 \pm 200$ & 4470 \\
 & \ion{He}{1}~$\rm 1.701$/Br10 & $0.22 \pm 0.06$ & $36400 \pm 1600$ & \\
B2 & \ion{He}{1}~$\rm 2.058$/Br$\gamma$ &
$0.52 \pm 0.02$ & $35600 \pm 200$ & 6170 \\
 & \ion{He}{1}~$\rm 1.701$/Br10 & $0.22 \pm 0.06$ & $36400 \pm 1600$ & \\
3D field & \ion{He}{1}~$\rm 2.058$/Br$\gamma$ &
$0.55 \pm 0.03$ & $36000 \pm 300$ & $8.3 \times 10^{4}$ \\
 & \ion{He}{1}~$\rm 1.701$/Br10 & $0.20 \pm 0.09$ & $36000 \pm 2300$ & \\
\enddata
\tablenotetext{a}
{
The ratios are corrected for extinction and, when appropriate,
for aperture size.
The uncertainties include those of the continuum subtraction, systematic 
effects, and extinction correction.
}
\tablenotetext{b}
{
$N_{\rm O8.5\,V}$ is the number of equivalent \ion{O8.5}{5} stars
required to produce the derived intrinsic Lyman continuum photon
emission rates (see section~\ref{Sub-NOB}).
}
\tablenotetext{c}
{
The \teffob\ and representative type for the SWS field of view is
adopted for the entire starburst core (see section~\ref{Sub-NOB}).
}
\end{deluxetable}

\clearpage

\begin{deluxetable}{lcccccc}
\tablecolumns{7}
\tablewidth{0pt}
\tablenum{10}
\tablecaption{Properties derived from the near-infrared continuum emission
              for selected regions  \label{tab-NIRcont}}
\tablehead{
\colhead{Property} & Units &
\colhead{Central 35~pc} & \colhead{B1} & \colhead{B2} & 
\colhead{3D field} & \colhead{Starburst core}
}
\startdata
Spectral type & ... & \ion{K5}{1} & \ion{K2}{1} &
\ion{K4}{1} & \ion{K4}{1} & \ion{K4}{1} \\
$D_{1.6}$ & \% & 0 & 0 & 0 & 0 & 0 \\
$D_{2.3}$ & \% & 0 & 25 & 15 & 10 & 0 \\
$D_{H}^{\rm OB}$ & \% & 0 & 2 & 1 & 0 & 0 \\ 
$D_{K}^{\rm OB}$ & \% & 0 & 1 & 1 & 0 & 0 \\ 
$D_{H}^{\rm Neb}$ & \% & $< 1$ & $< 8$ & $< 5$ & $< 3$ & 0 \\
$D_{K}^{\rm Neb}$ & \% & $< 1$ & $< 7$ & $< 4$ & $< 3$ & 0 \\
$A_{V}^{\rm UFS}$ & mag & 
$10 \pm 4$ & $6 \pm 3$ & $8 \pm 3$ & $6 \pm 3$ & 9 \\
$A_{V}^{\rm MIX}$ & mag & 
$45 \pm 20$ & $25 \pm 10$ & $28 \pm 10$ & $17 \pm 7$ & 21 \\
$L_{K}$ & $\rm 10^{8}~L_{\odot}$ &
$0.56 \pm 0.23$ & $0.074 \pm 0.024$ & $0.15 \pm 0.05$ &
$2.7 \pm 0.8$ & $13 \pm 4$ \\
$N_{\star}$ & ... &
$1.1 \times 10^{4}$ & 4400 & 5800 & $1.0 \times 10^{5}$ & $5.0 \times 10^{5}$ \\
\enddata
\tablecomments
{
\dilh\ and \dilk\ are the total amounts of dilution near 1.6\micr\
and 2.3\micr, with typical uncertainties of $\pm 10\%$.
$D_{H}^{\rm OB}$ and $D_{K}^{\rm OB}$ are the contributions
to the $H$- and $K$-band flux densities from OB stars while
$D_{H}^{\rm Neb}$ and $D_{K}^{\rm Neb}$ are those from
nebular free-free and free-bound processes.
The extinction values are those derived from the spectral fits
except for the starburst core, which correspond to the effective
extinction inferred as described in appendix~\ref{App-constraints}.
We adopt the results for the mixed model throughout this work.
$L_{K}$ is the intrinsic $K$-band luminosity from the evolved stars
corrected for dilution and extinction while $N_{\star}$ is the number
of representative stars required to produce $L_{K}$.
}
\end{deluxetable}

\clearpage

\begin{deluxetable}{lcccc}
\tablecolumns{5}
\tablewidth{0pt}
\tablenum{11}
\tablecaption{Properties relevant in the determination
              of the effective ionization parameter \label{tab-logU}}
\tablehead{
\colhead{Property} & \colhead{Units} & 
\colhead{B1} & \colhead{B2} & \colhead{Starburst core} 
}
\startdata
\multicolumn{5}{c}{Geometrical properties} \\
Sphere radius & pc & 19.5 & 19.5 & 225 \\
Column/disk radius & pc & 19.5 & 19.5 & 200 \\
Column/disk length & pc & 230 & 365 & 200 \\
\hline
\multicolumn{5}{c}{Gas properties} \\
$M_{\rm cl}$ & $\rm M_{\odot}$ & 220 & 220 & 545 \\ 
$r_{\rm cl}$ & pc & 0.4 & 0.4 & 0.6 \\
$M_{\rm H_{2}}$ & $\rm M_{\odot}$ & 
           $1.1 \times 10^{6}$ & $2.3 \times 10^{6}$ & $1.8 \times 10^{8}$ \\ 
$M_{\rm H^{+}}$ & $\rm M_{\odot}$ & 
           $1.4 \times 10^{5}$ & $2.0 \times 10^{5}$ & $7.6 \times 10^{6}$ \\
$N_{\rm cl}$ & ... & 5000 & 10450 & $3.3 \times 10^{5}$ \\
$n_{\rm cl}$ & $\rm 10^{-2}~pc^{-3}$ & 
           $16 - 1.8$ & $34 - 2.4$ & $0.69 - 1.3$ \\
$d_{\rm cl - cl}$ & pc & 
           $2.3 - 4.7$ & $1.8 - 4.3$ & $6.5 - 5.3$ \\
$\mathcal{N}_{\rm cl}$ & $\rm cm^{-2}$ & 
           $5.5 \times 10^{22}$ & $5.5 \times 10^{22}$ & $6.0 \times 10^{22}$ \\
$r_{\rm i}$ & pc & 1.0 & 0.88 & 0.99 \\
\hline
\multicolumn{5}{c}{Ionizing clusters properties} \\
$Q^{0}_{\rm Lyc}$ & $\rm s^{-1}$ & 
        $2.34 \times 10^{52}$ & $3.24 \times 10^{52}$ & $1.23 \times 10^{54}$ \\
$N_{\star}$ & ... & 2660 & 3690 & $1.4 \times 10^{5}$ \\
$n_{\star}$ & $\rm 10^{-2}~pc^{-3}$ & 
           $8.6 - 0.97$ & $12 - 0.85$ & $0.29 - 0.56$ \\
$d_{\star - \star}$ & pc & 
           $2.8 - 5.8$ & $2.5 - 6.1$ & $8.7 - 7.0$ \\
\hline
\multicolumn{5}{c}{Structural and nebular properties} \\
$\log U$\ \tablenotemark{a} & ... & $-1.2$ & $-1.1$ & $-1.7$ \\
$\log U_{\rm eff}$\ \tablenotemark{a} & ... & $-2.3$ & $-2.4$ & $-2.4$ \\
\enddata
\tablecomments
{
Two limiting cases are considered for the volume of each region:
a sphere or an edge-on disk for the starburst core,
and spheres or columns through the edge-on disk for B1 and B2.
The ranges for the space density and mean separation of the
gas clouds and stellar clusters correspond to these limiting cases.
}
\tablenotetext{a}
{
The $\log U$ is computed from the direct application of equation~(\ref{Eq-U}) 
with the total intrinsic Lyman continuum photon emission rates and assuming
spherical volumes.  The $\log U_{\rm eff}$ is derived from 
equation~(\ref{Eq-Ueff})
as appropriate for the randomized distribution of clouds and clusters 
within the starburst regions of \mqd.
}
\end{deluxetable}

\clearpage

\begin{deluxetable}{lccc}
\tablecolumns{4}
\tablewidth{0pt}
\tablenum{12}
\tablecaption{Bolometric luminosity estimates for
              selected regions in M\,82  \label{tab-Lbol}}
\tablehead{
\colhead{Region} & 
\colhead{$L_{\rm IR}$} &
\colhead{$L_{\rm bol}^{\rm evolved~stars}$} & 
\colhead{$L_{\rm bol}^{\rm tot}$} \\
 & \colhead{($\rm 10^{8}~L_{\odot}$)} & 
\colhead{($\rm 10^{8}~L_{\odot}$)} & 
\colhead{($\rm 10^{8}~L_{\odot}$)}
}
\startdata
Central 35~pc & $5.1 \pm 1.8$ & $11 \pm 7$ & $18 \pm 9$ \\
B1 & $5.7 \pm 2.0$ & $1.5 \pm 0.9$ & $8.9 \pm 3.1$ \\
B2 & $7.1 \pm 2.5$ & $2.9 \pm 1.7$ & $12 \pm 4$ \\
3D field & $130 \pm 50$ & $55 \pm 31$ & $220 \pm 90$ \\ 
Starburst core & $300 \pm 30$ & $270 \pm 80$ & $660 \pm 120$ \\ 
\enddata
\tablecomments
{
\lir\ is the infrared luminosity between $\rm 5~\mu m$ and $\rm 300~\mu m$ 
and $L_{\rm bol}^{\rm evolved~stars}$ is the bolometric luminosity from the 
evolved stellar population.
$L_{\rm bol}^{\rm tot}$ includes $L_{\rm IR}$, 
$L_{\rm bol}^{\rm evolved~stars}$, and an additional 30\% of
$L_{\rm IR}$ ($\pm 10\%$) to account for light escaping in
directions perpendicular to the galactic plane of M\,82.
The uncertainties account for those on the absolute calibration of the 
data used to derive the quantities, on the conversion factors between
the 12.4\micr\ flux density and $L_{\rm IR}$, and between $L_{K}$ and 
$L_{\rm bol}^{\rm evolved~stars}$, on the extinction, on the
hot dust contribution to $L_{K}$, and on the fraction of escaping light.
} 
\end{deluxetable}

\clearpage

\begin{deluxetable}{lcccc}
\tablecolumns{5}
\tablewidth{0pt}
\tablenum{13}
\tablecaption{Dynamical, gaseous, and stellar mass estimates in M\,82
              \label{tab-Mass}}
\tablehead{
\colhead{Region} & 
\colhead{$M_{\rm dyn}$} &
\colhead{$M_{\rm H_{2}}$} &
\colhead{$M_{\rm H^{+}}$\ \tablenotemark{a}} &
\colhead{$M^{\star}$} \\
 & \colhead{($\rm 10^{6}~M_{\odot}$)} & 
\colhead{($\rm 10^{6}~M_{\odot}$)} & 
\colhead{($\rm 10^{6}~M_{\odot}$)} & 
\colhead{($\rm 10^{6}~M_{\odot}$)} 
}
\startdata
Central 35~pc & $80 \pm 20$ & $0.82_{-0.41}^{+1.64}$ & 
$0.066_{-0.049}^{+0.25}$ & $79_{-21}^{+22}$ \\ 
Starburst core & $800 \pm 200$ & $180 \pm 50$ &
$7.6_{-5.4}^{+27}$ & $610_{-250}^{+270}$ \\ 
\enddata
\tablenotetext{a}
{
The uncertainties include those on the intrinsic Lyman continuum luminosity,
and on the electron density and temperature.
}
\end{deluxetable}


\setcounter{figure}{0}

\clearpage

\begin{figure*}[p]
\resizebox{\hsize}{!}{\includegraphics{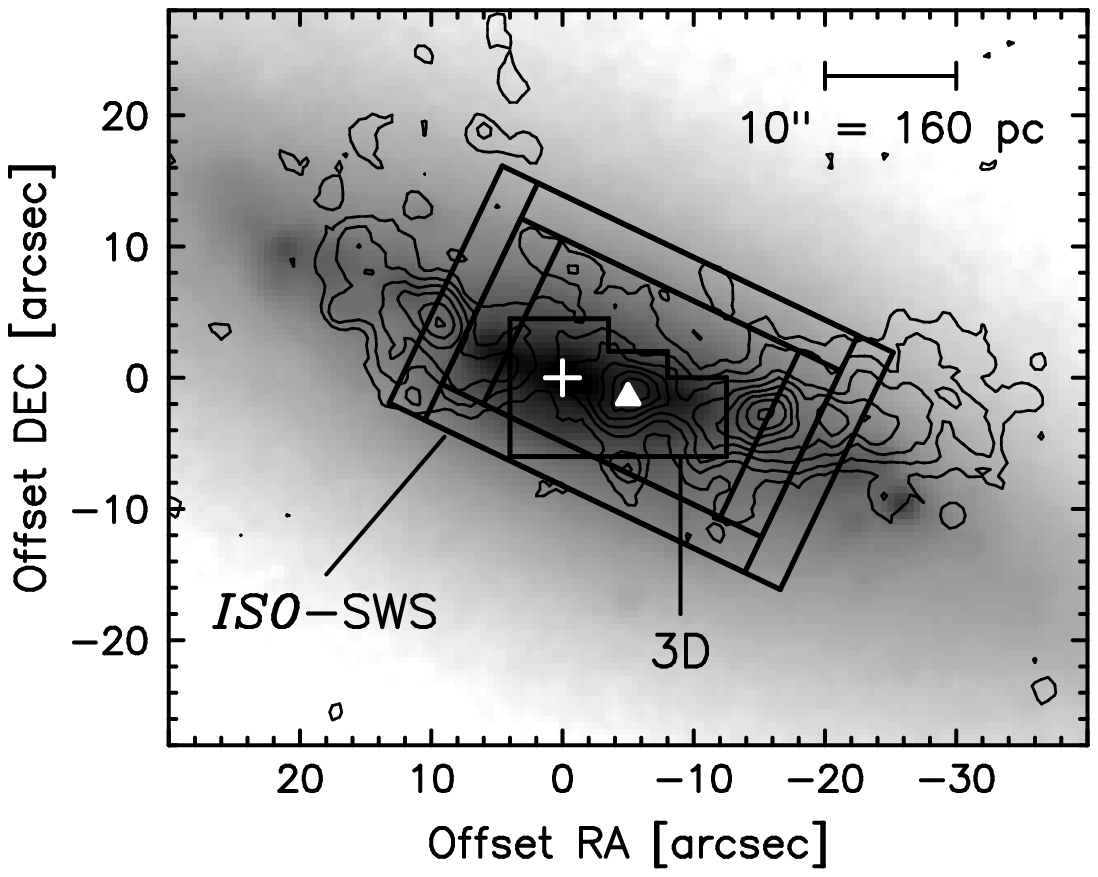}}
\vspace{-4.0cm}
\figcaption[]{}
\end{figure*}

\clearpage

\begin{figure*}[p]
\resizebox{\hsize}{!}{\includegraphics{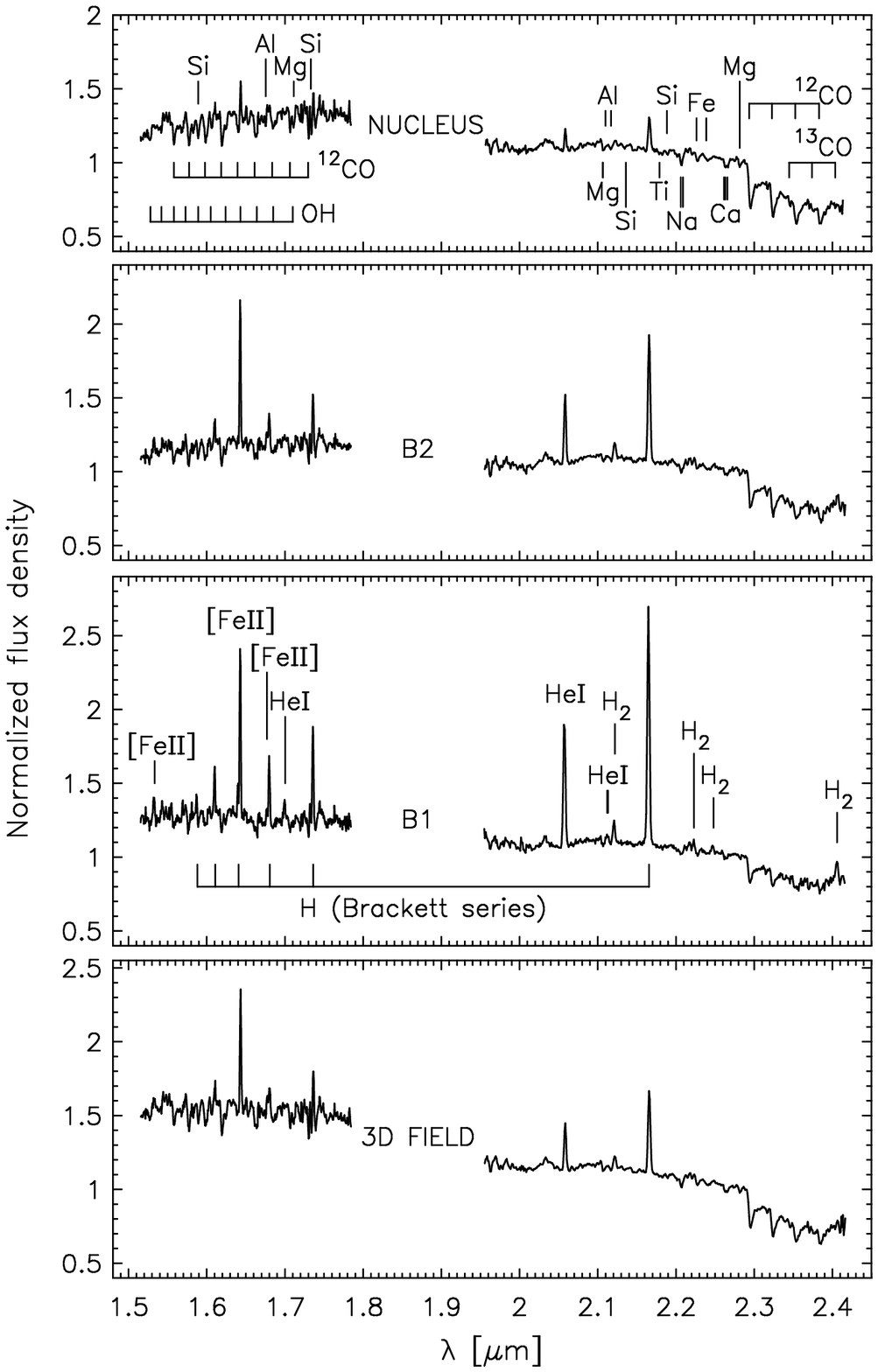}}
\vspace{-2.0cm}
\figcaption[]{}
\end{figure*}

\clearpage

\begin{figure*}[p]
\resizebox{\hsize}{!}{\includegraphics{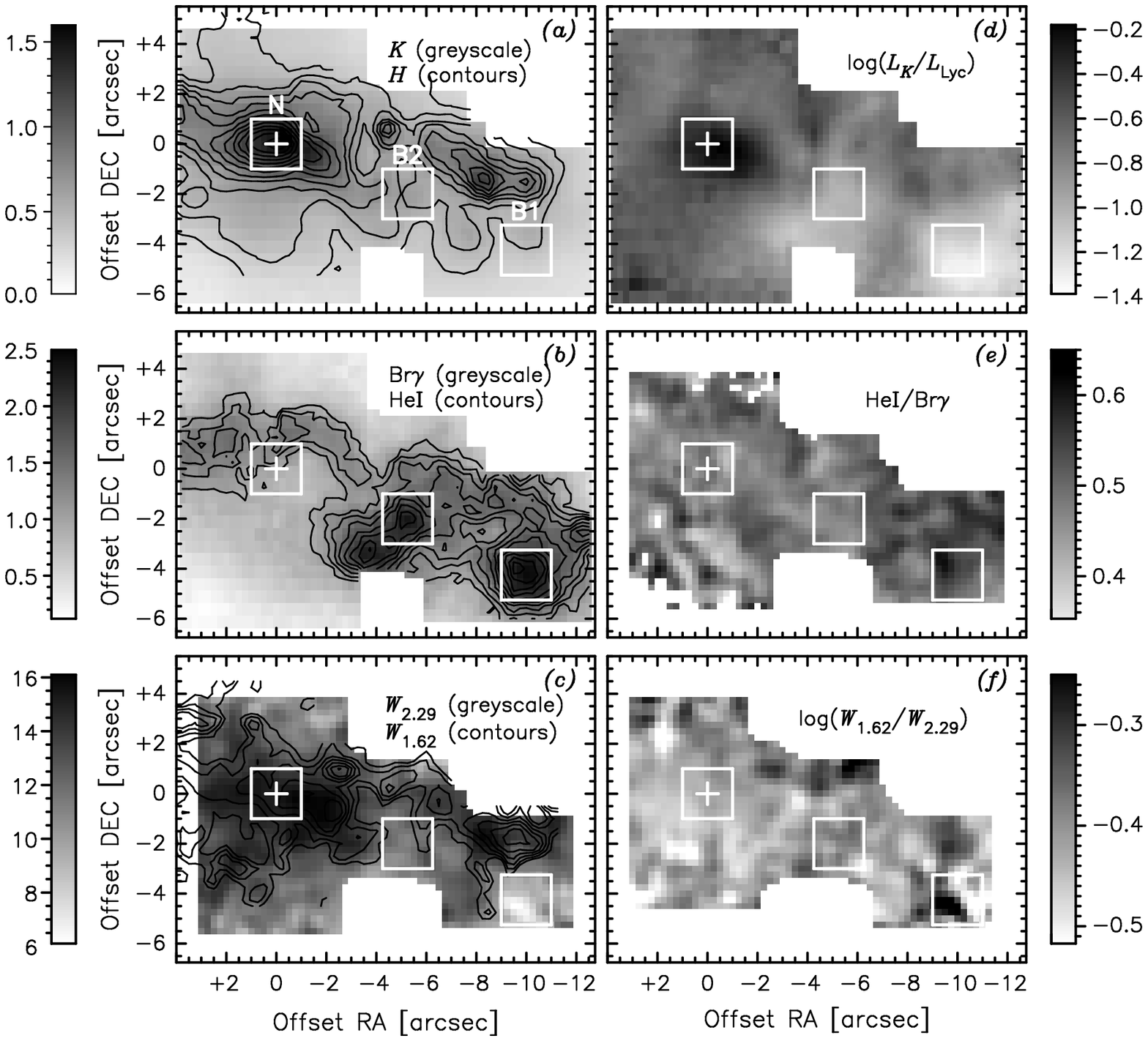}}
\vspace{-2.5cm}
\figcaption[]{}
\end{figure*}

\clearpage

\begin{figure*}[p]
\resizebox{\hsize}{!}{\includegraphics{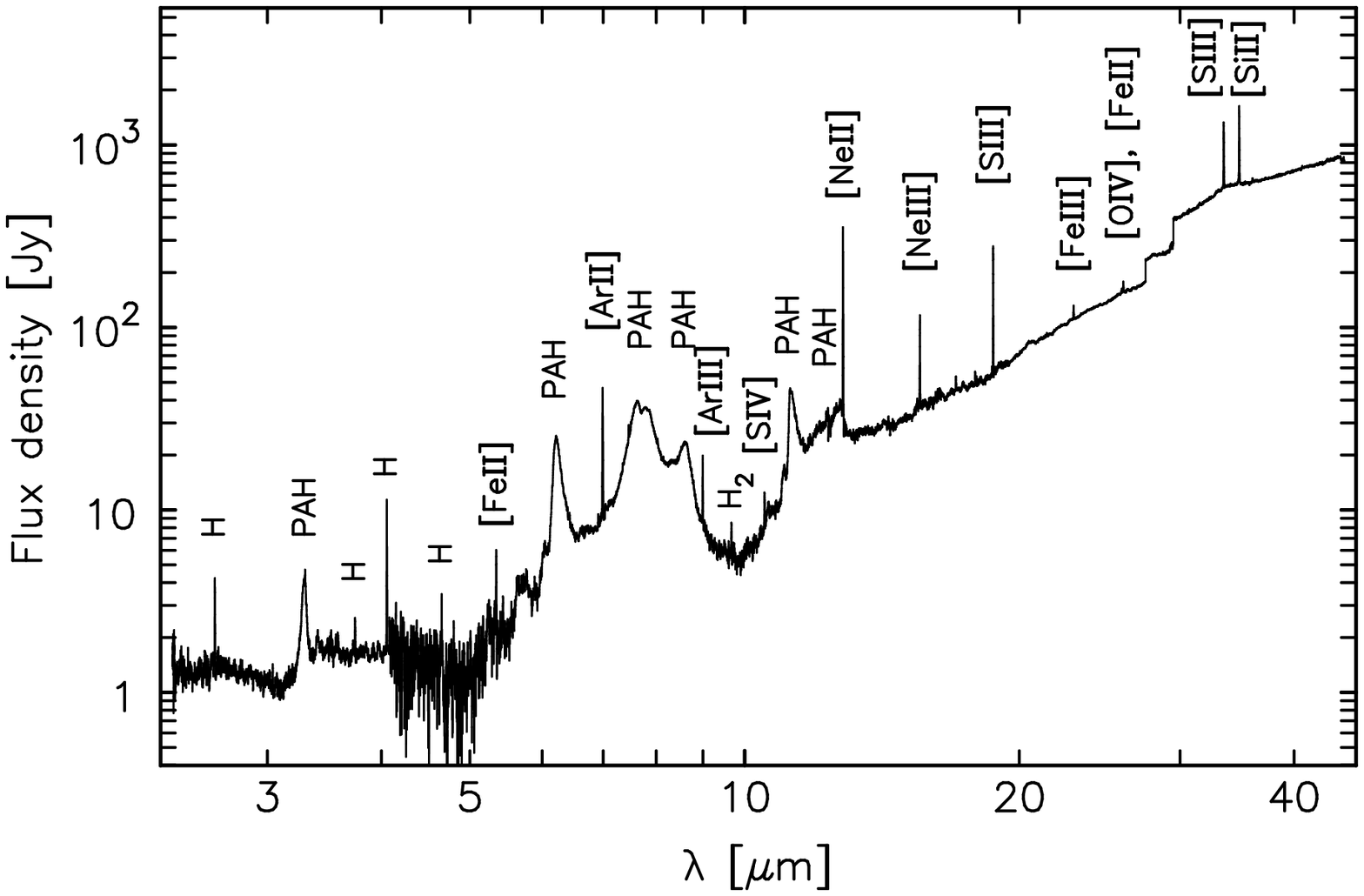}}
\vspace{-3.5cm}
\figcaption[]{}
\end{figure*}

\clearpage

\begin{figure*}[p]
\resizebox{\hsize}{!}{\includegraphics{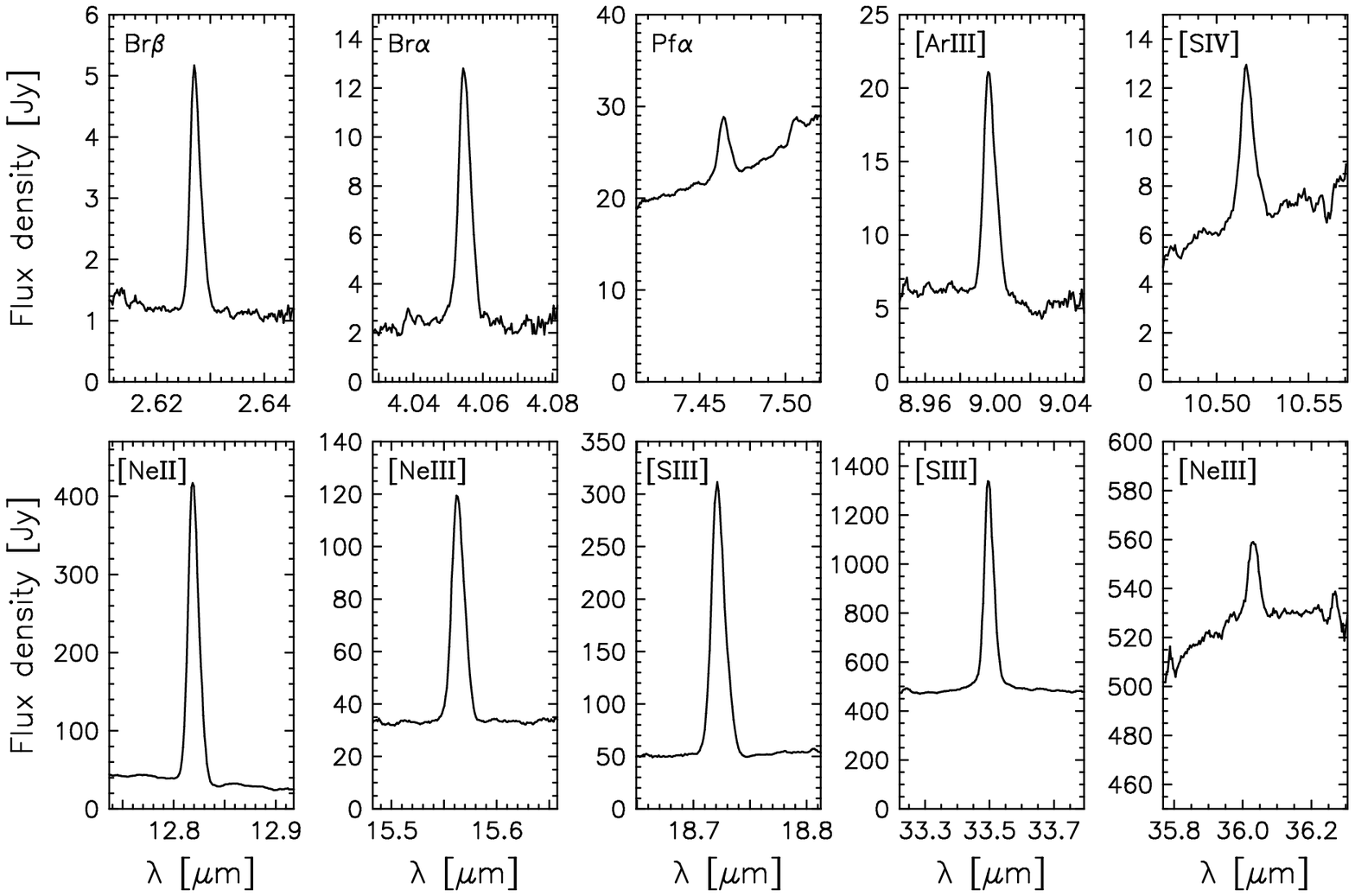}}
\vspace{-3.5cm}
\figcaption[]{}
\end{figure*}

\clearpage

\begin{figure*}[p]
\resizebox{\hsize}{!}{\includegraphics{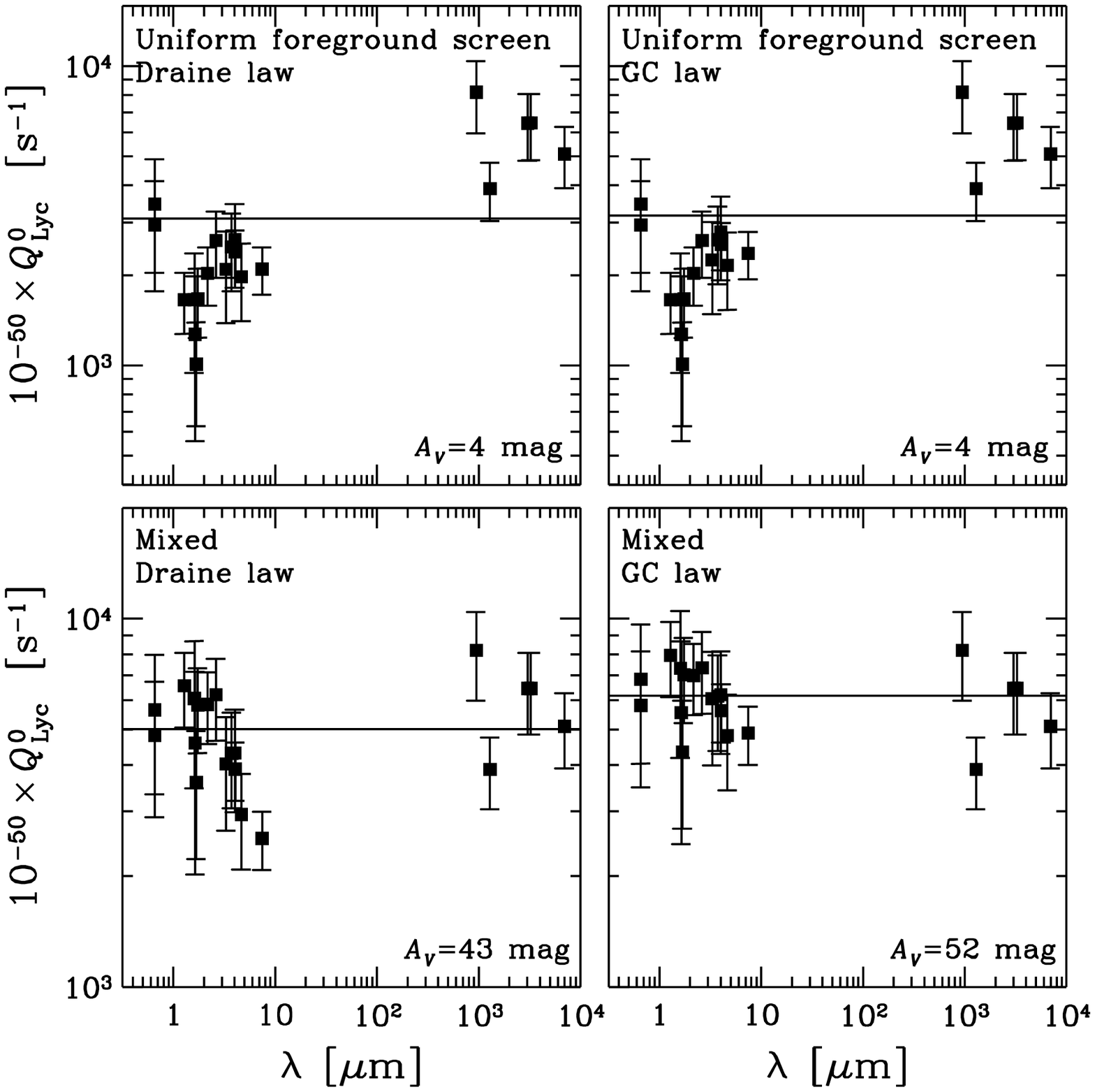}}
\vspace{1.0cm}
\figcaption[]{}
\end{figure*}

\clearpage

\begin{figure*}[p]
\resizebox{\hsize}{!}{\includegraphics{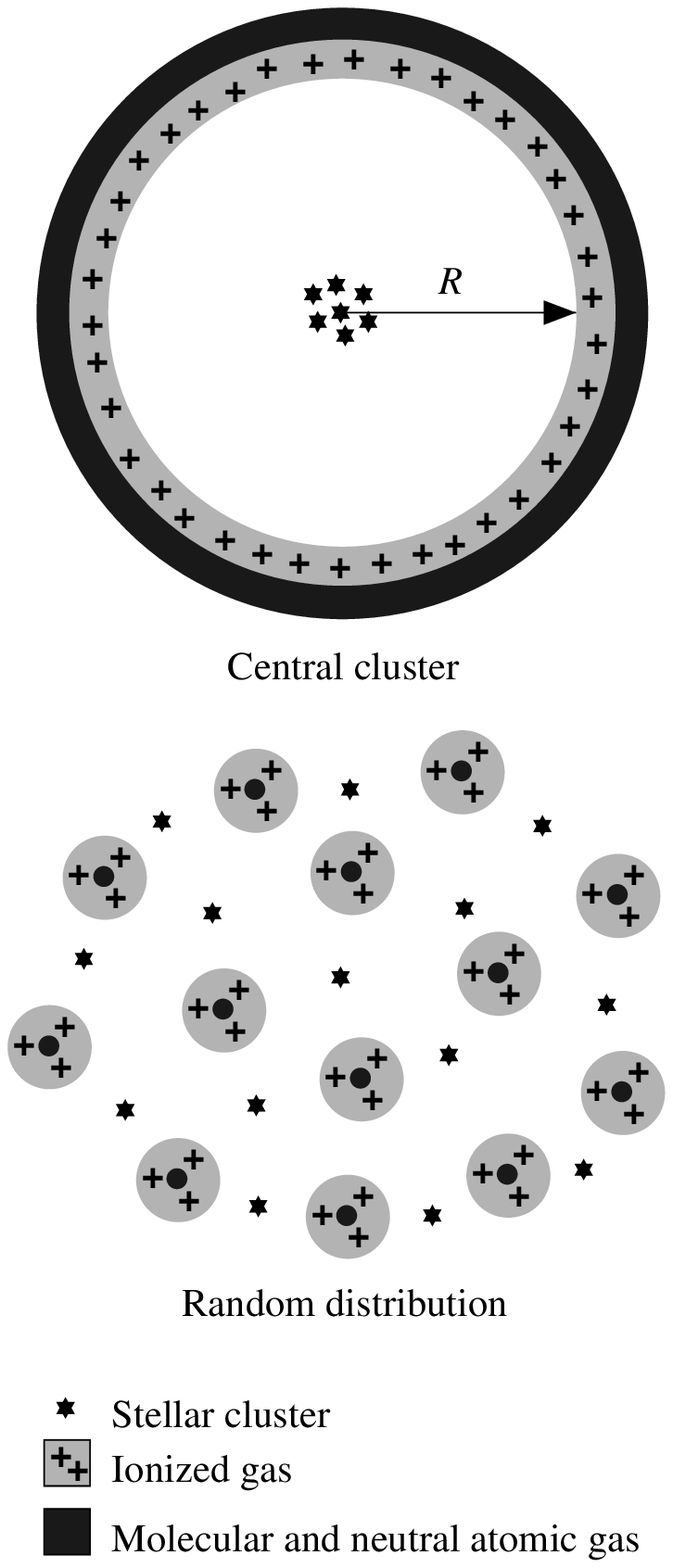}}
\vspace{-4.0cm}
\figcaption[]{}
\end{figure*}

\clearpage

\begin{figure*}[p]
\resizebox{\hsize}{!}{\includegraphics{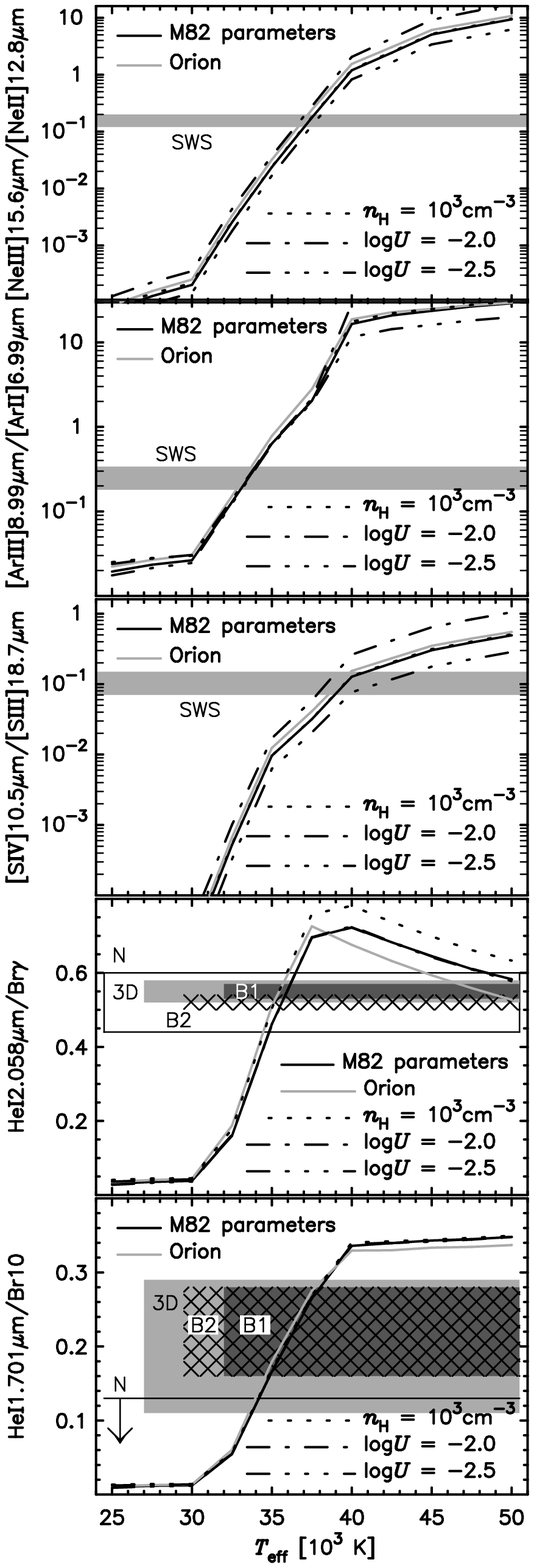}}
\vspace{-1.2cm}
\figcaption[]{}
\end{figure*}

\clearpage

\begin{figure*}[p]
\resizebox{\hsize}{!}{\includegraphics{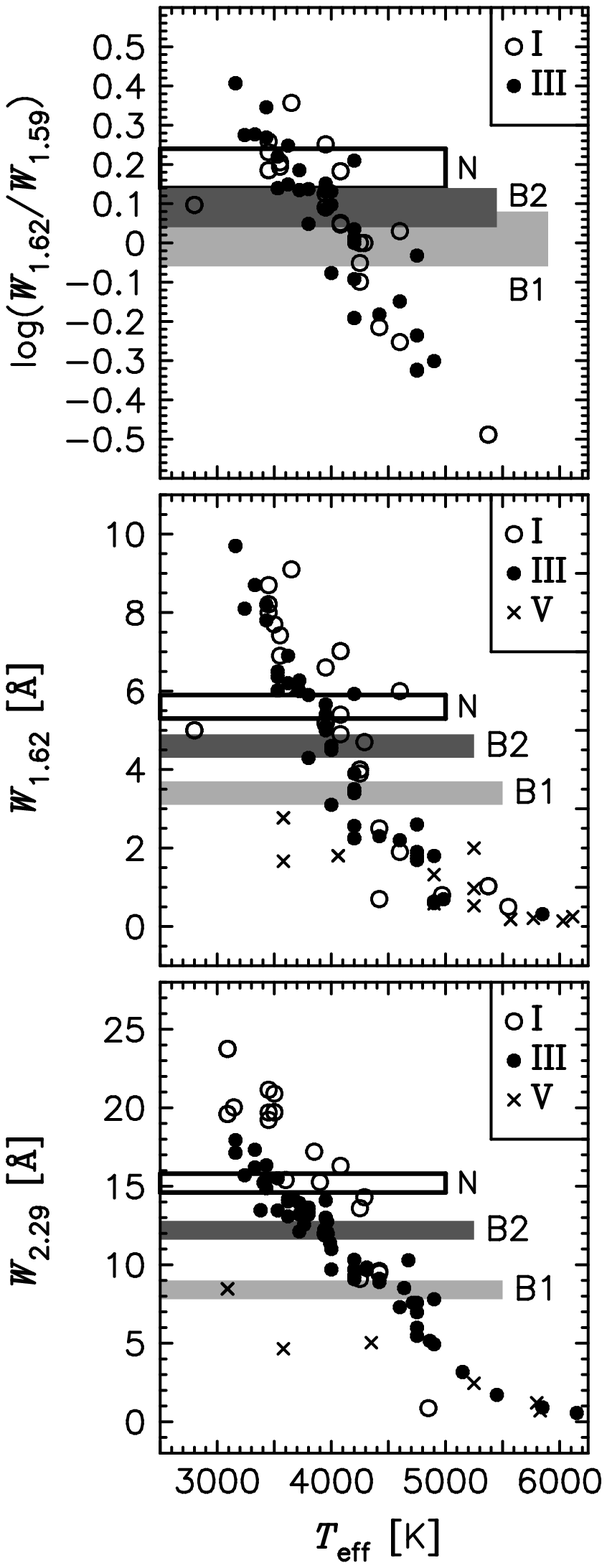}}
\vspace{-1.5cm}
\figcaption[]{}
\end{figure*}

\clearpage

\begin{figure*}[p]
\resizebox{\hsize}{!}{\includegraphics{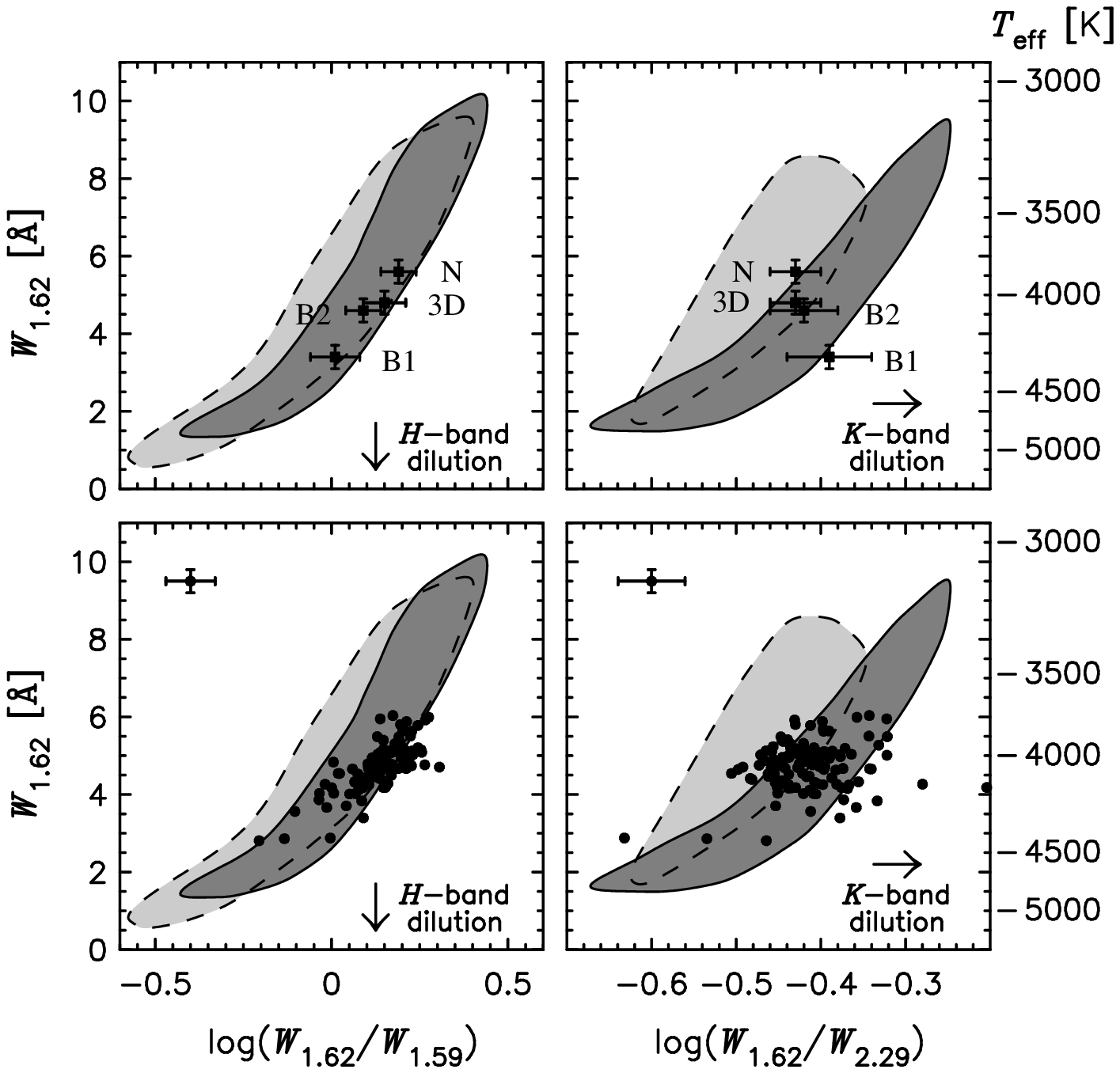}}
\vspace{-2.5cm}
\figcaption[]{}
\end{figure*}

\clearpage

\begin{figure*}[p]
\resizebox{\hsize}{!}{\includegraphics{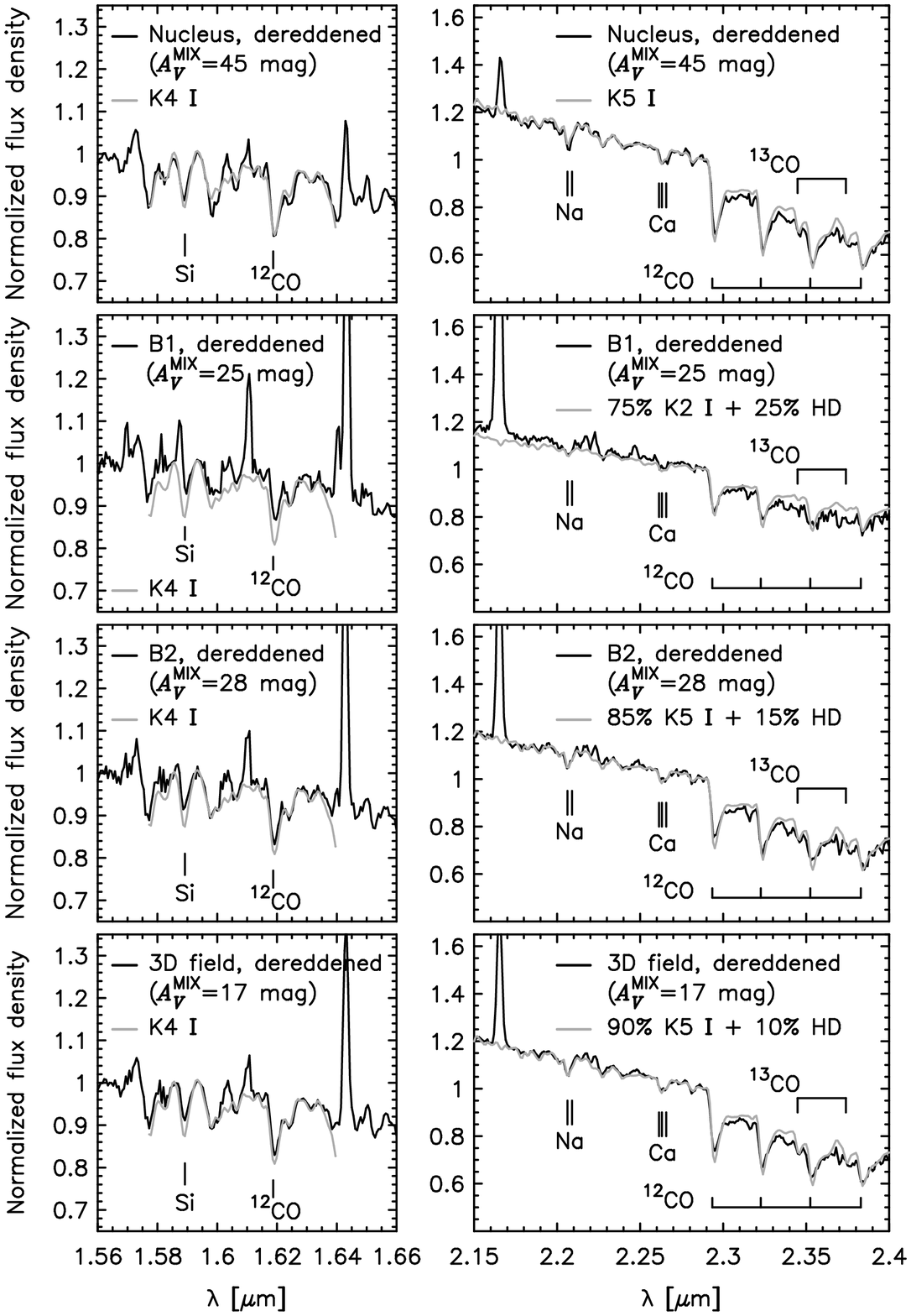}}
\vspace{-1.0cm}
\figcaption[]{}
\end{figure*}

\clearpage

\begin{figure*}[p]
\resizebox{\hsize}{!}{\includegraphics{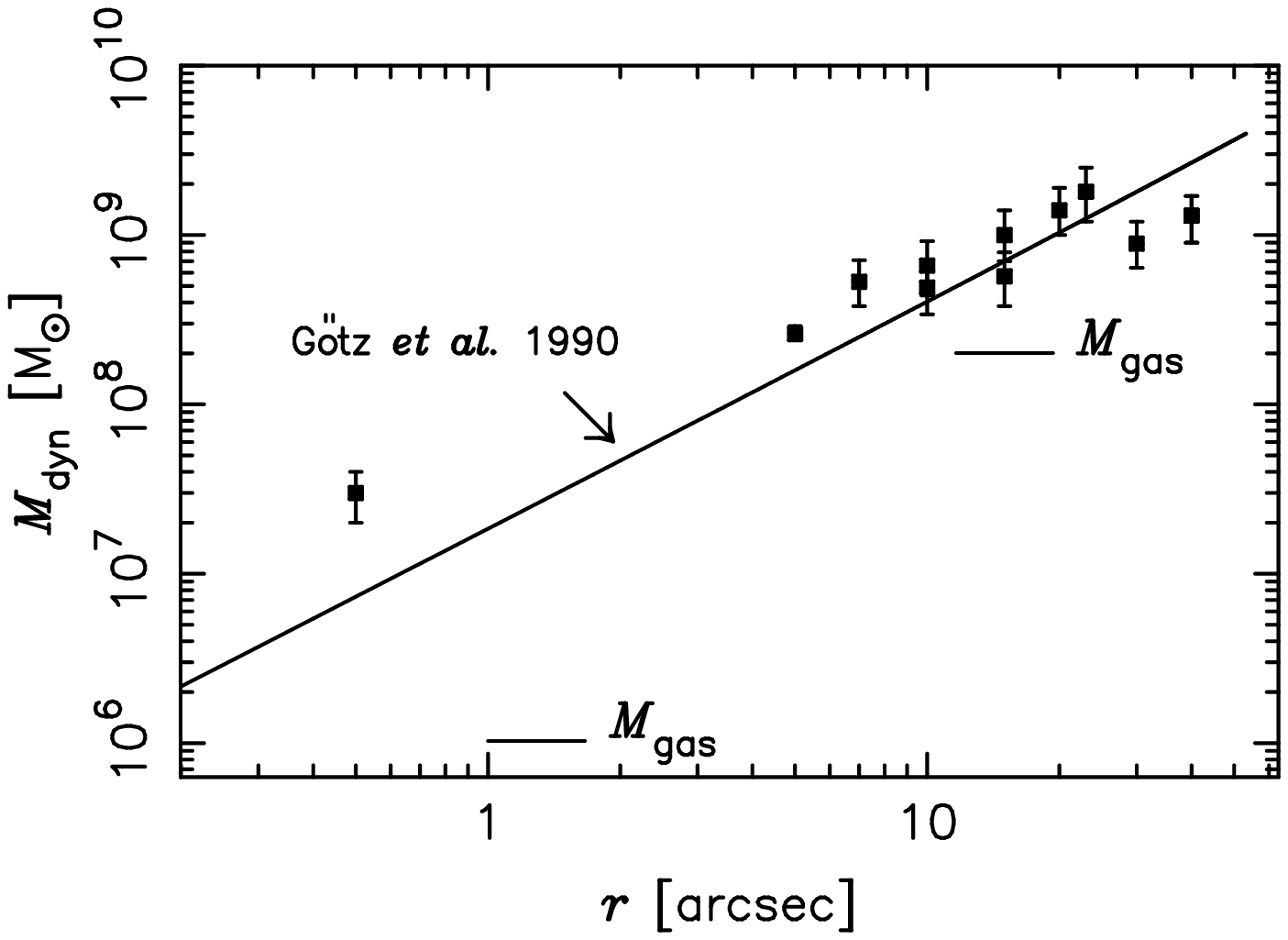}}
\vspace{-3.0cm}
\figcaption[]{}
\end{figure*}

\end{document}